\documentclass{article}
\usepackage [latin1]{inputenc}
\usepackage{authblk}
\usepackage{blindtext}
\usepackage[english]{babel}

\usepackage[letterpaper,top=2cm,bottom=2cm,left=1in,right=1in,marginparwidth=1.75cm]{geometry}

\usepackage{amsmath,amssymb, amsfonts,mathrsfs,bm}
\usepackage{graphicx}
\usepackage{subcaption, caption}

\usepackage[colorlinks=true, allcolors=blue]{hyperref}

\usepackage{siunitx}

\usepackage[normalem]{ulem}
\usepackage{hyperref}

\usepackage[]{natbib} %sort&compress
\newcommand{\sign}{\mathrm{sign}}

\title{How vortices enhance heat transfer from an oscillating plate}

\author[ ]{Xiaojia Wang}
\author[ ]{Silas Alben}
\affil[ ]{Department of Mathematics, University of Michigan, Ann Arbor, MI 48109, USA}
\affil[ ]{\textit {\{xiaojia@umich.edu ; alben@umich.edu\}}}

\begin{document}

\maketitle

\begin{abstract}
    Oscillations of a heated solid surface in an oncoming fluid flow can increase heat transfer from the solid to the fluid. Previous studies have investigated the resulting heat transfer enhancement for the case of a circular cylinder undergoing translational or rotational motions.  Another common geometry, the flat plate, has not been studied as thoroughly. The flat plate sheds larger and stronger vortices that are sensitive to the plate's direction of oscillation. To study the effect of these vortices on heat transfer enhancement, we compute the heat transfer from a flat plate with different orientations and oscillation directions in an oncoming flow with Reynolds number 100. We consider plates with fixed temperature and fixed heat flux, and find large heat transfer enhancement in both cases. We investigate the effects of the plate orientation angle and the plate oscillation direction, velocity, amplitude, and frequency, and find that the plate oscillation velocity and direction have the strongest effects on global heat transfer. The other parameters mainly affect the local heat transfer distributions through shed vorticity distributions. We also discuss the input power needed for the oscillating plate system and the resulting Pareto optimal cases.
\end{abstract}

\section{Introduction}\label{sec:intro}

Many studies have considered how oscillations of a heated solid surface in an oncoming fluid flow can increase heat transfer from the solid to the fluid. Motions of the boundary alter the flow patterns from the scale of the viscous boundary layer to the larger scale of the outer flow. Previous experimental and computational studies have quantified the resulting heat transfer enhancement for the case of a circular cylinder undergoing translational or rotational oscillations in a two-dimensional crossflow \citep{SL1978, FT2002, MA2018}.

After the circular cylinder, perhaps the next most common geometry is the flat plate. When a flat plate is held fixed in a flow, the flow transitions from steady flow to von K\'{a}rm\'{a}n vortex shedding at a lower Reynolds number (Re) than a circular cylinder \citep{TRRSH2014}, and undergoes a more complex series of wake transitions as Re increases. The flat plate also sheds stronger and larger vortices into the wake. For an oscillating body, one may anticipate similarly significant differences in the flow patterns and heat transfer.

%nuriev papers on flows with small and large amplitude oscillations

Heat transfer by oscillating thin plates has important applications in the biological world. Elephants' ears are thin surfaces that hold large networks of blood vessels, and ear flapping was speculated to make a large contribution to thermoregulation \citep{Sikes1971}. \citet{KAJ2014, KAJ2017} studied the enhancement of heat transfer due to oscillations in computational and experimental models of elephant ears, characterized vortical flow structures, and noted an additional heat transfer enhancement in a flexible model relative to a rigid one. Bats vary the blood flow to their wings for thermoregulation, and the constraints of this process affect bats' survival in cold climates \citep{RGVC2022, RC1951}. Heat transfer may have also played a role in the evolution of insect wings \citep{KK1985, Douglas1981}. Heat stress is a major factor in plant survival and reproductive success \citep{JWS2021}. Many studies have considered the effect of wind-induced oscillations on the cooling of plant leaves, both with and without water evaporation at the leaf surface \citep{MK1977, RP1993, Vogel2009, Schuepp}. 

In this work we focus on heat transfer under prescribed oscillations, but there is also a large body of work on heat transfer when the body motion results from fluid-structure interaction; \cite{MB2022} reviewed the application of immersed boundary methods in this area.

Recently, \cite{RT2020} studied the heat transfer enhancement due to sinusoidal oscillations of a heated flat plate in an oncoming flow. The plate was aligned with the oncoming flow direction and oscillated transversely, with a fixed Reynolds number of 100 based on the oncoming flow speed. They varied the plate oscillation amplitude and frequency and computed the Nusselt number averaged over a time interval [$T$, $2T$], where $T$ ranged from 8 to 16 oscillation periods. They identified the ``plunge velocity," i.e.~the product of dimensionless amplitude and frequency of the oscillation, as a key parameter for controlling the average Nusselt number. The average Nusselt number increased monotonically with the plunge velocity as it varied from 1/2$\pi$, close to the limit of no oscillation (a static plate), to 8/$\pi$, a rapid oscillation. When the oscillation amplitude and frequency were varied simultaneously while keeping their product (the plunge velocity) fixed, there was a smaller variation in the average Nusselt number.

In this paper we study the same system as \cite{RT2020} but we now consider the full range of plate orientations and oscillation directions relative to the oncoming flow velocity (see figure~\ref{fig:schematic}), instead of just a streamwise orientation and transverse oscillation direction. 
Like \cite{RT2020}, we find that the global (i.e. time- and space-averaged) heat transfer is strongly influenced by the ``oscillation velocity" (a generalization of the plunge velocity). 
%Nu rises roughly linearly with AoverU. 
 The next most important parameter is the plate oscillation direction $\alpha$. The global and local heat transfer typically improves as we move from in-plane to transverse oscillations, i.e. as $\alpha$ increases from $0^\circ$ to $90^\circ$.
%Nu rises almost monotonically as alpha increases from $0^\circ$ to $90^\circ$, and by a factor of about 2 at the AoverU = 0.4. 
The oncoming flow direction $\gamma$ and the oscillatory frequency (assuming fixed oscillation velocity) have noticeable but generally weaker effects on the global heat transfer. They strongly alter the flows and spatial distributions of heat transfer quantities, but when taking an average over the plate, these variations are reduced to a surprising extent.

The structure of the paper is as follows. In section \ref{sec:method} we present the governing equations and numerical method. Section \ref{sec:SteadyCases} presents the flow fields and the local heat transfer distributions in the benchmark cases of non-oscillating plates. Section \ref{sec:GlobalLocalHeatTransfer} shows the effect of vorticity patterns on global and local heat transfer for oscillating plates with different orientations and oscillation directions. In section \ref{sec:Power} we discuss the input power for this system, including the power to oscillate the plates and to drive the oncoming flow. Section \ref{sec:conclusion} presents the conclusions.

\section{Governing equations and numerical method}\label{sec:method}

\begin{figure}[ht]
    \centering
    \includegraphics[trim={0, 5cm, 4cm, 5cm}, clip, width = \textwidth]{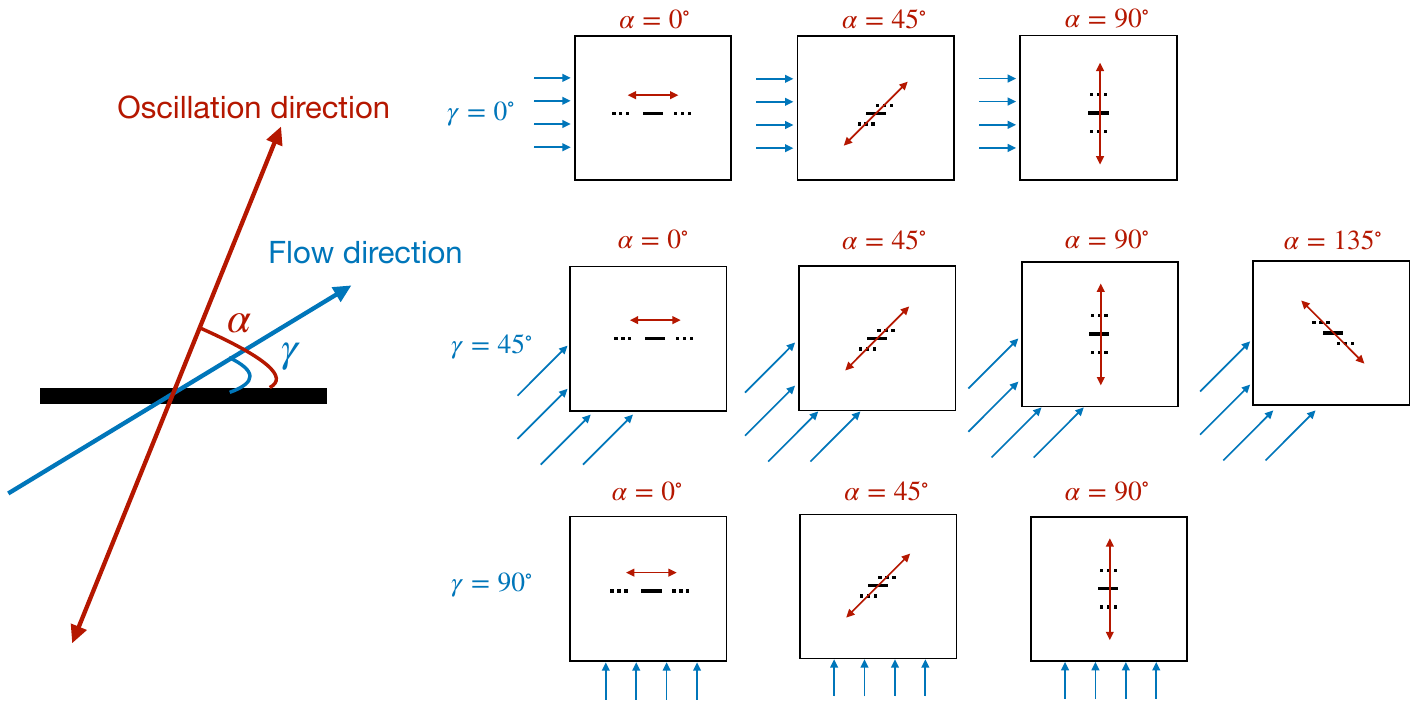}
    \caption{Schematic of the plate orientation and oscillation direction (angle $\alpha$) and the background flow direction (angle $\gamma$).}
    \label{fig:schematic}
\end{figure}

We consider a heated zero-thickness flat plate oscillating sinusoidally in a uniform oncoming flow at angle $\gamma$ relative to the plane of the plate (see figure~\ref{fig:schematic}). The plate oscillates in a direction with angle $\alpha$ relative to the plane of the plate, different from $\gamma$ in general.
We nondimensionalize all variables using the plate length $\ell^\star$ as the characteristic length scale and the plate oscillation period $1/f^\star$ as the characteristic time scale, with $f^\star$ the plate oscillation frequency. We use the fluid density $\rho_f^\star$ as the characteristic mass density.

We prescribe the plate's oscillation velocity, $\boldsymbol{U}_b(t) = (U_b(t), V_b(t))$ as follows:
\begin{equation}
        (U_b,V_b) = \left(\frac{U_b^\star}{f^\star \ell^\star},\frac{V_b^\star}{f^\star \ell^\star}\right) = 2 \pi A\sin(2\pi t)e^{-(t/t_0)^2} \left(\cos(\alpha), \sin(\alpha)\right),\label{eq:BodyVel}
\end{equation}
where the starred variables and parameters are dimensional and the unstarred variables and parameters are dimensionless. The latter include $A$, the plate's oscillation amplitude (relative to its length). The exponential term makes the simulated flow start smoothly from rest. We use $t_0=0.2$ for all the simulations.

The governing equations for an incompressible viscous flow in the (noninertial) frame of reference attached to the moving plate are \citep{LSB2002, alben2021a}
\begin{subequations}\label{eq:flow}
    \begin{equation}
        \frac{\partial \bm{u}}{\partial t} + \bm{u\cdot \nabla u} = -\bm{\nabla} p + \frac{1}{Re_f}\nabla^2 \bm{u} - \frac{\mathrm{d}\bm{U}_b}{\mathrm{d}t},
    \end{equation}
    \begin{equation}
        \bm{\nabla\cdot u} = 0,
    \end{equation}
\end{subequations}
with $\bm{u}(x, y, t)=(u(x, y, t), v(x, y, t))$ and $p(x, y, t)$ the flow velocity and pressure, respectively. $Re_f$ is the frequency-based Reynolds number, included in the list of important dimensionless parameters for this problem {\color{red} }:
\begin{equation}\label{eq:param}
    Re_f = \frac{f^\star {\ell^\star}^2}{\nu^\star}, \qquad \bm{U}_{\infty}=\frac{\bm{U}_{\infty}^\star}{f^\star\ell^\star}, \qquad Re_U = \frac{|\bm{U}_{\infty}^\star|\ell^\star}{\nu^\star}=Re_f\times |\bm{U}_{\infty}| ,\qquad A = \frac{A^\star}{\ell^\star}, \qquad \frac{A}{|\bm{U}_{\infty}|} =\frac{f^\star A^\star}{|\bm{U}_{\infty}^\star|} .
\end{equation}
$\bm{U}_{\infty} = (U_{\infty}, V_{\infty}) = (|\bm{U}_{\infty}|\cos(\gamma), |\bm{U}_{\infty}|\sin(\gamma))$ is the steady far-field oncoming flow velocity. Besides the frequency-based Reynolds number $Re_f$, we can also define a Reynolds number based on $\bm{U}_{\infty}$, which is $Re_U$ in equation~\eqref{eq:param}, also given by the product of $Re_f$ and $|\bm{U}_{\infty}|$. In this study, we vary both $Re_f$ and $\bm{U}_{\infty}$ but fix $Re_U=100$. The physical interpretation is that the plate length, fluid viscosity, and background flow speed are considered fixed, but the plate oscillation frequency varies (as do the oscillation amplitude and direction, and the plate orientation). Another important parameter is $A/|\bm{U}_{\infty}|$, which we term the ``oscillation velocity." It is proportional to the ratio between the plate's oscillation velocity amplitude, $2\pi f^\star A^\star$, and the oncoming flow velocity magnitude, $|\bm{U}^\star_{\infty}|$. For transverse oscillation ($\alpha = 90^\circ$) in an in-plane flow ($\gamma = 0^\circ$), the ``plunge velocity" \citep{RT2020} or ``Strouhal number" (twice the oscillation velocity) \citep{TTG1991, TTG1993} are often used, particularly in studies of locomotion and fluid-structure interaction. Because we consider a wide range of oscillation and flow directions, we use ``oscillation velocity" for this class of motions.
\cite{RT2020} found that the rate of heat transfer is increased at higher $A/|\bm{U}_{\infty}|$. In this study, we mainly study flows with $A/|\bm{U}_{\infty}|=0.2$ and $0.3$.

\begin{figure}[ht]
    \centering
    \includegraphics[width=0.45\linewidth]{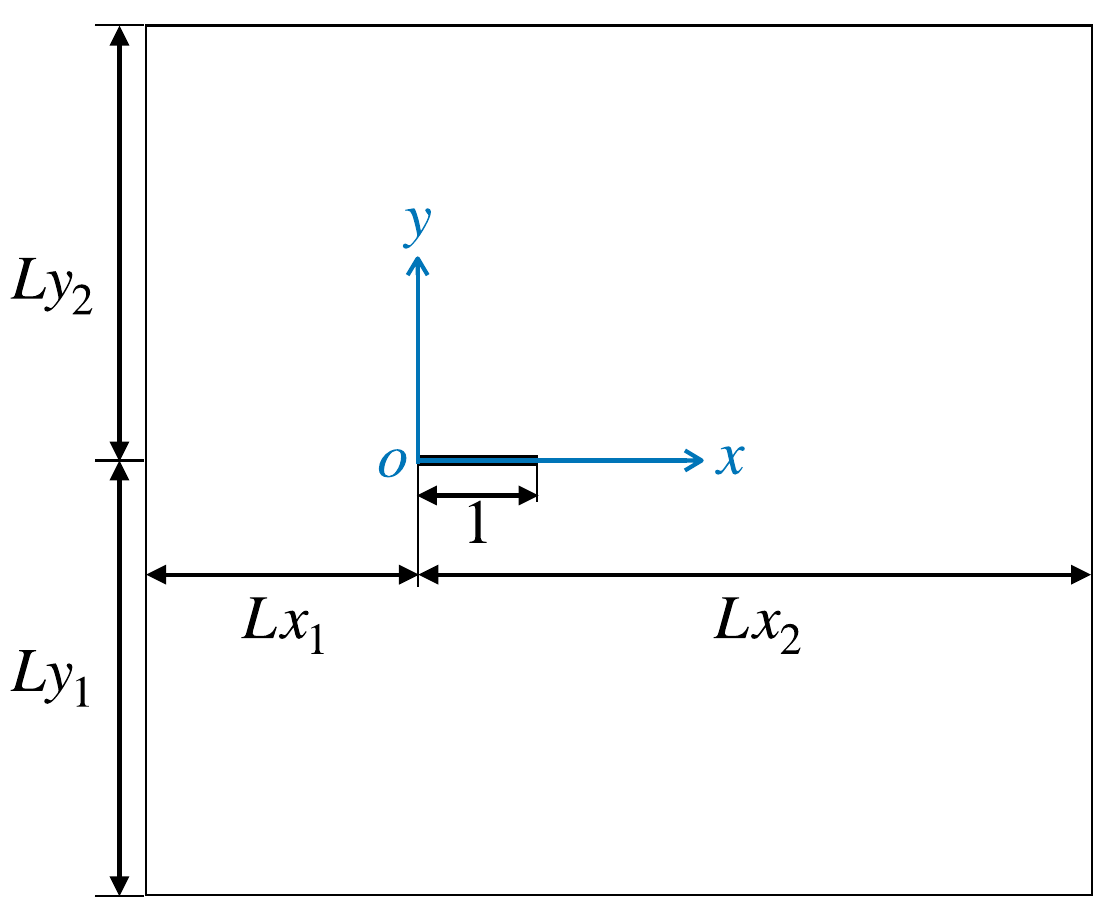}
    \caption{Schematic of the computational domain. All the parameters shown are dimensionless.}
    \label{fig:ComputeDomain}
\end{figure}

The temperature field $T(x,y,t)$ is governed by the unsteady advection-diffusion equation:
\begin{equation}\label{eq:temp}
    \frac{\partial T}{\partial t} + \bm{u}\cdot\bm{\nabla}T = \frac{1}{Re_f Pr}\nabla^2 T,
\end{equation}
where $Pr = \nu^\star/\kappa^\star$ is the Prandtl number, with $\kappa^\star$ the thermal diffusivity. In this study, we take air as the fluid, with $Pr=0.7$. As shown in figure~\ref{fig:ComputeDomain}, the heated plate is positioned at $0\leq x \leq 1, y=0$. For the temperature boundary conditions on the plate, we consider two cases:
\begin{align}
\mbox{\textbf{Fixed temperature:}} &\quad T|_{y=0^+} = T|_{y=0^-} = 1 \quad \text{for}\ 0\leq x \leq 1. \label{eq:fixtemp-bc} \\
\mbox{\textbf{Fixed heat flux:}} &\quad -\left.\frac{\partial T}{\partial y}\right|_{y=0^+}+\left.\frac{\partial T}{\partial y}\right|_{y=0^-} = 2, \quad T|_{y=0^+} = T|_{y=0^-}, \quad \text{for}\ 0\leq x \leq 1.\label{eq:fixflux-bc}
\end{align}
In (\ref{eq:fixtemp-bc}) the plate temperature is fixed, while in (\ref{eq:fixflux-bc}) the heat flux per unit plate length is fixed. In the latter case we have also equated the temperatures of the top and bottom surfaces of the plate, as we assume the plate is a thin conducting material. If instead of (\ref{eq:fixflux-bc}) we used $-\partial_y T(y=0^+) = \partial_y T(y=0^-) = 1$, the top and bottom surfaces of the plate would have different temperatures in general.

\begin{table}[h]
    \centering
    \begin{tabular}{|m{2.6em}|m{13.5em}|m{9.3em}|m{13.5em}|}
    \hline
         &  $\gamma = 0^\circ$ & $\gamma = 45^\circ$ & $\gamma = 90^\circ$ \\
    \hline
    $Lx_1$ & 8 & 12 & 6.5\\
    $Lx_2$ & 10 & 14 & 7.5\\
    $Ly_1$ & 5 & 12 & 12\\
    $Ly_2$ & 5 & 14 & 14\\
    $m$ & 432 & 420 & 336 \\
    $n$ & 240 & 416 & 624 \\
    \hline 
    left & \shortstack{$u = U_{\infty}-U_b$, $v = -V_b$\\ $T=0$} & \shortstack{$u = U_{\infty}-U_b$, \\$v = V_{\infty}-V_b$, $T=0$} & \shortstack{$u=-U_b$, $\displaystyle \frac{\partial v}{\partial x} =0$ \\ $\displaystyle b_1T + b_2\left( \frac{\partial T}{\partial t} -\bm{U_b\cdot\nabla}T \right) = 0 $} \\
    \hline 
    bottom & \shortstack{$\displaystyle\frac{\partial u}{\partial y}=0$, $v=-V_b$ \\ $\displaystyle a_1T + a_2\left( \frac{\partial T}{\partial t} -\bm{U_b\cdot\nabla}T \right) = 0 $ } & \shortstack{$u = U_{\infty}-U_b$, \\ $v = V_{\infty}-V_b$, $T=0$} & \shortstack{$u=-U_b$, $v = V_{\infty}-V_b$ \\ $T = 0$ } \\
    \hline
    right & \shortstack{$\displaystyle \frac{\partial u}{\partial x} = 0$, $\displaystyle \frac{\partial v}{\partial x} = 0$ \\  $\displaystyle \frac{\partial T}{\partial x} = 0$} & \shortstack{$\displaystyle \frac{\partial u}{\partial x} = 0$, $\displaystyle \frac{\partial v}{\partial x} = 0$ \\  $\displaystyle \frac{\partial T}{\partial x} = 0$} & \shortstack{$u=-U_b$, $\displaystyle \frac{\partial v}{\partial x} =0$ \\ $\displaystyle b_2 T + b_1\left( \frac{\partial T}{\partial t} -\bm{U_b\cdot\nabla}T \right) = 0 $}\\
    \hline
    top & \shortstack{$\displaystyle\frac{\partial u}{\partial y}=0$, $v=-V_b$ \\ $\displaystyle a_2T + a_1\left( \frac{\partial T}{\partial t} -\bm{U_b\cdot\nabla}T \right) = 0 $ } & \shortstack{$\displaystyle \frac{\partial u}{\partial y} = 0$, $\displaystyle \frac{\partial v}{\partial y} = 0$ \\  $\displaystyle \frac{\partial T}{\partial y} = 0$} &  \shortstack{$\displaystyle \frac{\partial u}{\partial y} = 0$, $\displaystyle \frac{\partial v}{\partial y} = 0$ \\  $\displaystyle \frac{\partial T}{\partial y} = 0$} \\
    \hline
    \end{tabular}
\caption{The dimensions of the computational domain, the number of grid cells along $x$ and $y$ (denoted $m$ and $n$, respectively), and the boundary conditions, with $a_1 = (1+\sign(-V_b))/2$, $a_2 = (1-\sign(-V_b))/2$,  $b_1 = (1+\sign(-U_b))/2$, and $b_2 = (1-\sign(-U_b))/2$. }
    \label{tab:domainBC}
\end{table}

\begin{figure}[h]
    \centering
    \includegraphics[width=0.7\linewidth]{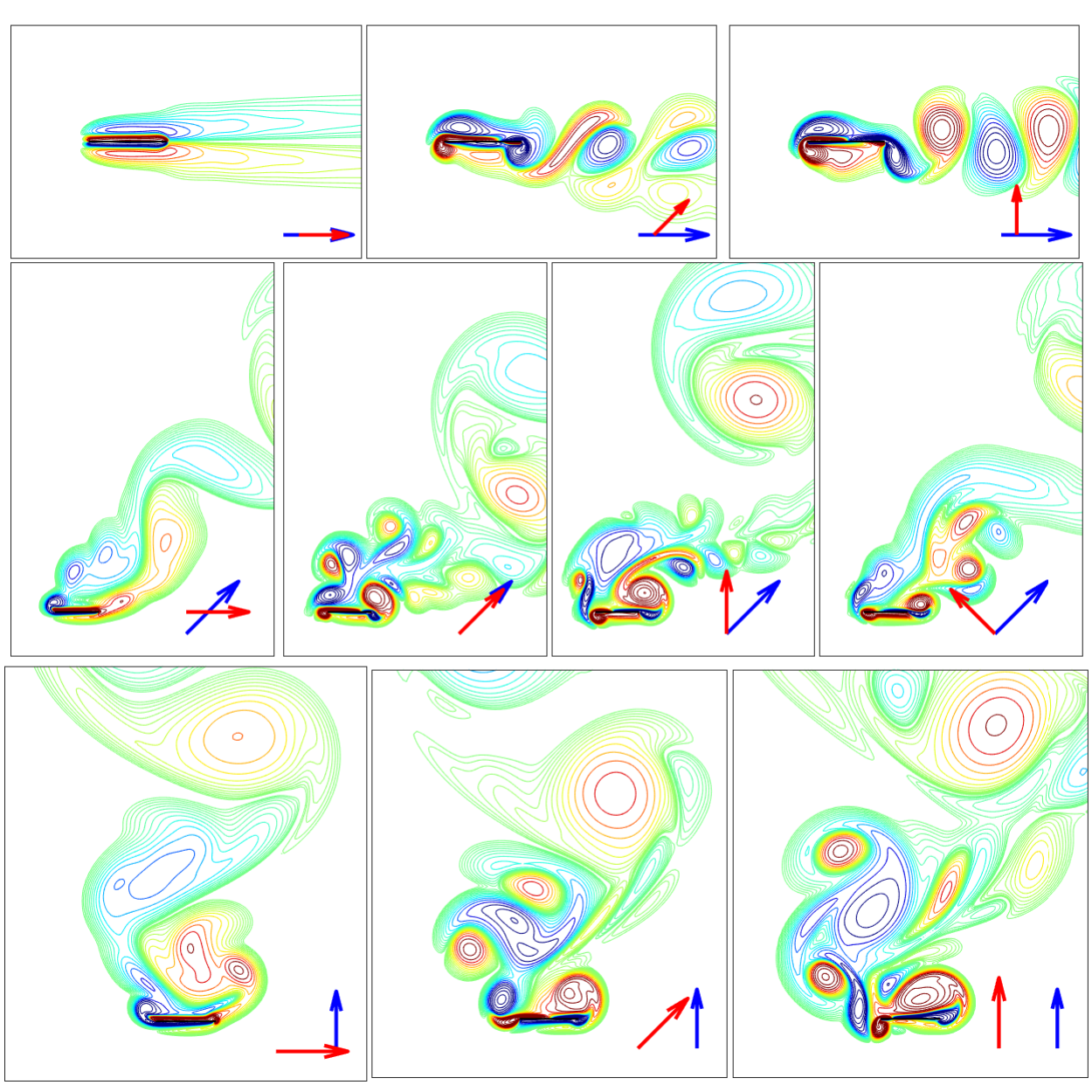}
    \caption{Examples of the vorticity fields for different plate orientations and oscillation directions. The rows from top to bottom have $\gamma = 0^\circ$, $45^\circ$, and $90^\circ$, respectively. The blue arrow indicates the oncoming flow direction, while the red arrow indicates instantaneous plate oscillation direction given by $\alpha$. Within each row, $\alpha$ increases in increments of $45^\circ$ moving rightward, starting at $0^\circ$ in the leftmost panel. All of the examples have $Re_f=100$, $Re_U=100$, $A=0.2$, and $A/|\bm{U}_{\infty}|=0.2$.}
    \label{fig:ExampleVorticity}
\end{figure}

We solve equation~\eqref{eq:flow} as a fully coupled system for $\bm{u}$ and $p$, using essentially the same method as in \cite{alben2021a}, a second-order finite difference method on the MAC grid. The grid spacing is larger toward the boundary of the computational domain and smaller near the plate, where the vorticity is large. The distances from the plate to the outer boundary (given by $L_{x1}$, $L_{x2}$, $L_{y1}$, $L_{y2}$ in figure~\ref{fig:ComputeDomain}) and the outer boundary conditions are chosen differently for each value of $\gamma$ to limit the influence of the outer boundary. We use Dirichlet boundary conditions at the inflow sides and Neumann boundary conditions at the outflow sides. For sides parallel to the mean oncoming flow, we set the normal velocity to the far-field value, and impose zero shear, to avoid vorticity generation \citep{SMB2009, TTW1994}. These boundary conditions are listed in table~\ref{tab:domainBC}. A comparison of results with different domain and mesh sizes is given in appendix~\ref{app:GridIndependence}.  On the plate surface, we apply no-slip boundary conditions. At $t=0$, we assume steady uniform flow with $\bm{u} = \bm{U}_{\infty}$ (at points off of the plate), except in one case, $\alpha = \gamma = 90^\circ$, where we use an asymmetric initial flow described in section \ref{subsubsec:gamma90}. In figure~\ref{fig:ExampleVorticity} we provide examples of typical vorticity fields, with $\gamma$ increasing from the top row to the bottom row, and $\alpha$ increasing from left to right within each row.
The overall orientation of the wake relative to the body is set by $\gamma$; both $\gamma$ and $\alpha$ affect the vorticity patterns around the body and particularly within the wake. A main focus of the current study is the effect of these vorticity patterns on heat transfer.

After computing the flow velocity $\bm{u}$, we solve for the temperature using a second-order discretization of equation~\eqref{eq:temp} at the same grid points as $p$ (the centers of the MAC grid cells). Either equation~(\ref{eq:fixtemp-bc}) or~(\ref{eq:fixflux-bc}) is applied in one-sided finite-difference formulas for the $y$-derivatives in equation~\eqref{eq:temp} on either side of the plate, depending on whether we fix the plate temperature or the heat flux. As with $\bm{u}$, we apply different outer boundary conditions for $T$ depending on $\gamma$, i.e. which sides of the computational domain are the inflow and outflow boundaries. These are listed in table~\ref{tab:domainBC}. We apply $T=0$ at inflow boundaries and $\partial T/\partial n = 0$ at outflow boundaries, with $\partial/\partial n$ the normal derivative. For sides parallel to the mean oncoming flow, we switch between Dirichlet and outflow boundary conditions for $T$ depending on whether the flow is inward or outward at a given time, as shown in table~\ref{tab:domainBC}. At $t = 0$ the temperature is set to zero at all points in the fluid.

The main heat transfer performance metric in this study is the Nusselt number, the local heat flux relative to the difference between the plate temperature $T_{\text{plate}}$ and the far-field fluid temperature $T_\infty$, set to zero here \citep{Incropera-Book}: 
\begin{align}
Nu = \frac{\partial T/\partial y}{T_{\text{plate}} - T_\infty}. \label{Nudef}   
\end{align}
A larger rate of heat transfer corresponds to larger $Nu$.
With the fixed-plate-temperature boundary condition, the denominator of (\ref{Nudef}) is fixed at unity while the numerator varies along the plate. 
We consider two local and one global Nusselt numbers in this case:
\begin{equation}\label{eq:Nu}
    Nu_{\text{top}}(x, t) = -\left.\frac{\partial T}{\partial y}\right|_{y=0^+}, \quad Nu_{\text{bot}}(x,t) = \left.\frac{\partial T}{\partial y}\right|_{y=0^-}, \quad \overline{Nu}(t) = \displaystyle\int_0^1 -\left.\frac{\partial T}{\partial y}\right|_{y=0^+} + \left.\frac{\partial T}{\partial y}\right|_{y=0^-}\, \mathrm{d}x.
\end{equation}
The first two Nusselt numbers are the local heat fluxes from the plate's top and bottom surfaces respectively, while the third is the global Nusselt number, the total heat flux from both plate surfaces. 

With the fixed-heat-flux boundary condition, the numerator of (\ref{Nudef}) summed over both sides of the plate is fixed at each $x$ while $T_{\text{plate}}$ in the denominator varies with $x$. In this case $Nu$ is inversely proportional to $T_{\text{plate}}$, so lower plate temperature corresponds to better heat transfer performance. In this case, instead of $Nu$, we use the plate temperature distribution $T_{\text{plate}}(x, t)$, the maximum plate temperature $T_{\max}(t)$ and the spatially-averaged plate temperature $T_{\text{avg}}(t)$ as performance metrics:
\begin{equation}\label{eq:T}
    T_{\text{plate}}(x, t) = T|_{y=0}, \quad T_{\max}(t) = \max(T_{\text{plate}}), \quad T_{\text{avg}}(t) = \displaystyle\int_0^1 T_{\text{plate}}(x, t)\, \mathrm{d}x.
\end{equation}
In the following sections, we will also 
examine the time averages of quantities in equations~\eqref{eq:Nu} and \eqref{eq:T}, denoted as $\langle \cdot \rangle$. In section~\ref{subsec:GlobalHeatTransfer} we explain how we choose appropriate time spans for the time averages.
%
%
%
%start here
%
%
\section{Non-oscillating plate in steady oncoming flows}
\label{sec:SteadyCases}

\begin{figure}[ht]
    \centering
    \begin{subfigure}[b]{0.49\textwidth}
        \caption{$\gamma = 0^\circ$}
        \includegraphics[width =\textwidth]{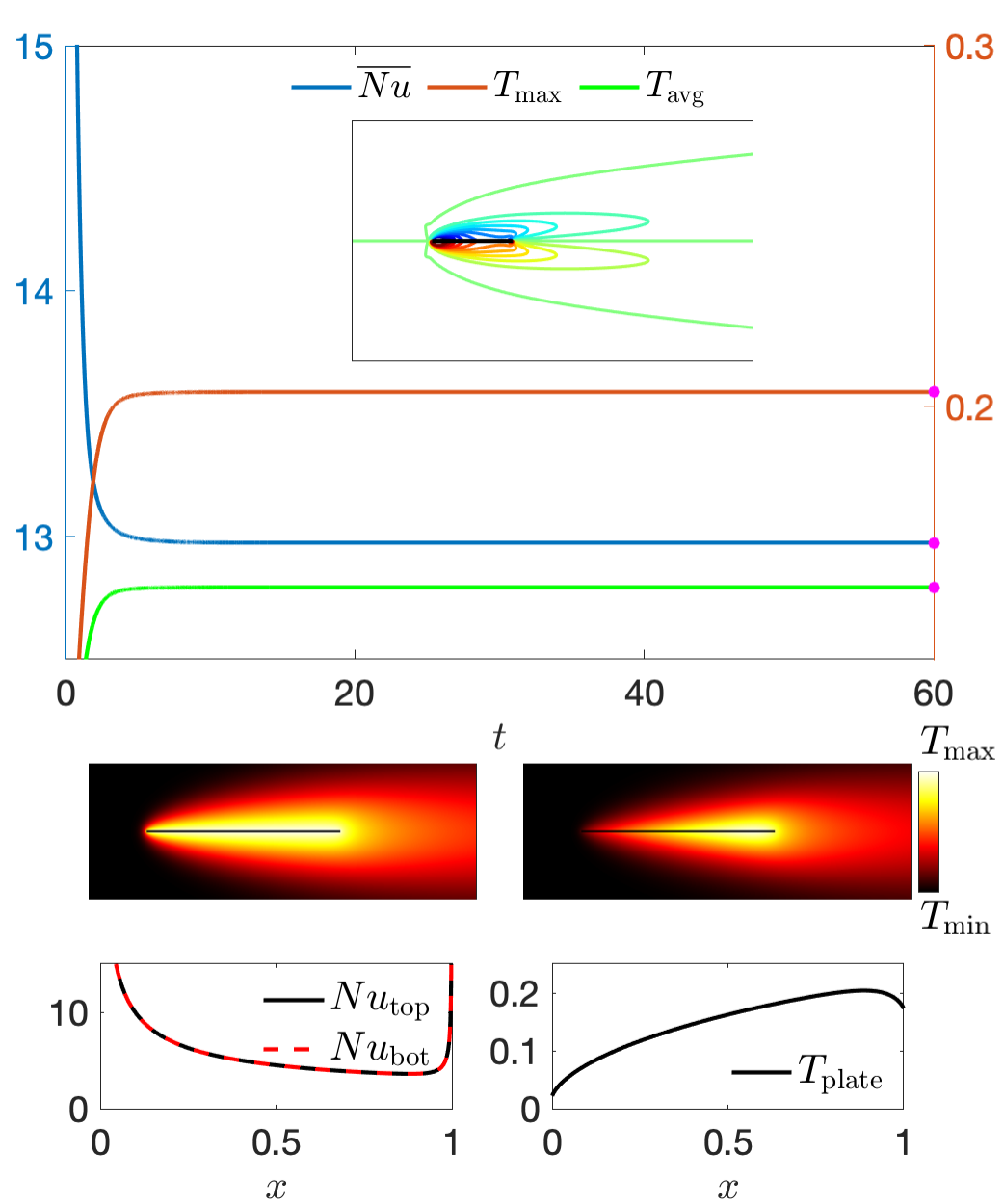}
        \label{fig:steady-0}
    \end{subfigure}
    \begin{subfigure}[b]{0.49\textwidth}
        \caption{$\gamma = 45^\circ$}
        \includegraphics[width =\textwidth]{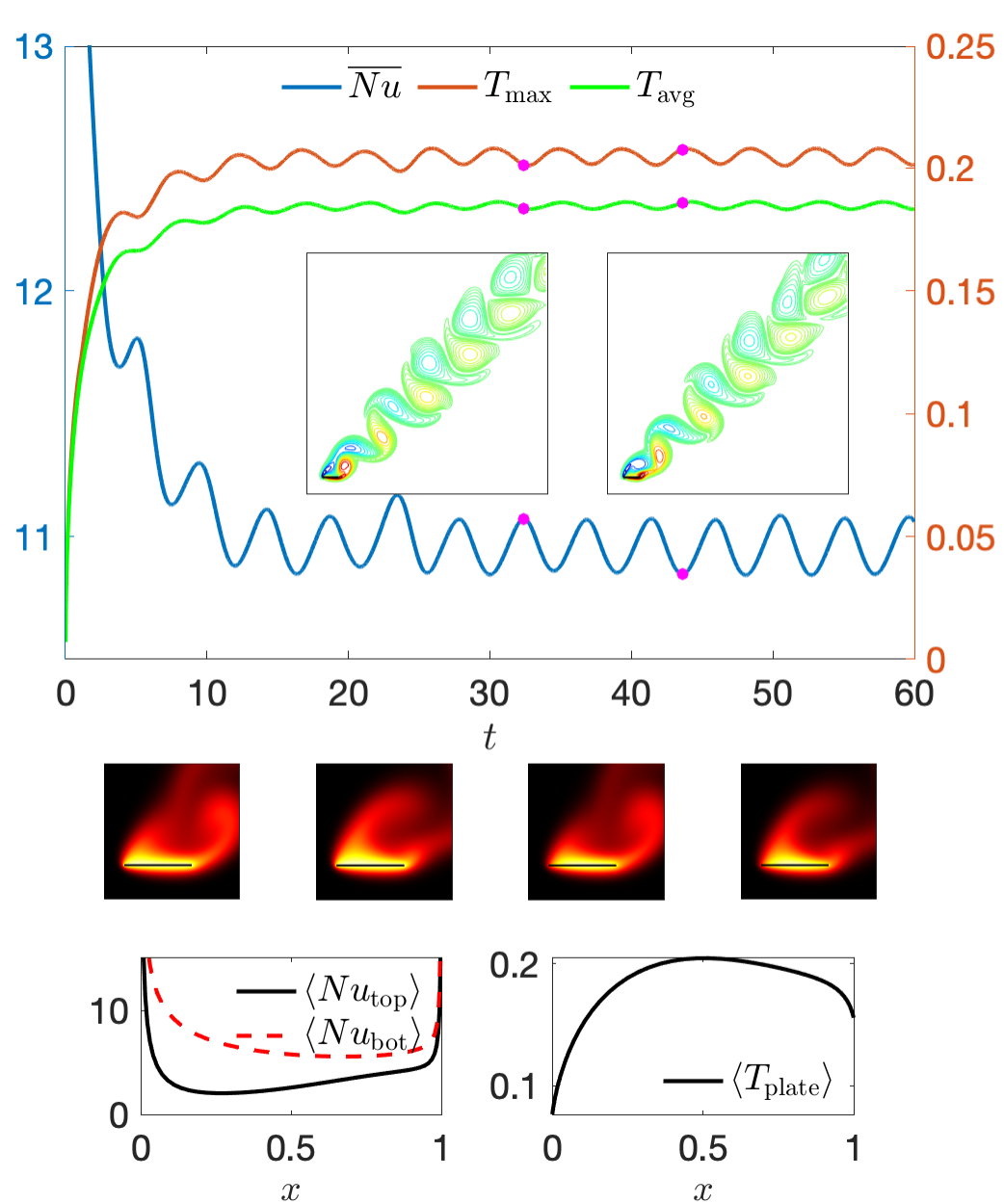}
        \label{fig:steady-45}
    \end{subfigure}
    \caption{Heat transfer from a non-oscillating plate in steady oncoming flows angled at $\gamma = 0^\circ$ (panel (a)), and $\gamma = 45^\circ$ (panel (b)). The time series of $\overline{Nu}$, $T_{\max}$, and $T_{\text{avg}}$ with selected snapshots of the vorticity field and temperature field (at times marked with pink dots) are shown for each case. In panel (a), the temperature field snapshots show the isothermal-plate and fixed-flux cases, on the left and right respectively. In panel (b) the same quantities are shown but with four temperature snapshots, the leftmost two for the isothermal-plate case, and the rightmost two for the fixed-flux case. At the bottom of each panel, the local $Nu$ and plate temperatures are shown, steady in panel (a) and averaged over the last four periods in panel (b).}
    \label{fig:steady0and45}
\end{figure}
\begin{figure}[ht]
    % \ContinuedFloat
    % \begin{subfigure}[b]{0.99\textwidth}
    %     \centering
    %     \caption{$\gamma = 90^\circ$}
    \centering
    \includegraphics[width =0.8\textwidth]{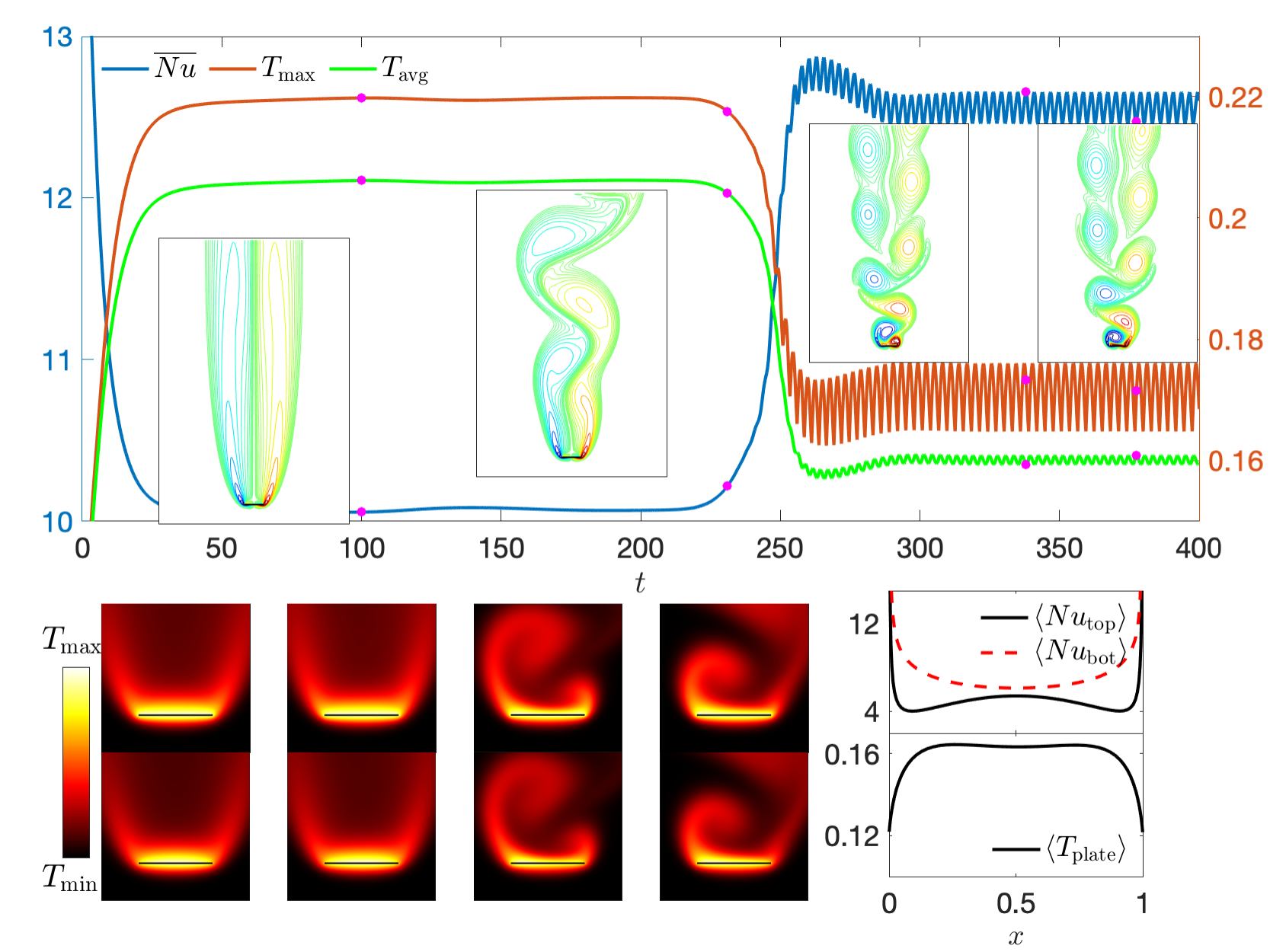}
    % \end{subfigure}
    \caption{Heat transfer from a non-oscillating plate in a steady oncoming flow angled at $\gamma = 90^\circ$. The time series of $\overline{Nu}$, $T_{\max}$, and $T_{\text{avg}}$ with selected snapshots of the vorticity field and temperature field (at times marked with pink dots) are shown at the bottom left. The first row of the temperature snapshots is for the isothermal case and the second row is for the fixed-flux case. The local $Nu$ and plate temperature, averaged over the last four periods, are shown at the bottom right.}
    \label{fig:steady-90}
\end{figure}

In figure~\ref{fig:steady0and45} and \ref{fig:steady-90}, we show results for a non-oscillating plate in steady oncoming flows, which can be taken as a reference state for investigating how plate oscillations influence heat transfer. As in the cases with oscillating plates, the Reynolds number of the oncoming flow is fixed at $Re_U=100$ for different plate orientations. However, since the plate is not oscillating, $t$ is nondimensionalized using $\ell^\star/|\bm{U}^\star_{\infty}|$ instead of $1/f^\star$. 

In figure~\ref{fig:steady-0} the plate is aligned with the flow ($\gamma =0^\circ$). We have rapid convergence to a steady flow with symmetrical boundary layers along both sides of the plate. The temperature field is also up-down symmetric, and the thickness of the thermal boundary layer increases toward the trailing edge (the right edge). For the isothermal plate (left sides of the bottom rows of panel (a)), the local Nusselt number is equal on the top and bottom surfaces, and larger near the leading edge than the trailing edge. When instead the heat flux from the plate is fixed (right sides of the bottom rows of panel (a)), the hottest spot is located slightly upstream of the trailing edge. 

In figure~\ref{fig:steady-45}, with an oblique oncoming flow relative to the plate ($\gamma=45^\circ$), we observe a periodic flow with a von K\'arm\'an vortex street wake instead of the steady flow at $\gamma =0^\circ$. During one complete period, a negative vortex (in blue) and a positive vortex (in red) form and shed at the left and right edges of the top wall alternately. Because the flow is oblique, the negative vortex covers a larger portion of the top wall, shown in the rightmost vorticity snapshot in panel (b). The rotation of the negative vortex sweeps hot fluid along the top surface from the right edge toward the left edge, which makes the thermal boundary layer thicker near the left edge. Therefore, the time averaged local Nusselt number of the top wall $\langle Nu_{\text{top}} \rangle$ (black solid line), which is roughly the inverse of the temperature boundary layer thickness, has a minimum near the left edge. There are no vortices adjacent to the bottom wall, and the thermal boundary layer thickness is more uniform along the bottom wall.  The thermal boundary layer is much thinner along the bottom wall, so $\langle Nu_{\text{bot}} \rangle > \langle Nu_{\text{top}} \rangle$. The temperature fields in the case of fixed heat flux from the plate (two rightmost snapshots) are qualitatively similar to those for the isothermal plate (two leftmost snapshots). The time averaged plate temperature $\langle T_{\text{plate}} \rangle$ rises sharply from the left edge to a maximum near the center, and decreases more gradually on the right side.  

For the vertical oncoming flow ($\gamma=90^\circ$) in figure~\ref{fig:steady-90}, the vortex wake instead has a long transition from a symmetric initial state to the time-periodic alternating-vortex-shedding state. Initially, there is a pair of recirculation bubbles behind the plate, and the corresponding temperature field is symmetric about the middle of the plate, as shown in the first column of the temperature snapshots. The thermal boundary layer thickness is fairly uniform and thinner along the bottom wall as before. Small asymmetric perturbations (from numerical round-off error) grow exponentially during this time and reach O(1) amplitude at $t \approx 200$. During this instability, which has been studied more extensively for circular cylinders than for flat plates (\cite{TRRSH2014}), the long recirculation bubbles become wavy and shed vortices, as shown in the second vorticity field snapshot. However, the corresponding temperature field close to the plate does not change much from the first stage, as shown in the second column of the temperature snapshots, because the flow is still fairly symmetric close to the plate. At $t \approx 300$, the system reaches a periodic state, with alternating shedding of positive and negative vortices from the edges, giving rise to a von K\'{a}rm\'{a}n vortex street. The near-wake flow and temperature fields (third and the fourth columns of the temperature snapshots) are now very asymmetric. The top panel shows jumps in the global Nusselt number $\overline{Nu}$, $T_{\text{avg}}$ and $T_{\max}$ during the  transition at $t \approx 250$. Similarly to the case of $\gamma=45^\circ$, the vortices behind the plate make the thermal boundary layer thinner along the top wall, increasing heat transfer, though $\langle Nu_{\text{bot}} \rangle$ is still larger than $\langle Nu_{\text{top}} \rangle$. The plots of $\langle Nu_{\text{top}} \rangle$, $\langle Nu_{\text{bot}} \rangle$, and $\langle T_{\text{plate}} \rangle$ are symmetric about the plate center when $\gamma=90^\circ$, reflecting the symmetry of the plate with respect to the oncoming flow. The vortices at the two edges take the same time to develop and have the same strength. The vortices alternately sweep hot fluid toward the two edges, so the thermal boundary layer is thinnest in the middle of the plate, where $\langle Nu_{\text{top}} \rangle$ is largest and $\langle T_{\text{plate}} \rangle$ is smallest. \cite{TRRSH2014} studied the transient wake development for elliptical cylinders ranging from a flat plate to a circular cylinder at a Reynolds number of $150$. Using a much larger domain than we use here, they reported not only the von K\'{a}rm\'{a}n street behind the plate but also secondary vortex shedding further downstream. We believe that the secondary vortex shedding has negligible influence on heat transfer as the near wake reported by \cite{TRRSH2014} is very similar to what we show in figure~\ref{fig:steady-90} and very close to periodic, despite the existence of the secondary vortex shedding. 

Our discussion so far has shown that the local heat transfer is quite different for $\gamma=0^\circ, 45^\circ$, and $90^\circ$. The time-averaged global heat transfer quantities $\langle\overline{Nu} \rangle$ and $\langle T_{\text{avg}} \rangle$ can be inferred from the graphs in figures 
\ref{fig:steady0and45} and \ref{fig:steady-90} at the latest times (in the final periodic steady state for $\gamma=90^\circ$, after the symmetry breaking). For the isothermal plate, $\gamma=0^\circ$ has the best heat transfer---i.e., the largest
$\langle\overline{Nu} \rangle$, 13.0, and the smallest $\langle T_{\text{avg}} \rangle$, 0.15. Slightly worse is $\gamma=90^\circ$, with $\langle\overline{Nu} \rangle = 12.6$ and $\langle T_{\text{avg}} \rangle=0.16$. Third best is $\gamma=45^\circ$ with $\langle\overline{Nu} \rangle = 11.0$ and $\langle T_{\text{avg}} \rangle=0.19$. At $\gamma=45^\circ$, local heat transfer is decreased particularly at the left edge of the top wall, as shown by $\langle Nu_{\text{top}} \rangle$ and the temperature snapshots in figure~\ref{fig:steady-45}. Considering the possibility of device failure above a temperature threshold \citep{WY2021-book},
%\textcolor{red}{cite Advances in Heat Transfer and Thermal Engineering, Chuang Wen, Yuying Yan, 2021},
it is also important to consider the maximum temperature over space and time, $\max(T_{\max})$. By this measure $\gamma=90^\circ$ is best with $\max(T_{\max}) = 0.18$. The other cases are close, with $\max(T_{\max})$ 0.20 and 0.21 for $\gamma=0^\circ$ and $45^\circ$ respectively. The temperature snapshots of figures~\ref{fig:steady-0} and \ref{fig:steady-45} show that the highest temperatures occur near the trailing edge for $\gamma=0^\circ$ and near the top left edge for $\gamma=45^\circ$, where the mean flow tends to advect hot fluid in each case. By contrast, the alternating vortex shedding for $\gamma=90^\circ$ leads to a more uniform temperature distribution, lowering the temperature of the hottest spots, as shown in the last two columns of the temperature snapshots in figure~\ref{fig:steady-90}. 
%{\color{blue} The direct comparison of the global heat transfer of the non-oscillating plate at the three cases of $\gamma$ is also given as the dashed lines in figure~\ref{fig:GlobalAllData} of section~\ref{sec:GlobalLocalHeatTransfer}.}
%start here
\section{Vorticity patterns and their effects on heat transfer}
\label{sec:GlobalLocalHeatTransfer}

Having quantified heat transfer in the baseline non-oscillating-plate cases at the three $\gamma$ values, we now examine the same heat transfer quantities {\it with} plate oscillation, in the direction $\alpha$, at frequency $Re_f$, and oscillation velocity $A/|\bm{U}_{\infty}|$, and at the same three $\gamma$ (flow orientation) values.

For $\gamma=0^\circ$ and $90^\circ$, we consider five $\alpha$ values: $0^\circ,\ 30^\circ,\ 45^\circ,\ 60^\circ$, and $90^\circ$. For $\gamma=45^\circ$, we consider eight $\alpha$ values: $0^\circ,\ 30^\circ,\ 45^\circ,\ 60^\circ,\ 90^\circ,\ 120^\circ,\ 135^\circ$, and $150^\circ$. (There is no need to consider $\alpha>90^\circ$ for $\gamma=0^\circ$ and $90^\circ$ since $\alpha = \alpha_1$ and $\alpha = 180^\circ-\alpha_1$ give the same flows after a reflection.) For each combination of $\gamma$ and $\alpha$, we use two values of the plate's oscillation velocity $A/|\bm{U}_{\infty}|$, 0.2 and $0.3$, and vary $Re_f$ from 50 to 250 in increments of 50. Note that
$Re_U=Re_f\times A \times |\bm{U}_{\infty}|/A$ by equation~\eqref{eq:param}. At a given $A/|\bm{U}_{\infty}|$ (0.2 or 0.3) and at fixed $Re_U$ (always 100), the product of the oscillatory frequency and amplitude $Re_f \times A$ must be constant. Therefore, at each $A/|\bm{U}_{\infty}|$ we vary $Re_f$ with the understanding that simultaneously the amplitude $A$ is varied inversely.

\subsection{Global heat transfer}
\label{subsec:GlobalHeatTransfer}

\begin{figure}[t]
    \centering
    \includegraphics[width=\textwidth]{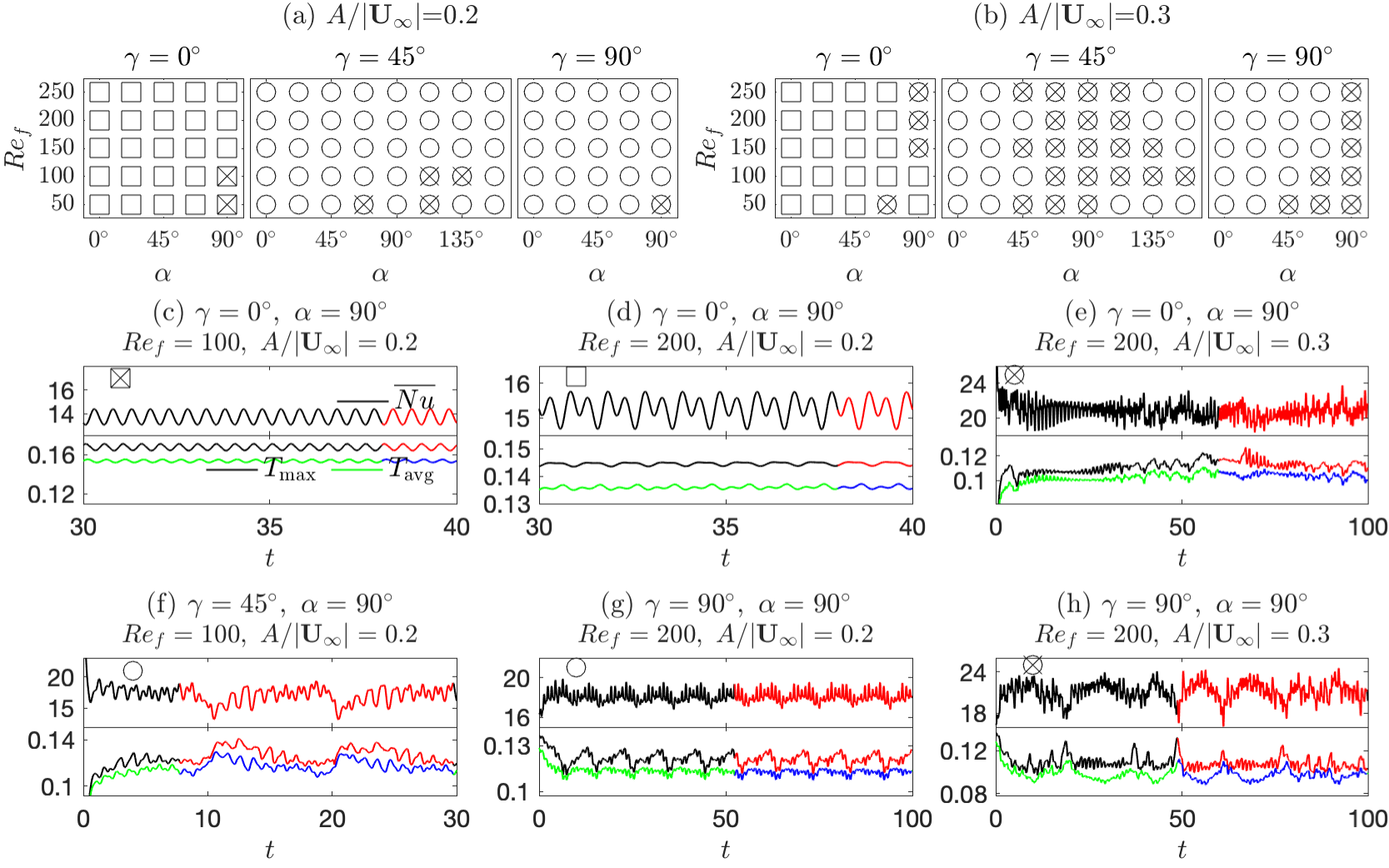}
    \caption{The four classes of the time series of $\overline{Nu}$, $T_{\text{avg}}$ and $T_{\max}$. The three panels in (a) and (b) correspond to $\gamma =0^\circ$, $\gamma=45^\circ$ and $\gamma =90^\circ$, respectively. The open and filled squares denote period-1 and period-1/2 cases, respectively. The open and filled circles represent almost-periodic and nonperiodic cases, respectively, with periods typically $\gg 1$ in the almost-periodic cases. Panel (c)--(h) show examples of the four classes, with the class label at the upper left. The portion of the curves in red and blue is used to compute the time average.}
    \label{fig:HistrOverview}
\end{figure}

We investigate the global heat transfer using $\overline{Nu}(t)$, $T_{\text{avg}}(t)$ and $T_{\max}(t)$ from equations~\eqref{eq:Nu} and \eqref{eq:T}. Although the plate oscillates periodically, these three quantities are not necessarily time-periodic. We use the time series of $\overline{Nu}(t)$, $T_{\text{avg}}(t)$, and $T_{\max}(t)$ to classify the types of periodic and nonperiodic dynamics in figure~\ref{fig:HistrOverview}. Panels (a) and (b) give the classifications for $A/|\bm{U}_{\infty}|=0.2$ and 0.3 respectively. Each panel is divided into three subpanels, one each for $\gamma=0^\circ, 45^\circ$, and $90^\circ$. Each subpanel classifies the dynamics in the space of $\alpha$ and $Re_f$ as one of four types using circles and squares, with and without crosses. 

The first major class corresponds to periodic time series, shown with squares. The squares are open for period-1 dynamics and filled with crosses for period-1/2 dynamics.
Most cases with $\gamma=0^\circ$ (left sides of panels (a) and (b)) have period 1, i.e. the period of the plate oscillation. Period 1/2 occurs when the flow is symmetric on the two half-strokes of the plate, and this is possible only when the two half-strokes are symmetric with respect to the oncoming flow, i.e. for  $\alpha=90^\circ$. Two cases with
$\gamma=0^\circ$, $\alpha=90^\circ$, and $A/|\bm{U}_{\infty}|=0.2$ have period 1/2 (filled squares), and the remainder have period 1. 

A period-1/2 example is shown in figure~\ref{fig:HistrOverview}(c), and we will show the corresponding up-down symmetry of the vorticity field in section~\ref{subsec:LocalHeatTransfer}. Panel (d) shows an example with period 1 but not 1/2. Although the plate motion is up-down symmetric here, the flow is not. For all the periodic cases, we compute time averages using the last two plate oscillation cycles, e.g.~the intervals marked in red and blue in panels (c) and (d). 

The second major class corresponds to nonperiodic time series, labeled with circles. All cases with $\gamma=45^\circ$ and $90^\circ$, and several cases with $\gamma=0^\circ$ at $A/|\bm{U}_{\infty}|=0.3$ belong to this class. Within this class, open circles denote cases that are almost periodic with a period longer than 1. Figure~\ref{fig:HistrOverview}(f) and (g) show examples of these cases. The time series have a component with the frequency of the plate oscillation and a much lower frequency component, corresponding to 10 plate oscillations.
The low frequency component is caused by vortex grouping dynamics over several plate oscillation cycles, at almost the same location each time. We take time averages using at least two of these longer cycles of vortex grouping, shown in red and blue in panels (f) and (g). There are subtle but noticeable differences between the longer cycles that make these cases nonperiodic: although the vortex configuration near the plate is nearly the same from one long period to the next, the configuration far from the plate differs. 

The last subclass, shown by circles with crosses, corresponds to strongly nonperiodic dynamics. In these cases vortex grouping still occurs over long cycles, but the long cycles vary in duration, the positions of the groups vary from one long cycle to the next, and the vortex shedding can be irregular. These cases are more prevalent at $A/|\bm{U}_{\infty}|=0.3$ than 0.2, and with $\alpha$ close to $90^\circ$, i.e. with more transverse plate motions. Such cases correspond to stronger vortices with more complicated dynamics.
Example time series are shown in figure~\ref{fig:HistrOverview}(e) and (h). For this subclass we compute  time averages over time intervals that are long enough that further increases  give only minor changes, as shown in appendix~\ref{app:GridIndependence}. The beginnings and ends of the intervals are chosen so that the vorticity distributions near the plate approximately match. 

We can interpret the periodic and almost-period cases by considering the frequency of vortex shedding due to the plate oscillation together with the natural wake shedding frequency seen in the steady cases of section \ref{sec:SteadyCases} at $\gamma = 45^\circ$ and $90^\circ$. At $\gamma = 0^\circ$ there is no natural wake shedding frequency, so here the dynamics are mostly periodic with the period of the plate oscillation. At $\gamma = 45^\circ$ and $90^\circ$ there is the additional wake frequency, but many cases are close to periodic, particularly at $A/|\bm{U}_{\infty}|$ = 0.2. Such cases resemble frequency locking between the wake and plate oscillation, which has been studied most often for circular cylinders \citep{Koopmann1967, WR1988, CF1998, ABG1986}.
% \textcolor{red}{cite The vortex wakes of vibrating cylinders at low Reynolds numbers Koopman 1967; Vortex formation in the wake of an oscillating cylinder
% CHK Williamson, A Roshko - Journal of fluids and structures, 1988; LOCK-ON OF VORTEX SHEDDING DUE TO ROTATIONAL OSCILLATIONS OF A FLAT PLATE IN A UNIFORM STREAM
% J.M. CHEN
% Y.-C. FANG;
% The effect of a perturbation on the flow over a bluff cylinder
% BJ Armstrong, FH Barnes, I Grant}. 
In these studies, lock-in was observed when the body was forced at a period within about 20\% of the natural wake period, while in our case the natural wake period. Locking of the flow to a multiple of the period of the oscillating body was found in experiments by \cite{BH1964}
% \textcolor{red}{The lift and drag forces on a circular cylinder oscillating in a flowing fluid
% RED Bishop, AY Hassan} 
and in computations of an oscillating ellipse that is free to translate horizontally \citep{Alben2008}.
% \textcolor{red}{An implicit method for coupled flow-body dynamics
% S. Alben, Journal of Computational Physics}.

\begin{figure}[t!]
    \centering
    \begin{subfigure}[b]{0.99\textwidth}
        \centering
        \caption{$\langle\overline{Nu}\rangle$}
        \medskip
        \includegraphics[width = 0.5\textwidth]{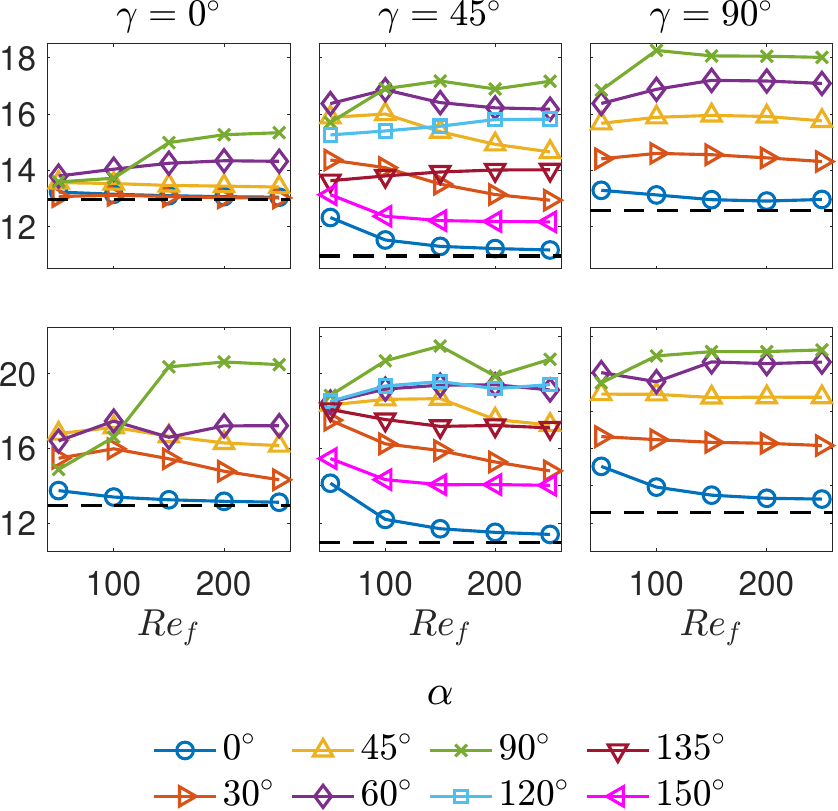}
    \end{subfigure}
    \begin{subfigure}[b]{0.49\textwidth}
        \centering
        \caption{$\langle T_{\text{avg}} \rangle$}
        \medskip
        \includegraphics[width = \textwidth]{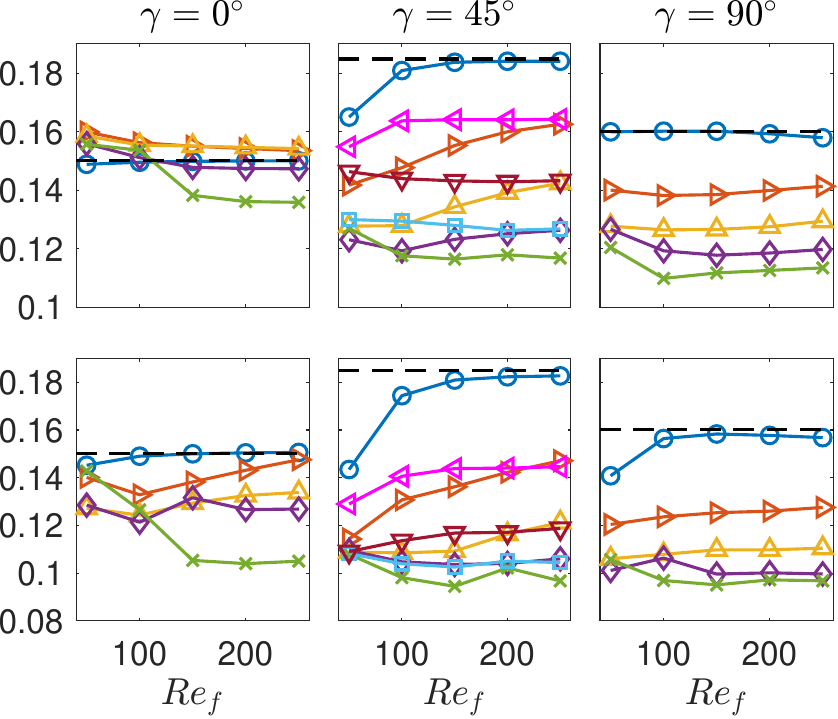}
    \end{subfigure}
    \begin{subfigure}[b]{0.49\textwidth}
        \centering
        \caption{$\max(T_{\max})$}
        \medskip
        \includegraphics[width = \textwidth]{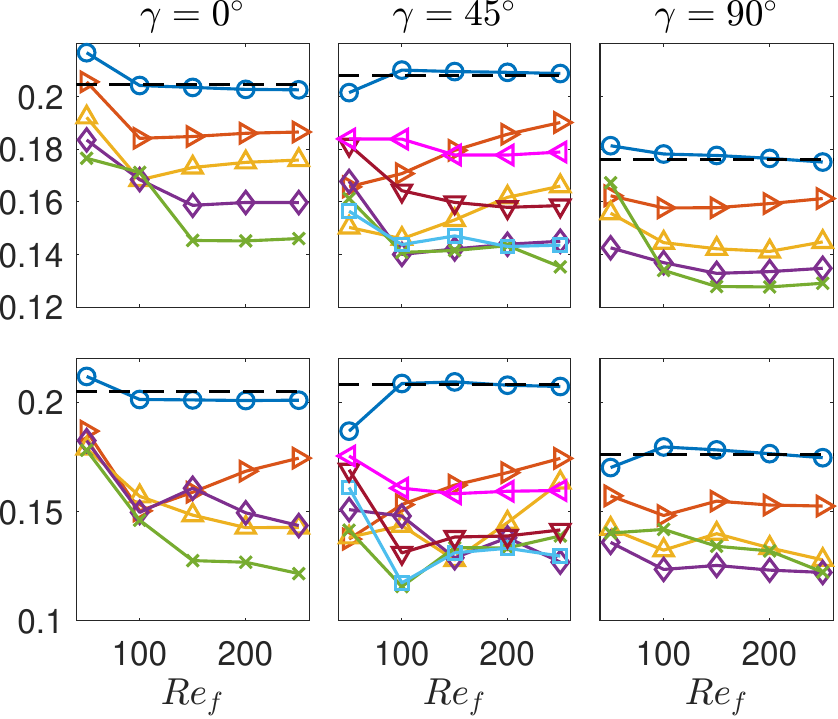}
    \end{subfigure}
    \caption{(a) The time averaged global Nusselt number, $\langle\overline{Nu}\rangle$, for the isothermal plate, (b) the time averaged plate temperature, $\langle T_{\text{avg}} \rangle$, and (c) the maximum plate temperature, $\max(T_{\max})$, when the heat flux from the plate is fixed. The first and second rows in each panel correspond to $A/|\boldsymbol{U}_{\infty}|=0.2$ and 0.3 respectively. The dashed line in each panel gives the values for the non-oscillating plate.}
    \label{fig:GlobalAllData}
\end{figure}

Having classified the types of dynamics over the full parameter space in figure
~\ref{fig:HistrOverview}(a) and (b),  
we plot the key global measures of heat transfer in the same parameter space in figure~\ref{fig:GlobalAllData}. Panels (a) and (b) show the time- and space-averaged Nusselt number $\langle\overline{Nu} \rangle$ and plate temperature $\langle T_{\text{avg}} \rangle$ in the fixed-temperature and fixed-flux cases respectively. In the latter case, panel (c) shows the maximum plate temperature over space and time, $\max(T_{\max})$. Instead of 2D plots in the ($Re_f$, $\alpha$) space of figure~\ref{fig:HistrOverview},   figure~\ref{fig:GlobalAllData} plots each quantity versus $Re_f$ with a separate line for each $\alpha$, so that we can compare the values of the quantities clearly by referring to the vertical axes. 

Comparing figure~\ref{fig:GlobalAllData}(a) and (b) we see that $\langle\overline{Nu} \rangle$ and $\langle T_{\text{avg}} \rangle$ are generally anticorrelated as $Re_f$ and $\alpha$ vary. I.e. the two quantities give the same ordering of parameters in terms of optimality for heat transfer.  
$\max(T_{\max})$ is generally consistent with this ordering but shows more deviations, because it is more sensitive to the details of the velocity and temperature field evolution near the particular instant when the maximum temperature occurs.

Now we discuss the effect of each of the four key parameters---$A/|\bm{U}_{\infty}|$, $\alpha$, $\gamma$, and $Re_f$---on global (time- and space-averaged) heat transfer, and the role of vortex dynamics. In the next subsection, we will present detailed descriptions of the vorticity distributions across space and time, and the corresponding local distributions of heat transfer quantities that underlie the global patterns.

Consistent with \cite{RT2020}, we find $A/|\bm{U}_{\infty}| = 0.3$ (bottom rows of each panel) gives better global heat transfer---larger $\langle\overline{Nu} \rangle$ and smaller $\langle T_{\text{avg}} \rangle$---than 0.2 (top rows), if the other parameters are the same. The extent of the improvement varies with $\alpha$, $\gamma$, and $Re_f$. Larger $A/|\bm{U}_{\infty}|$ means larger plate velocity and stronger vorticity generation, which tend to bring colder fluid close to the plate.

By comparing different colored lines within each panel we see the effect of varying the plate oscillation direction $\alpha$. 
At $\gamma = 45^\circ$ and $90^\circ$, the three metrics vary by a factor of 1.5--2 as $\alpha$ is varied, with smaller variation at $\gamma = 0^\circ$. In most cases $\alpha$ has a stronger effect than $A/|\bm{U}_{\infty}|$ with other parameters fixed. Transverse oscillation ($\alpha=90^\circ$) generally gives the best heat transfer (highest $\langle\overline{Nu} \rangle$ and lowest $\langle T_{\text{avg}}\rangle$) and in-plane oscillation ($\alpha=0^\circ$) is generally worst, except for a few cases at low $Re_f$. The lowest    $\max(T_{\max})$ generally occurs at $\alpha = 90^\circ$ also, or at nearby values ($60^\circ$ or $120^\circ$).

These dependences on $\alpha$ and other, more subtle effects, are connected to vortex dynamics. First, the plate generates stronger vortices when $\alpha$ is closer to $90^\circ$ (i.e. the oscillation is more transverse than in-plane). This generally enhances local heat transfer in the wake as discussed in the next section, \ref{subsec:LocalHeatTransfer}.
Some exceptions are seen, e.g.~in panel (b), the fixed-heat-flux cases, in the top left subpanel ($\gamma=0^\circ$ and $A/|\bm{U}_{\infty}|=0.2$). Here $\alpha=30^\circ$ and $45^\circ$ have higher $\langle T_{\text{avg}} \rangle$ than $\alpha=0^\circ$, while $\max(T_{\max})$ is more monotonic with $\alpha$. The vorticity fields at $\alpha=30^\circ$ and $45^\circ$ create a more uniform thermal boundary layer with a lower maximum temperature but a higher average temperature than $\alpha=0^\circ$. 

At $\gamma=45^\circ$, $\alpha$ spans the range $[0^\circ, 180^\circ)$, unlike $\gamma=0^\circ$ and $90^\circ$ which have the ($\alpha \leftrightarrow 180^\circ-\alpha$) symmetry. For $\gamma=45^\circ$, the eight $\alpha$ values form five groups in order of proximity to 90$^\circ$: 90$^\circ$, 60$^\circ$/120$^\circ$, 45$^\circ$/135$^\circ$, 30$^\circ$/150$^\circ$, and 0$^\circ$.
We have already mentioned that heat transfer increases with proximity to 
90$^\circ$, but there is an additional variation within the three pairs. The plate oscillation is more in-plane with the oncoming flow for the first member of each pair ($\alpha < 90^\circ$) and more transverse for the second ($\alpha > 90^\circ$), e.g. $\alpha = 45$ versus $135^\circ$ with $\gamma = 45^\circ$ in figure \ref{fig:schematic}. Comparing the graphs for each pair in figure \ref{fig:GlobalAllData}, we usually have better global heat transfer for in-plane oscillation, but the difference becomes smaller as $Re_f$ or $A/|\bm{U}_{\infty}|$ increases. For example, $\langle\overline{Nu} \rangle$ is larger and $\langle T_{\text{avg}} \rangle$ is smaller with $\alpha=60^\circ$ than $\alpha=120^\circ$ at $\gamma=45^\circ$ and $A/|\bm{U}_{\infty}|=0.2$ (top center subpanels of (a) and (b)), and the difference between the two becomes smaller at higher $Re_f$, and is almost nonexistent at $A/|\bm{U}_{\infty}|$ = 0.3 (bottom center subpanels of (a) and (b)). In the next section we will discuss the differences between the vortex dynamics  that underlie these results.

The plate orientation $\gamma$ generally has a weaker effect on heat transfer than $\alpha$ and $A/|\bm{U}_{\infty}|$, particularly when we focus on the variations with $\gamma$ when $\alpha = 90^\circ$ in figure~\ref{fig:HistrOverview} (the green curves), usually the best $\alpha$ for heat transfer. At  $A/|\bm{U}_{\infty}|$ = 0.2 (top rows), $\langle\overline{Nu} \rangle$ is lowest at $\gamma = 0^\circ$, where the flow 
is periodic with period 1 and has an attached vortex group near the leading edge of the bottom wall that decreases local heat transfer. By contrast, the flows
for $\gamma=45^\circ$ and $90^\circ$ are nonperiodic, with alternating formations of vortex groups at both edges of the plate's top wall over many plate oscillation cycles, which improves the local (and global) heat transfer. We will give examples in section~\ref{subsec:LocalHeatTransfer}.

At $A/|\bm{U}_{\infty}| = 0.3$, by contrast, the maximum $\langle\overline{Nu} \rangle$ and minimum $\langle T_{\text{avg}} \rangle$ are about the same for all three $\gamma$ values, as shown in the bottom rows of each subpanel. All of these cases are nonperiodic and involve long-lasting groups of strong vortices at the plate edges. Here different $\gamma$ yield different vortex group positions that enhance heat transfer in different locations, but the global heat transfer performance is similar,
as we will show in section \ref{subsec:LocalHeatTransfer}.

Surprisingly, the global heat transfer quantities do not vary much with $Re_f$ in most cases---i.e. most curves in figure~\ref{fig:GlobalAllData} are nearly flat, particularly at $Re_f$ above 100. Recall from the beginning of section~\ref{sec:GlobalLocalHeatTransfer} that we vary $Re_f$ by keeping the oscillation velocity $A/|\bm{U}_{\infty}|$ fixed, so for each graph in figure~\ref{fig:GlobalAllData}, increasing the frequency $Re_f$ decreases the amplitude $A$ such that the product is fixed. As we will see in section~\ref{subsec:LocalHeatTransfer}, the vortex dynamics and the resulting local heat transfer do not vary much as $Re_f$ increases. An exception is $\gamma=0^\circ$ and $\alpha=90^\circ$ (green curves). In the leftmost subpanels, the green curves jump when $Re_f$ increases from 100 to 150, showing improved heat transfer. These jumps correspond to flow transitions: the transition from period-1/2 to period-1 dynamics at $A/|\bm{U}_{\infty}|=0.2$ due to symmetry breaking, and the transition from period-1 to nonperiodic dynamics at $A/|\bm{U}_{\infty}|=0.3$. With each transition, local and global heat transfer are enhanced, as described in  section~\ref{subsec:LocalHeatTransfer}. At other parameters (e.g. $\gamma = 45^\circ$ and $90^\circ$), the curves also jump between $Re_f$ = 50 and 100. These changes also correspond to transitions in vortex dynamics, generally from one nonperiodic state to another.

% \begin{figure}[ht]
%     \centering
%     \begin{subfigure}[b]{0.99\textwidth}
%     \centering
%         \caption{$\gamma = 0^\circ$}
%         \includegraphics[width=0.7\textwidth]{Gamma0AtoV02TimeAvgOverview.pdf}
%     \end{subfigure}
%     \begin{subfigure}[b]{0.99\textwidth}
%     \centering
%         \caption{$\gamma = 45^\circ$}
%         \includegraphics[width=0.7\textwidth]{Gamma45AtoV02TimeAvgOverview.pdf}
%     \end{subfigure}
%     \caption{Time averaged heat transfer performance at $A/\boldsymbol{U}_{\infty}=0.2$.}
%     \label{fig:HeatTransferAtoU02}
% \end{figure}
% %
% %
% \begin{figure}[ht]
%     \centering
%     \begin{subfigure}[b]{0.99\textwidth}
%     \centering
%         \caption{$\gamma = 0^\circ$}
%         \includegraphics[width=0.7\textwidth]{Gamma0AtoV03TimeAvgOverview.pdf}
%     \end{subfigure}
%     \begin{subfigure}[b]{0.99\textwidth}
%     \centering
%         \caption{$\gamma = 45^\circ$}
%         \includegraphics[width=0.7\textwidth]{Gamma45AtoV03TimeAvgOverview.pdf}
%     \end{subfigure}
%     \caption{Time averaged heat transfer performance at $A/\boldsymbol{U}_{\infty}=0.3$.}
%     \label{fig:HeatTransferAtoU03}
% \end{figure}

\subsection{Local heat transfer }
\label{subsec:LocalHeatTransfer}

\begin{figure}[h]
    \centering
    \includegraphics[width=0.95\textwidth]{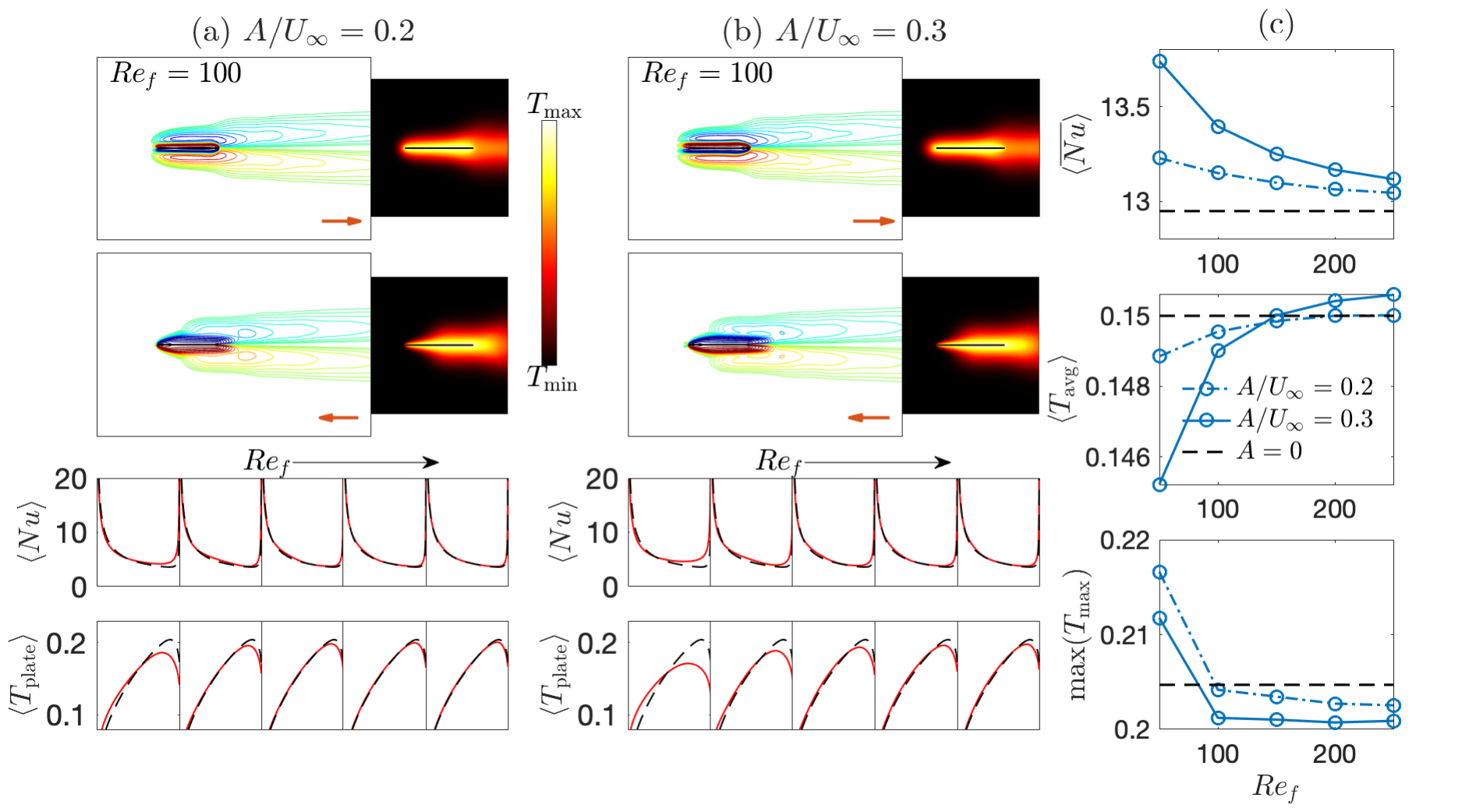}
    \caption{Flows and heat transfer for in-plane plate oscillation ($\alpha = 0^\circ$) in an in-plane oncoming flow ($\gamma=0^\circ$). Panels (a) and (b) show vorticity and temperature fields (in the first two rows) at $Re_f = 100$, the time-averaged local $Nu$ with fixed plate temperature (red lines in the third row), and the time-averaged plate temperature with fixed heat flux (red lines in the fourth row). The red arrows in the vorticity snapshots indicate the instantaneous plate velocity direction. The motion in panel (b) is shown in the supplementary movie ``movie1," accessible \href{https://drive.google.com/drive/folders/1AUg5BWYRKvcFZnJcTIGqaiImf02z0a8a?usp=drive_link}{here}. Panel (c) shows time-averaged global heat transfer quantities. In each panel the black dashed lines are the non-oscillating-plate results.}
\label{fig:gamma0alpha0vorticitytemp}
\end{figure}

In the previous section we summarized the general trends of the vorticity dynamics and heat transfer in terms of time- and space-averaged (global) quantities. Now we show the vorticity and temperature fields and the heat flux and temperature distributions on the plate that explain these trends, in the space of the four key parameters:  $A/|\bm{U}_{\infty}|$, $\alpha$, $\gamma$, and $Re_f$. Since the temperature fields are similar for the isothermal and fixed-heat-flux plates (also seen in figures~\ref{fig:steady-0}, \ref{fig:steady-45} and \ref{fig:steady-90}), we only show the temperature fields for the isothermal plate in this section.

\subsubsection{In-plane oncoming flow: \texorpdfstring{$\gamma=0^\circ$}{}}
\label{subsubsec:gamma0}

We start with in-plane plate oscillation ($\alpha = 0^\circ$) in an in-plane oncoming flow ($\gamma=0^\circ$). These are the only cases without significant vortex shedding and thus have the smallest change in heat transfer due to plate motion.

The top panels of figure~\ref{fig:gamma0alpha0vorticitytemp}(a) and (b) show the vorticity and temperatures fields for $A/|\bm{U}_{\infty}| =0.2$ and $A/|\bm{U}_{\infty}| =0.3$ respectively, at the instants of maximum rightward and leftward velocity, shown by the red arrows. 
The attached shear layers and vortex wakes are slightly different from
those without plate motion (figure~\ref{fig:steady-0}), but still up-down symmetric. 
When the plate moves to the right (top row), the thermal boundary layer is more uniform than for the steady plate, and heat flux is reduced near the leading edge. When the plate moves to the left (second row), the larger velocity difference with the oncoming flow reduces the thermal boundary thickness near the leading edge, increasing the heat flux there.

The third and fourth rows of panels (a) and (b) show that the time-averaged quantities $\langle Nu \rangle$ and $\langle T_{\text{plate}} \rangle$ (red lines) are slightly worse near the leading edge and slightly better near the trailing edge than the steady case (black dashed lines), and the difference diminishes as $Re_f$ increases. Here $Nu$ denotes $Nu_{\text{top}}$ and $Nu_{\text{bot}}$, equal by symmetry.
As $Re_f$ increases, the vorticity fields and $\langle\overline{Nu} \rangle$ (panel (c)) are closer to the steady values. The lower two subpanels of (c) show that interestingly, $\langle T_{\text{avg}} \rangle$ and $\max(T_{\max})$ have opposite trends with $Re_f$, so smaller, higher frequency oscillations mitigate the maximum temperature better, though the overall differences are small for both quantities.

\begin{figure}[t!]
    \centering
    \includegraphics[width=0.8\textwidth]{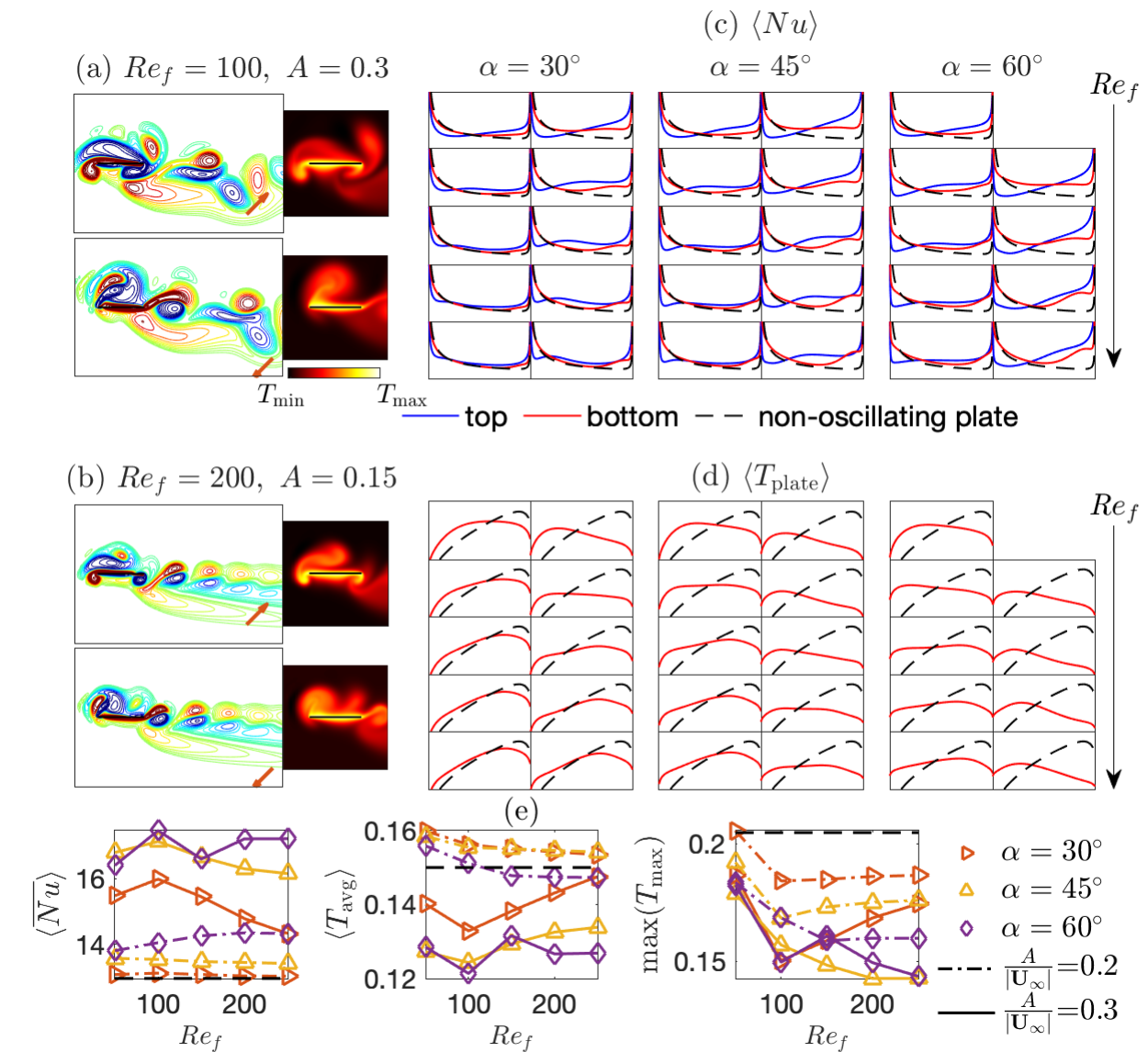}
    \caption{Flows and heat transfer for oblique plate oscillation ($\alpha = 30^\circ, 45^\circ$, and $60^\circ$) in an in-plane oncoming flow ($\gamma=0^\circ$). Panels (a) and (b) show snapshots of the vorticity and temperature field at $\alpha = 45^\circ$ and $\gamma=0^\circ$. The motion in panel (b) is shown in the supplementary movie ``movie2," accessible \href{https://drive.google.com/drive/folders/1AUg5BWYRKvcFZnJcTIGqaiImf02z0a8a?usp=drive_link}{here}. The red arrows in the vorticity snapshots show the instantaneous plate velocity direction. Panels (c) and (d) show the time-averaged local $Nu$ and plate temperature distributions versus $x$ (the horizontal axis) when the plate temperature or heat flux is fixed, respectively. The graphs are arranged in three columns, one for each $\alpha$, and five rows, one for each $Re_f$ from 50 to 250, increasing downward. Pairs of subpanels are shown in each column, corresponding to $A/|\bm{U}_{\infty}| =0.2$ and 0.3 on the left and right respectively. Panel (e) shows the global heat transfer quantities; here $\langle \overline{Nu} \rangle$ and $\langle T_{\text{avg}} \rangle$ are the spatial averages of the graphs in panels (c) and (d) (and summing the top and bottom values of $\langle Nu \rangle$ for $\langle \overline{Nu} \rangle$). In all cases the black dashed lines show the results for the non-oscillating plate.} 
\label{fig:gamma0alpha304560vorticitytemp}
\end{figure}

\begin{figure}[t]
    \centering
    \includegraphics[width=0.8\textwidth]{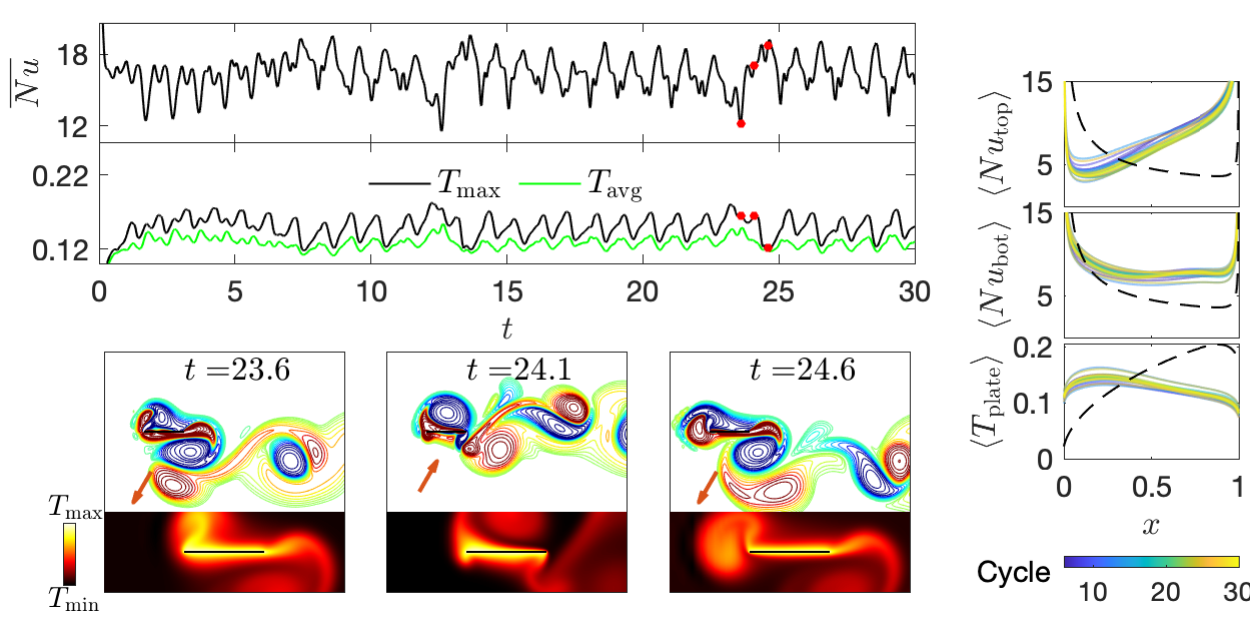}
    \caption{The nonperiodic case with $\gamma=0^\circ$, $\alpha = 60^\circ$, $Re_f=50$, and $A/{|\bm{U}|_{\infty}}=0.3$. The three snapshots of the vorticity and temperature field correspond to times marked with red dots on the graphs of $ \overline{Nu}$ and $\langle T_{\text{max}} \rangle$. The red arrows in the vorticity snapshots show the instantaneous direction of plate motion. The right three panels show the cycle-averaged $\langle Nu \rangle$ of the top and bottom wall for the isothermal plate, and the cycle-averaged $\langle T_{\text{plate}} \rangle$ with fixed heat flux. The first 5 cycles are omitted to avoid transient effects. The black dashed line shows the results for the non-oscillating plate. This motion is shown in the supplementary movie ``movie3," accessible \href{https://drive.google.com/drive/folders/1AUg5BWYRKvcFZnJcTIGqaiImf02z0a8a?usp=drive_link}{here}.}
    \label{fig:gamma0alpha60}
\end{figure}

Next, we discuss oblique plate oscillation ($\alpha = 30^\circ, 45^\circ$, and $60^\circ$) in an in-plane oncoming flow ($\gamma=0^\circ$), with results shown in figure~\ref{fig:gamma0alpha304560vorticitytemp}. As shown in figure \ref{fig:HistrOverview}, unlike $\gamma=45^\circ$ and $90^\circ$, these flows are time-periodic at both $A/|\bm{U}_{\infty}|$ and all $Re_f$, except for a single case ($\alpha=60^\circ$, $Re_f=50$, and $A/|\bm{U}_{\infty}| = 0.3$) that is deferred to figure~\ref{fig:gamma0alpha60}. The vortex dynamics and temperature fields of the cases in figure~\ref{fig:gamma0alpha304560vorticitytemp} have a common form. Panels (a) and (b) show two examples with $\alpha=45^\circ$, at a lower $Re_f$ (100) and a higher $Re_f$ (200), respectively. 
In both cases, negative (blue) vortices shed from the left edge merge into a group that stays attached to the top of the edge. Meanwhile a series of dipoles sheds from the right edge and advects downstream. At the lower $Re_f$ in panel (a), the vortical structures are larger than in panel (b) because they have more time to grow.
In both cases, the rotation of the negative vortex group advects hot fluid leftward along the top wall, making the thermal boundary layer thicker than that of the non-oscillating plate near the left edge but thinner along the rest of the top wall. The dipoles shed from the right edge also enhance heat transfer near the right edge because they disrupt the layer of hot fluid that forms there in the non-oscillating-plate case. These effects are seen in the $\langle Nu_{\text{top}} \rangle$ distributions of panel (c) (the blue curves), which vary along $x$, the horizontal axis, and are time averaged. These curves are lower than the non-oscillating-plate curves (black dashed lines) near the left edge but higher on the rest of the wall. 
Meanwhile $\langle Nu_{\text{bot}} \rangle$ (red curves) is increased from the non-oscillating case on the right sides and less changed on the left sides.
The net effect, shown in $\langle \overline{Nu} \rangle$ in panel (e), is a maximum increase of 
$1$-$11\%$ at $A/|\bm{U}_{\infty}|=0.2$ and 
$23$-$35\%$ at $A/|\bm{U}_{\infty}|=0.3$ as $\alpha$ ranges from 30 to 60$^\circ$.
For the fixed-heat-flux plate, panel (d) shows that $\langle T_{\text{plate}} \rangle$ is higher than the steady values on the left sides and lower and more uniform on the right sides. The net result, $\langle T_{\text{avg}} \rangle$ in panel (e), ranges from worse to slightly better than the steady plate as $\alpha$ ranges from 30 to 60$^\circ$ at $A/|\bm{U}_{\infty}|=0.2$. There is always improvement at $A/|\bm{U}_{\infty}|=0.3$ however, up to 11-19\% decreases in average temperature in the best cases over this $\alpha$ range. Panel (e) shows generally larger improvements in $\max(T_{\max})$. The improvements are larger at larger $\alpha$ because stronger vorticity is shed when the oscillation is more transverse (closer to 90$^\circ$) than in-plane (0$^\circ$, where essentially no shedding occurs). Improvements are also larger at the larger $A/|\bm{U}_{\infty}|$, where stronger vorticity is shed.

The effect of $Re_f$ is more subtle but still noticeable. The dips at the left side of the blue $\langle Nu_{\text{top}} \rangle$ curves in panel (c) become more concentrated as $Re_f$ increases (moving downward), because the shed vortices are smaller in this high-frequency-low-amplitude regime. Meanwhile the improvements in $\langle Nu_{\text{top}} \rangle$ and $\langle Nu_{\text{bot}} \rangle$ on the right sides become smaller. These effects are strongest at  $\alpha=30^\circ$ (left column of panel (c)), still significant at $\alpha=45^\circ$ (middle column), and less noticeable $\alpha=60^\circ$ (right column), particularly at $A/|\bm{U}_{\infty}|=0.3$ (right side of right column). In this case, higher $A/|\bm{U}_{\infty}|$ and more transverse oscillations make the vortices strong enough even at large $Re_f$ (small amplitude) to give strong heat transfer enhancement on the right side.
Panel (d) shows the effects of $Re_f$ on $\langle T_{\text{plate}} \rangle$ in the fixed-heat-flux case. The results are similar to those in panel (c), with increases and decreases in $\langle T_{\text{plate}} \rangle$ relative to the steady case in panel (d) occurring where there are relative decreases and increases in $\langle Nu_{\text{top}} \rangle$ in panel (c). With stronger vortices, the location of the largest 
$\langle T_{\text{plate}} \rangle$ shifts farther from the non-oscillating-plate location at the right side towards the left side, due to advection of cold fluid by the vortex group near the left edge as well as the dipoles shed at the right edge.

%%%%

A single case was omitted in figure~\ref{fig:gamma0alpha304560vorticitytemp}(c) and (d), in the upper right corners where blank spaces appear. This is the only nonperiodic case in that parameter range, as can also be seen in figure \ref{fig:HistrOverview}(b) where, at $\gamma=0^\circ$, $\alpha = 60^\circ$, and $Re_f = 50$, it is marked with a filled circle rather than the open square of the other cases from figure~\ref{fig:gamma0alpha304560vorticitytemp}, which have period 1. 
Figure \ref{fig:HistrOverview} shows that nonperiodicity is most common with $\gamma = 45^\circ$ and $90^\circ$, but among those cases, the dynamics are farther from periodic with  $\alpha$ close to $90^\circ$ and $A/|\bm{U}_{\infty}| = 0.3$. For the case omitted from figure~\ref{fig:gamma0alpha304560vorticitytemp}(c) and (d), $\gamma=0^\circ$ like the periodic cases discussed so far, but $\alpha = 60^\circ$ and $A/|\bm{U}_{\infty}| =0.3$, values associated with stronger shed vorticity and nonperiodicity. This case is shown in figure~\ref{fig:gamma0alpha60}, in a format we use for other nonperiodic cases shown subsequently. 

The vortex dynamics have similarities to the period-1 cases shown in figure~\ref{fig:gamma0alpha304560vorticitytemp}. However, the formation and shedding of vortices  
varies from one cycle to the next in an irregular fashion. The first and third vorticity snapshots of figure~\ref{fig:gamma0alpha60}, one oscillation period apart, have some similarities but differences strong enough to give large differences in $\overline{Nu}$ and $\langle T_{\text{max}} \rangle$, corresponding to the first and third red dots on the graphs.
Instead of long-time averages of their spatial distributions, as shown previously, in the three 
rightmost panels we show ensembles of distributions averaged over single oscillation cycles, from the 6th to the 30th cycles. The variations among the cycles are noticeable but not large compared to the mean (not shown). The distributions are close to those next to the omitted panels in figure~\ref{fig:gamma0alpha304560vorticitytemp}.
Like those cases, large negative vortex groups at the left edge and vortex shedding at the right edge increase $\langle Nu_{\text{top}} \rangle$ and $\langle Nu_{\text{bot}} \rangle$ near the right edge, and $\langle T_{\text{plate}} \rangle$ is maximum near the left edge.

\begin{figure}[htp]
    \centering
    \includegraphics[width=0.8\textwidth]{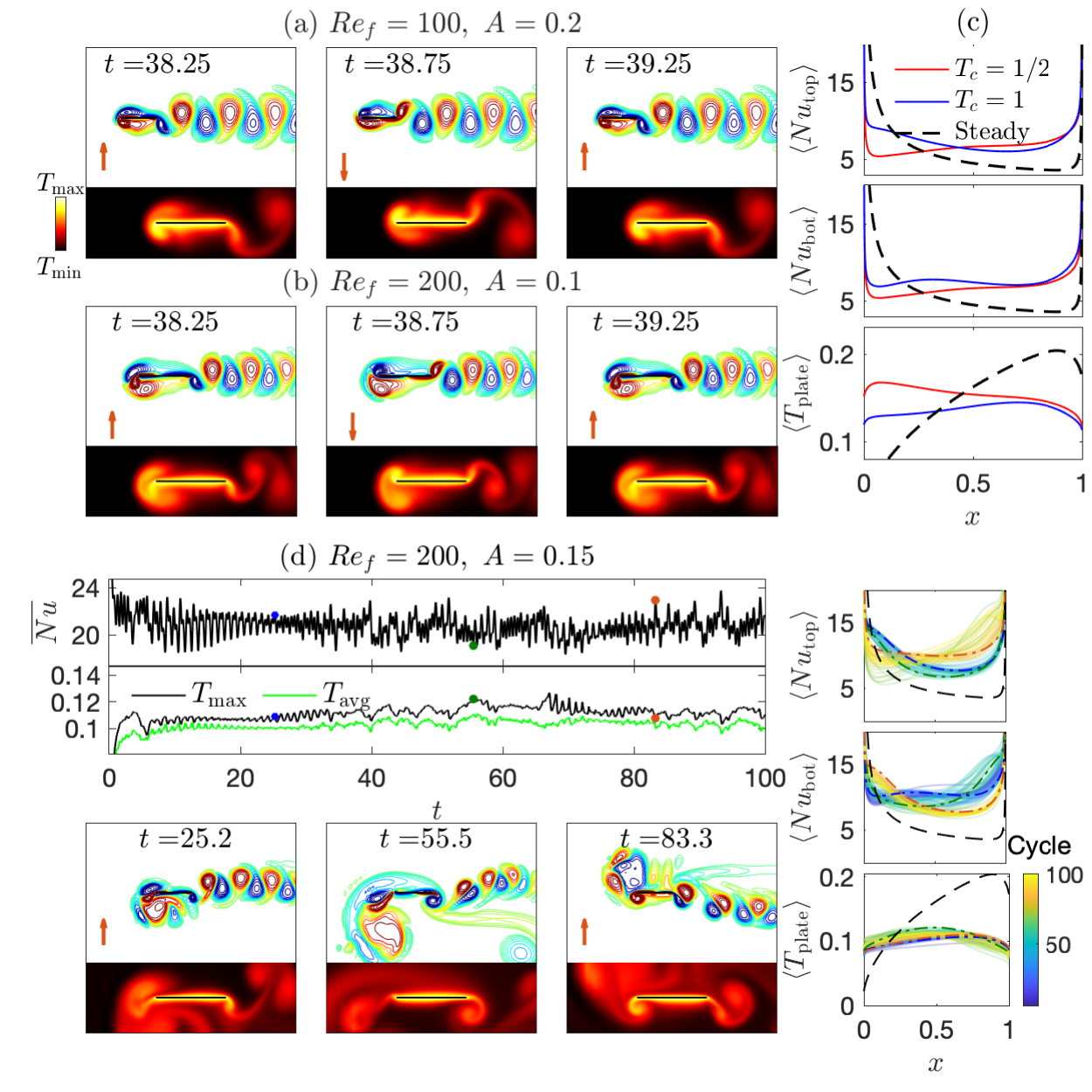}
    \caption{Flows and heat transfer for transverse plate oscillation ($\alpha = 90^\circ$) in an in-plane oncoming flow ($\gamma=0^\circ$). (a) The case with period $ T_c=1/2$ at $Re_f=100$ and $A=0.2$, also shown in the supplementary movie ``movie4," accessible \href{https://drive.google.com/drive/folders/1AUg5BWYRKvcFZnJcTIGqaiImf02z0a8a?usp=drive_link}{here}. (b) The case with period $T_c=1$ at $Re_f=200$ and $A=0.1$, also shown in the supplementary movie ``movie5," accessible \href{https://drive.google.com/drive/folders/1AUg5BWYRKvcFZnJcTIGqaiImf02z0a8a?usp=drive_link}{here}. (c) Comparison of time-averaged local $Nu$ and plate temperature for (a) and (b). (d) The nonperiodic case at $Re_f=200$ and $A=0.15$. The three snapshots at the bottom left occur during the 26th, 56th, and 84th cycles of oscillation respectively, and the corresponding cycle-averaged local heat transfer quantities are shown with dash-dotted curves blue, green, and orange curves, respectively, in the three rightmost subpanels. This motion is shown in the supplementary movie ``movie6," accessible \href{https://drive.google.com/drive/folders/1AUg5BWYRKvcFZnJcTIGqaiImf02z0a8a?usp=drive_link}{here}. The red arrows in the vorticity snapshots show the instantaneous direction of plate motion. In all cases the black dashed lines show the results for the non-oscillating plate.}
    \label{fig:gamma0alpha90vorticitytemp}
\end{figure}

For $\gamma = 0^\circ$ we have discussed $\alpha = 0^\circ, 30^\circ, 45^\circ,$ and $60^\circ$. The last $\alpha$, $90^\circ$, was previously studied for heat transfer (\cite{RT2020}) and is the most commonly studied case for propulsion (\cite{alben2021a}). At $A/|\bm{U}_{\infty}| = 0.2$, the dynamics transition from up-down symmetric with period 1/2 to asymmetric with period 1 as $Re_f$ increases above 100. Figure~\ref{fig:gamma0alpha90vorticitytemp}(a) and (b) show the vorticity and temperature fields before and after the transition, respectively. Panel (c) shows that the local heat transfer quantities improve after the transition, moving from the red to the blue lines. The improvement is mainly on the left side. This corresponds to enlarged positive vorticity and shrunken negative vorticity regions below and above the left edge, respectively, in panel (b). Panel (c) shows that the heat transfer is enhanced on both sides of the plate, with a smaller but broader region of enhancement on the bottom.
Near the right edge, similar reverse von K\'arm\'an vortex streets form and give similar distributions of heat transfer quantities.
Overall, $\langle\overline{Nu} \rangle$ increases by 11\% and (in the fixed-flux case) $\langle T_{\text{avg}} \rangle$ and $\max(T_{\max})$ decrease by 11\% and 15\% respectively moving from $Re_f=100$ to 200. 

At $A/|\bm{U}_{\infty}| = 0.3$, the dynamics instead transition from periodic to nonperiodic as $Re_f$ increases, again with significant improvement of the global heat transfer after the transition. 
Panel (d) shows examples of the vorticity and temperature fields at $Re_f=200$, after the transition. Here vortex grouping occurs in an irregular fashion, and the graphs of $\overline{Nu}$ and $T_{\text{max}}$ are not close to periodic.
We observe three different stages in the vortex dynamics and show one snapshot from each stage. The first snapshot somewhat resembles those in panel (b), with grouping of positive vortices near the left edge of the bottom wall and a reverse von K\'arm\'an vortex street emanating from the right edge. At this instant, blue dashed-dotted lines show the cycle-averaged local heat transfer quantities in the three subpanels on the right side of panel (d). We see some similarities with the blue lines in panel (c). The second snapshot shows the second stage, when the large positive vortex group detaches from the left edge. It continues absorbing positive vortices from the left edge, but is more diffuse than in the previous stage. The local heat transfer quantities for this snapshot are shown by the green dashed-dotted lines in the rightmost subpanels, and are somewhat worse at the middle and left side of the plate than the blue dashed-dotted lines of the first stage. At the right edge the vortex street is more strongly paired as dipoles, and $\langle Nu_{\text{bot}} \rangle$ is improved. The third snapshot approximates a mirror image of the first snapshot, with a region of negative vorticity above the left edge and an array of dipoles shed from the right edge, directed downward. The local heat transfer quantities are shown by the orange dashed-dotted line in the rightmost subpanels. $\langle Nu_{\text{top}}\rangle$ and $\langle Nu_{\text{bot}} \rangle$ for this snapshot are approximately $\langle Nu_{\text{bot}} \rangle$ and $\langle Nu_{\text{top}}\rangle$ for the first snapshot (blue dashed-dotted lines), while the $\langle T_{\text{plate}} \rangle$ distributions are approximately equal, consistent with a mirror-image symmetry of the flows and temperature fields. In these cases and in 
figures~\ref{fig:gamma0alpha304560vorticitytemp} and \ref{fig:gamma0alpha60}, $\langle Nu \rangle$ has a local minimum near the leading edge on the side facing the large region of merged vortices, and is larger (relative to the steady case, the black dashed line) near the trailing edge. On the other side of the plate, $\langle Nu \rangle$ is instead larger near the leading edge and smaller near the trailing edge. In figure~\ref{fig:gamma0alpha90vorticitytemp}(d) and in other nonperiodic cases discussed next, vortices merge near one edge into a large cloud over many cycles, then the cloud detaches, and then vortex merging occurs at the other edge. The switching between the edges occurs at somewhat irregular intervals. During these changes in the local heat transfer distribution, the global (spatially-averaged) heat transfer does not change much; increases in heat transfer in one region are balanced by decreases in other regions.
At lower $Re_f$, 50 and 100, the dynamics are periodic and similar to panel (b). In these two cases $\langle \overline{Nu} \rangle$ is 28\% and 20\% lower than than the nonperiodic $Re_f=200$ case, while $\langle T_{\text{avg}} \rangle$ is 37\% and 22\% higher than at $Re_f=200$, respectively. The merged vorticity near the leading edge is stronger in the nonperiodic case than in the periodic case, and hence more capable of bringing cold fluid towards the plate.

% \begin{figure}[ht]
%     \centering
%     \includegraphics[width=\linewidth]{Gamma45VorticityOpt.pdf}
%     \caption{The snapshots of the vorticity field of the best heat transfer performance at $\gamma=45^\circ$.}
%     \label{fig:gamma45vorticity}
% \end{figure}

\subsubsection{Oblique oncoming flow: \texorpdfstring{$\gamma=45^\circ$}{}}
\label{subsubsec:gamma45}
\begin{figure}[t!]
    \centering
    \begin{subfigure}[b]{\textwidth}
        \centering
        \caption{$\gamma=45^\circ$, $\alpha = 90^\circ$, $Re_f=150$, $A=0.1333$}
        \includegraphics[width=0.65\linewidth]{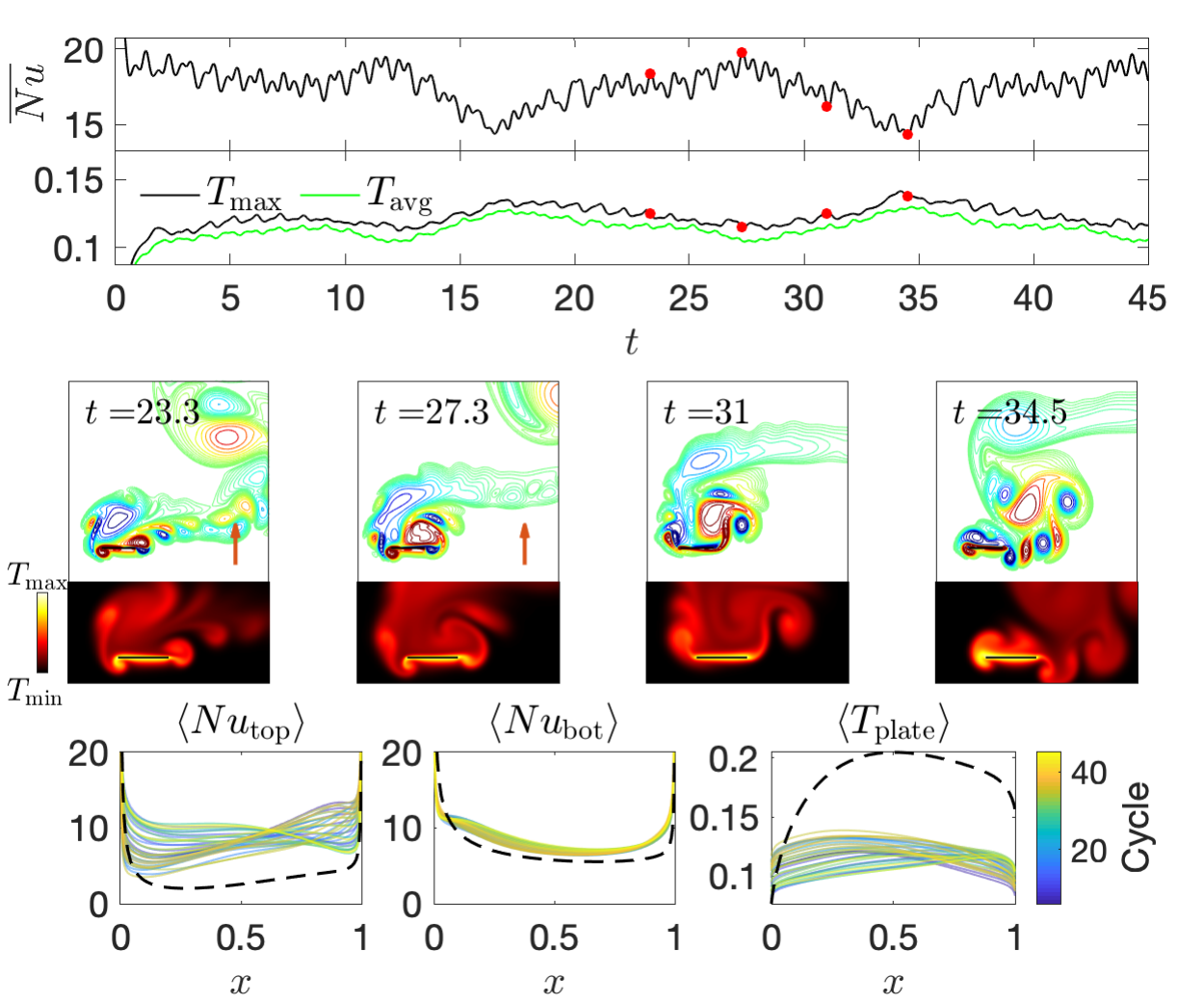}
    \end{subfigure}
    \begin{subfigure}[b]{\textwidth}
        \centering
        \caption{$\gamma=45^\circ$, $\alpha = 90^\circ$, $Re_f=150$, $A=0.2$}
        \includegraphics[width=0.65\linewidth]{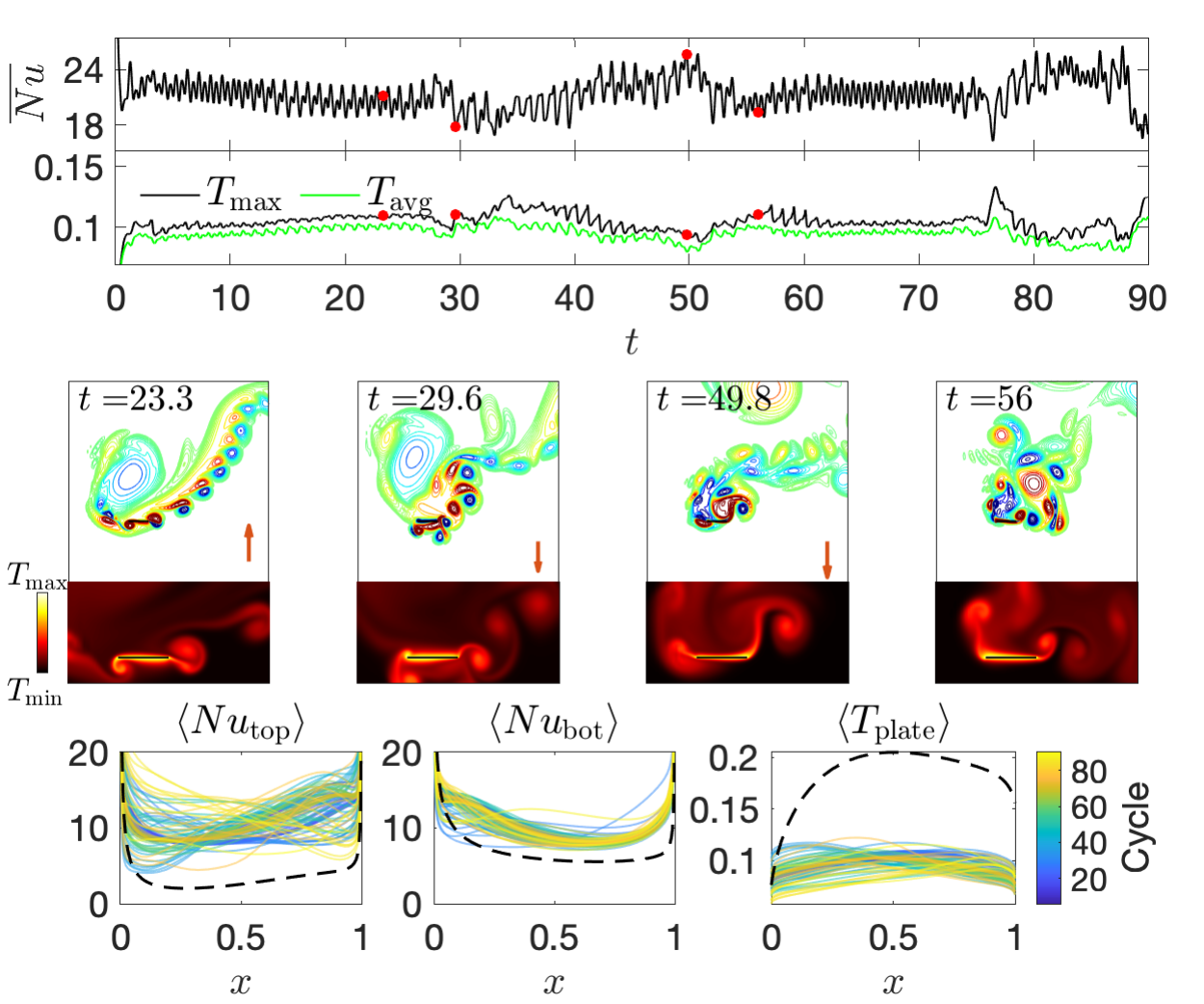}
    \end{subfigure}
    \caption{The cases that optimize global heat transfer ($\langle \overline{Nu} \rangle$ and $\langle T_{\text{avg}} \rangle$) at $\gamma=45^\circ$ for $A/|\boldsymbol{U}_{\infty}|=0.2$ (a) and 0.3 (b). The supplementary movies for panel (a) and (b) are named ``movie7'' and ``movie8," respectively, and can be found \href{https://drive.google.com/drive/folders/1AUg5BWYRKvcFZnJcTIGqaiImf02z0a8a?usp=drive_link}{here}. In each panel, the four snapshots of the vorticity and temperature field correspond to times marked with red dots on the graphs of $ \overline{Nu}$ and $\langle T_{\text{max}} \rangle$, with the red arrows indicating the instantaneous direction of plate motion. The bottom three panels show the cycle-averaged $\langle Nu \rangle$ of the top and bottom wall for the isothermal plate, and the cycle-averaged $\langle T_{\text{plate}} \rangle$ with fixed heat flux. The first 5 cycles are omitted to avoid transient effects. The black dashed line shows the time-averaged results for the non-oscillating plate in the final periodic state for $\gamma=45^\circ$.}
    \label{fig:gamma45alpha90vorticitytemplocal}
\end{figure}

The previous section considered the case of the background flow along the plane of the plate, $\gamma = 0^\circ$. We now consider heat transfer in the same system with $\gamma$ increased to $45^\circ$, an oblique background flow. For any $\gamma$ significantly different from $0^\circ$, the wake oscillates periodically (for a range of $Re_U$ near the present value of 100) even for a steady plate, as shown in section \ref{sec:SteadyCases}. When we oscillate the plate also but at a different period, the dynamics can be almost periodic on a longer time scale (i.e. the flow period locks to a multiple of the plate period) or clearly nonperiodic, as indicated by the open and filled circles respectively in figure~\ref{fig:HistrOverview} for $\gamma = 45^\circ$ and 90$^\circ$. Almost-periodic cases are the most common at $A/|\boldsymbol{U}_{\infty}|=0.2$, while nonperiodic cases are about equally common at $A/|\boldsymbol{U}_{\infty}|=0.3$, and more common with transverse oscillation, when the shed vorticity is stronger. In almost-periodic cases, vortex merging occurs over the same number of plate oscillation cycles and at the same location each time. In nonperiodic cases, the locations and dynamics of vortex merging are somewhat irregular.

In figure~\ref{fig:gamma45alpha90vorticitytemplocal}, we show the cases with the best global heat transfer ($\langle \overline{Nu} \rangle$) when $\gamma = 45^\circ$, for $A/|\boldsymbol{U}_{\infty}|=0.2$ (panel (a)) and 0.3 (panel (b)). Both occur with transverse oscillation 
($\alpha=90^\circ$), which is also optimal at $\gamma = 0^\circ$ and 90$^\circ$. Both optima occur at an intermediate $Re_f$ of 150, but other $Re_f$ give similar $\langle \overline{Nu} \rangle$ and $\langle T_{\text{avg}} \rangle$. Panel (a) is nearly periodic with a period of 17 plate oscillation cycles, while panel (b) is nonperiodic. In both cases a dipole sheds alternately from the left and right edge every half cycle; the stronger vortex of the dipole pair merges with the cloud of previously-shed vortices of the same sign and the weaker vortex diffuses. Over many cycles, negative vortices group at the left edge, detach, and then positive vortices group at the right edge, detach, and the process repeats. During this time, the graphs of the global quantities $\overline{Nu}$, $T_{\text{avg}}$, and $T_{\text{max}}$ show oscillations on the time scale of the plate oscillation and on much longer time scales. At the bottom of each panel, the graphs of the local quantities $\langle Nu_{\text{top}} \rangle$, $\langle Nu_{\text{bot}} \rangle$, and $\langle T_{\text{plate}} \rangle$ also oscillate, between distributions with maxima on the left and right alternately for the first and third quantities. These oscillations are correlated with the vortex dynamics shown in the snapshots in the middle of each panel.

In the first vorticity snapshot of panel (a), a region of negative merged vortices lies above the left edge, which makes $\langle Nu_{\text{top}} \rangle$ low on the left and high on the right and gives a maximum of $\langle T_{\text{plate}} \rangle$ near the left edge. The top subpanel shows $\overline{Nu}$ is generally increasing at this time. In the second vorticity snapshot, a positive vortex group forms and repels the negative vortex group. The maximum of $\langle Nu_{\text{top}} \rangle$ shifts from the right towards the left. The second red dot on the $\overline{Nu}$ time series above shows that the global heat transfer reaches a maximum at this time.
In the third snapshot, the positive vortex group detaches, while $\overline{Nu}$ decreases, reaching a minimum at the time of the fourth snapshot, when the negative vortex group begins to grow again. The graphs of $T_{\text{avg}}$ and $T_{\text{max}}$ below show nearly the inverse behavior for the fixed-flux case, as expected.
During this entire cycle $\langle Nu_{\text{bot}} \rangle$, which is less influenced by the vortex wake dynamics, has much smaller oscillations about a mean distribution slightly above that of the steady case (black dashed line).  

A few differences are seen in panel (b) ($A/|\boldsymbol{U}_{\infty}|=0.3$). The vortex groups form over more oscillation cycles than at $A/|\boldsymbol{U}_{\infty}|=0.2$. In the first snapshot of panel (b), the negative vortex group is more diffuse than in panel (a), with a longer chain of dipoles emanating from the right edge. Meanwhile the global heat transfer quantities ($\overline{Nu}$, $T_{\text{avg}}$, and $T_{\text{max}}$) are nearly steady. In the second snapshot, the positive vortices from the right edge have formed a loose group that repels the negative vortex group, and $\overline{Nu}$ drops to a local minimum. In the third snapshot the positive vortices have formed a more compact group attached to the right edge of the plate, during a long period of increase of $\overline{Nu}$. In the fourth snapshot, the positive vortex group has detached, the negative vortex group is reforming, and $\overline{Nu}$ drops to another local minimum. Meanwhile $\langle Nu_{\text{top}} \rangle$, $\langle Nu_{\text{bot}} \rangle$, and $\langle T_{\text{plate}} \rangle$ at the bottom oscillate between distributions similar to those in panel (a), but more erratically and with larger oscillation amplitudes.

\begin{figure}[ht!]
    \centering
    \includegraphics[width=\linewidth]{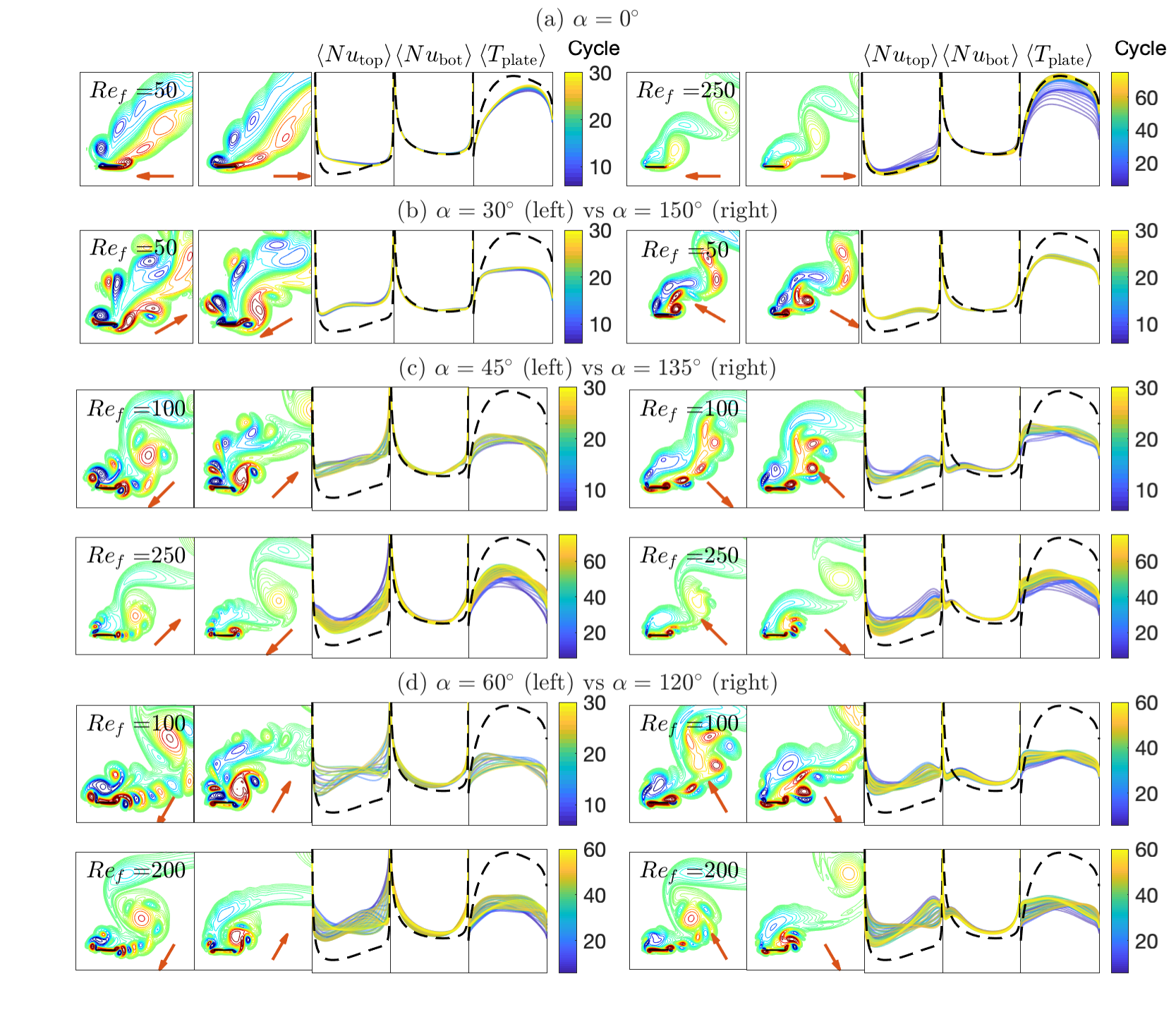}
    \caption{Vorticity fields and cycle-averaged local heat transfer quantities for an oscillating plate
    in an oblique flow ($\gamma=45^\circ$) with $A/|\boldsymbol{U}_{\infty}| = 0.2$ and $\alpha\neq90^\circ$, i.e. oscillation directions other than transverse. The red arrow in each vorticity panel indicates the instantaneous plate velocity direction. Supplementary movies are available at this \href{https://drive.google.com/drive/folders/1AUg5BWYRKvcFZnJcTIGqaiImf02z0a8a?usp=drive_link}{link} for the left case of panel (a) (``movie9''), the two cases in panel (b) (``movie10'' and ``movie11''), and the two cases in panel (c) (``movie12'' and ``movie13'').}
    \label{fig:gamma45vorticityLocalHeatTransfer}
\end{figure}

Figure~\ref{fig:gamma45vorticityLocalHeatTransfer} shows vorticity field snapshots and local heat transfer quantities for oblique flow ($\gamma=45^\circ$) and the remaining seven plate oscillation directions, i.e. $\alpha \neq 90^\circ$. We only discuss $A/|\boldsymbol{U}_{\infty}|=0.2$ because the phenomena are generally similar at $A/|\boldsymbol{U}_{\infty}|=0.3$. The top row (panel (a)) shows in-plane oscillation ($\alpha=0^\circ$), for which the largest heat transfer enhancement occurs at 
$Re_f=50$ (shown on the left side), and mainly on the top side of the plate near the left edge. As 
$Re_f$ increases and the amplitude decreases, the enhancement drops. The flow and local heat transfer quantities converge to those of the steady plate, as shown on the right side of panel (a), neglecting the dark blue graphs of the initial transient period before the steady state. With in-plane oscillations, each edge mainly sheds vorticity of a single sign, as for the static plate. Without strong interactions of oppositely-signed vorticity, there is relatively weak fluid mixing 
near the plate.

The remaining rows, panels (b)--(d), show $\alpha = 30^\circ, 45^\circ,$ and $60^\circ$ on the left paired with the supplementary angles $150^\circ, 135^\circ,$ and $120^\circ$ on the right. For each pair, the transverse velocity component is the same, ranging from smaller (more in-plane) at $\alpha = 30^\circ$ and $150^\circ$ (panel (b)) to larger (more transverse) at $\alpha = 60^\circ$ and $120^\circ$ (panel (d)). But the in-plane velocity components are reversed, so the left side ($\alpha < 90^\circ$) oscillates more {\it along} the oncoming flow direction while the right side ($\alpha > 90^\circ$) oscillates more {\it across} it, as seen from the plate velocity vectors (red arrows) in each panel. 
In figure~\ref{fig:GlobalAllData} we saw that oscillating along the flow is better, i.e. within each pair $\alpha < 90^\circ$ outperforms $\alpha > 90^\circ$, and the difference between the two becomes smaller as $Re_f$ increases. This is also seen in the local heat transfer graphs of figure~\ref{fig:gamma45vorticityLocalHeatTransfer}, for which $\langle Nu_{\text{top/bot}} \rangle$ is generally higher and $\langle T_{\text{plate}} \rangle$ is lower relative to the steady reference case (black dashed) on the left side, oscillating along the flow.

We observe differences in vortex dynamics for $\alpha < 90^\circ$ and $\alpha > 90^\circ$ that may explain the
differences in heat transfer enhancement. With $A/|\boldsymbol{U}_{\infty}|$ = 0.2, the ratio of the maximum plate velocity to the oncoming flow velocity is $2\pi\cdot 0.2 = 1.26$, so the plate oscillation is strong enough to reverse the oncoming flow velocity. This occurs for $\alpha = 45^\circ$, when the plate oscillates along the flow, and at nearby values, i.e. over most of the range $0^\circ < \alpha < 90^\circ$. In this case, oppositely-signed vortices are shed from each edge on each cycle, and they have relatively strong interactions and mix the fluid close to the plate, as can be seen in the vorticity snapshots on the left sides of panels (b)--(d). When $\alpha = 135^\circ$, the plate oscillates across the flow, and there is always a positive flow component in the oncoming flow direction. The same is true over most of the range $90^\circ < \alpha < 180^\circ$. In this case, the vorticity shed from each edge is predominantly (but not completely) single-signed, similarly to the in-plane oscillation of panel (a). Here there are weaker interactions between the vortices near the plate, before they are advected away by the oncoming flow, as seen in the vorticity snapshots on the right sides of panels (b)--(d). These differences may underlie some of the common features of the local heat transfer quantities on the left and right. For example, the right subpanels but not the left subpanels show a dip in $\langle Nu_{\text{bot}} \rangle$, and a rise in $\langle T_{\text{plate}} \rangle$, at the left edge relative to the steady graph (black dashed line). The difference between the two cases diminishes as $Re_f$ increases, because the vorticity fields become more similar. This is seen in the second rows of panels (c) and (d), where the vorticity snapshots are more similar on the left and right sides than in the first rows, at lower $Re_f$.

% \begin{figure}[ht]
%     \centering
%     \includegraphics[width=\linewidth]{Gamma90VorticityOpt.pdf}
%     \caption{The snapshots of the vorticity field of the best heat transfer performance at $\gamma=90^\circ$.}
%     \label{fig:gamma90vorticity}
% \end{figure}

\subsubsection{Transverse oncoming flow: \texorpdfstring{$\gamma=90^\circ$}{}}
\label{subsubsec:gamma90}
\begin{figure}[t!]
    \centering
    \begin{subfigure}[b]{\textwidth}
        \centering
        \caption{$\gamma=90^\circ$, $\alpha = 90^\circ$, $Re_f=100$, $A=0.2$}
        \includegraphics[width=0.65\linewidth]{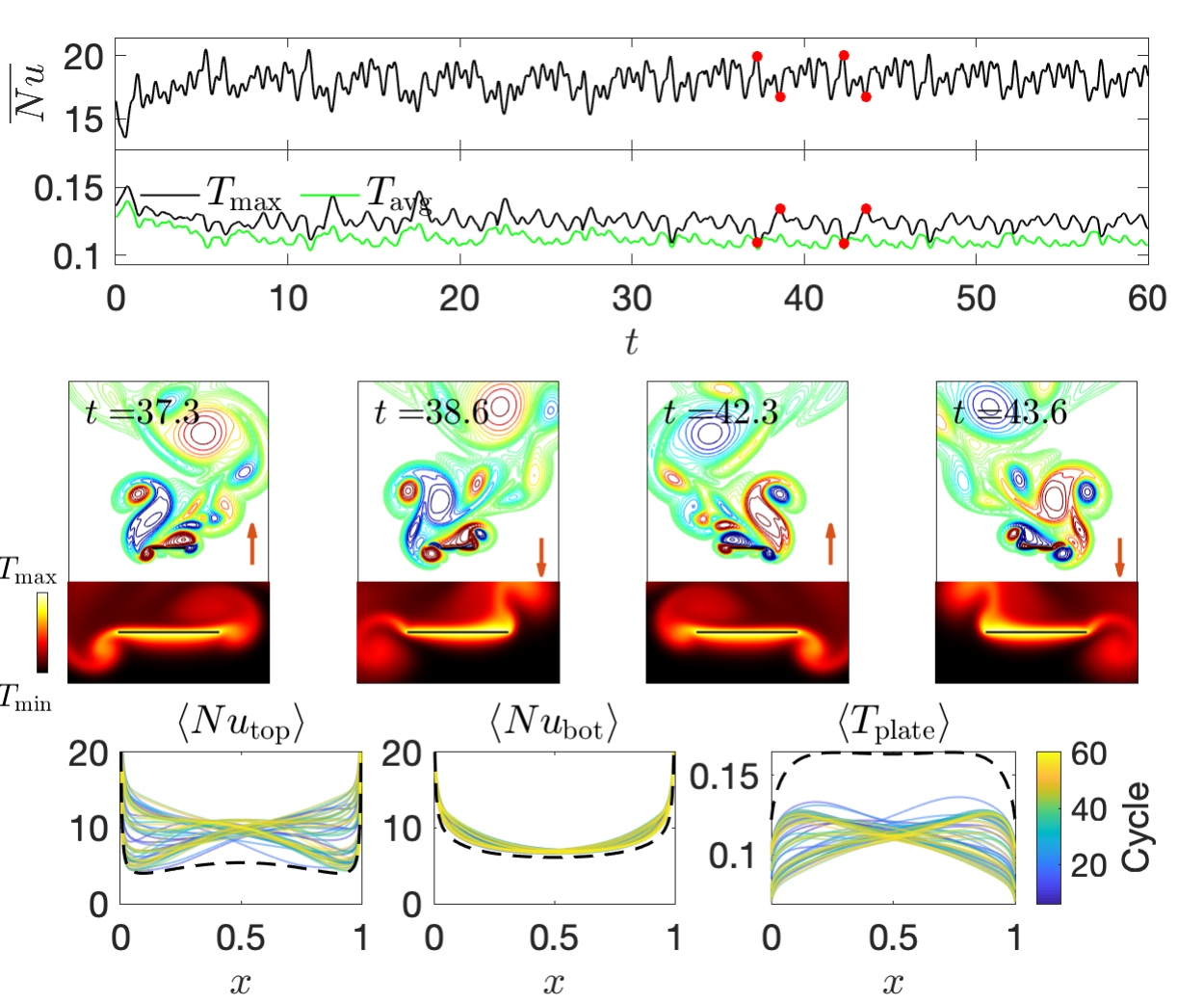}
    \end{subfigure}
    \begin{subfigure}[b]{\textwidth}
        \centering
        \caption{$\gamma=90^\circ$, $\alpha = 90^\circ$, $Re_f=250$, $A=0.12$}
        \includegraphics[width=0.65\linewidth]{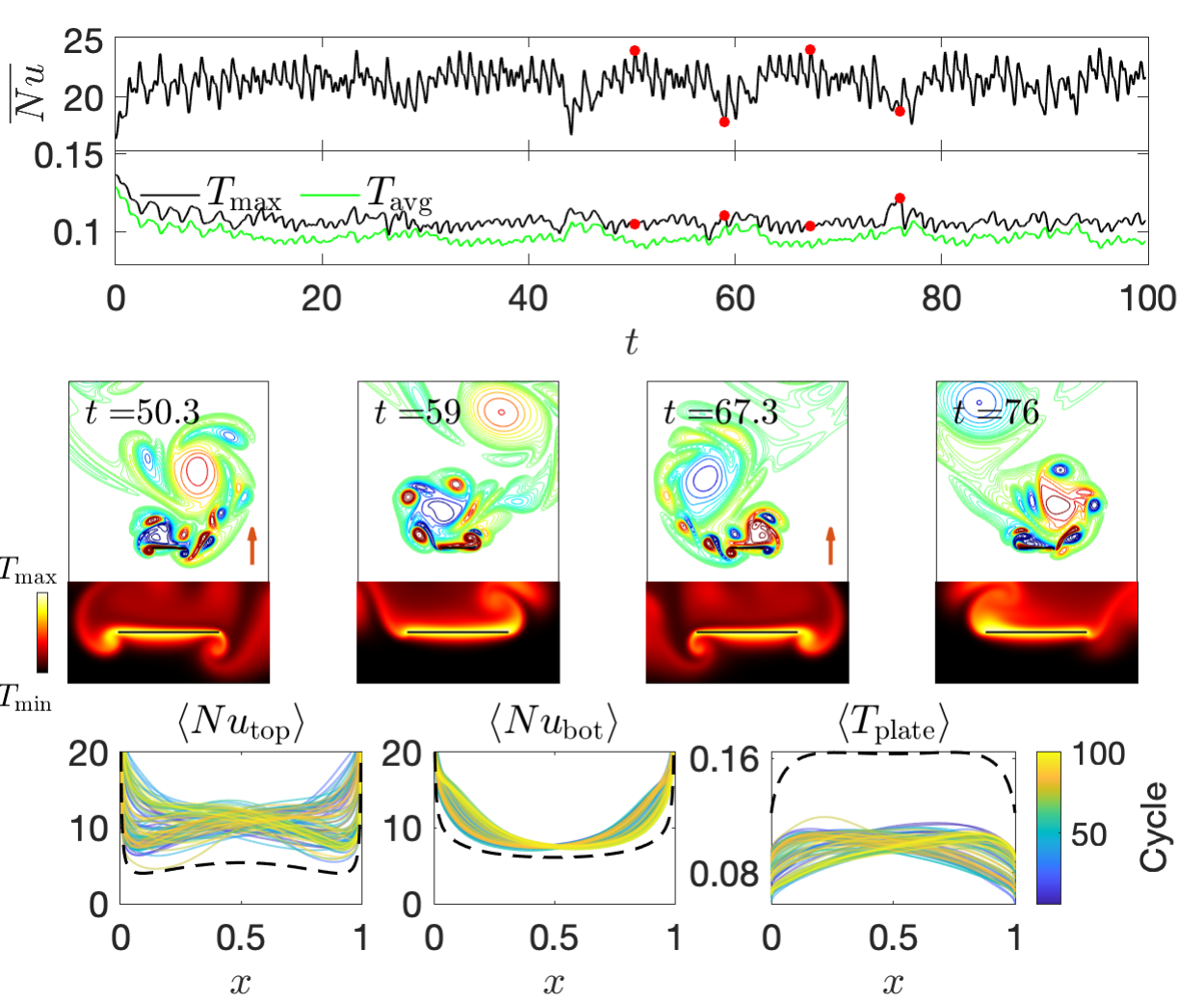}
    \end{subfigure}
    \caption{The cases that optimize global heat transfer ($\langle \overline{Nu} \rangle$ and $\langle T_{\text{avg}} \rangle$) at $\gamma=90^\circ$ for $A/|\boldsymbol{U}_{\infty}|=0.2$ (a) and 0.3 (b), with supplementary movies named ``movie14'' and ``movie15," respectively, linked \href{https://drive.google.com/drive/folders/1AUg5BWYRKvcFZnJcTIGqaiImf02z0a8a?usp=drive_link}{here}. In each panel, the four snapshots of the vorticity and temperature field correspond to times marked with red dots on the graphs of $ \overline{Nu}$ and $\langle T_{\text{max}} \rangle$, with the red arrows indicating the instantaneous direction of plate motion. The bottom three panels show the cycle-averaged $\langle Nu \rangle$ of the top and bottom wall for the isothermal plate, and the cycle-averaged $\langle T_{\text{plate}} \rangle$ with fixed heat flux. The first 5 cycles are omitted to avoid transient effects. The black dashed line shows the time-averaged results for the non-oscillating plate in the final periodic state for $\gamma=90^\circ$, after the symmetry breaking.}
    \label{fig:gamma90alpha90vorticitytemplocal}
\end{figure}
Having discussed in-plane and oblique oncoming flows, $\gamma=0^\circ$ and $45^\circ$, we
now consider transverse oncoming flow, $\gamma=90^\circ$. Due to symmetry, we can restrict to $0^\circ \leq \alpha \leq 90^\circ$. As before, the best global heat transfer occurs with transverse plate oscillations ($\alpha=90^\circ$). At this particular $\alpha$ and $\gamma$, as for the steady plate at $\gamma=90^\circ$, a uniform initial oncoming flow with small asymmetric perturbations reaches the asymmetric steady state slowly, after $\approx 300$ time units. To reach the steady state faster, we instead initialize the flow and temperature fields to be those at $t=400$ for the steady plate with $\gamma=90^\circ$, similar to the cases shown at the latest times in figure~\ref{fig:steady-90}, which have a von K\'{a}rm\'{a}n vortex wake. 

The optimal flows for heat transfer at $A/|\boldsymbol{U}_{\infty}|=0.2$ and 0.3 are shown in figure~\ref{fig:gamma90alpha90vorticitytemplocal}(a) and (b) respectively. With the combination of a plate oscillation frequency and the bluff-body wake shedding frequency, the vortex dynamics and resulting heat transfer for $\gamma=90^\circ$ are generally similar to those for $\gamma=45^\circ$ in a few ways. First, in both cases complex vortex shedding patterns occur in the wake, and greatly enhance heat transfer on the top of the plate (the side facing the wake). Second, shed vortices merge into large regions of single-signed vorticity over many oscillation periods, and then detach, alternately at one edge and then the other. Third, in most cases at $A/|\boldsymbol{U}_{\infty}|=0.2$ the vortex dynamics are close to periodic and locked to a multiple of the plate oscillation period (10 in panel (a)), while at  $A/|\boldsymbol{U}_{\infty}|=0.3$ more cases are nonperiodic, particularly near $\alpha = 90^\circ$. One difference is the bilateral symmetry of the $\gamma=\alpha = 90^\circ$ cases, e.g. in figure ~\ref{fig:gamma90alpha90vorticitytemplocal}, so that the positive and negative vortex groups are essentially mirror images in the two leftmost and rightmost snapshots of the periodic case in panel (a), and almost mirror images in the nonperiodic case of panel (b).
A corresponding left-right symmetry is seen in the ensemble of cycle-averaged $\langle Nu_{\text{top}} \rangle$, $\langle Nu_{\text{bot}} \rangle$ and $\langle T_{\text{plate}} \rangle$ distributions
in panel (a), and to a lesser extent in panel (b).
Key features of the graphs of the global heat transfer quantities are correlated with events in the vortex dynamics, similarly to $\gamma = 45^\circ$. Troughs in 
$\overline{Nu}$ and peaks in $T_{\text{avg}}$ and $T_{\text{max}}$ in panels (a) and (b) occur when a vortex group detaches from one edge and starts to form at the other edge, as shown in the second and fourth snapshots of each panel. During the process of vortex merging, the global heat transfer quantities first increase and then oscillate about fairly constant levels, as occurs during the first and third snapshot of panel (b). 

\begin{figure}[t]
    \centering
    \includegraphics[width=\linewidth]{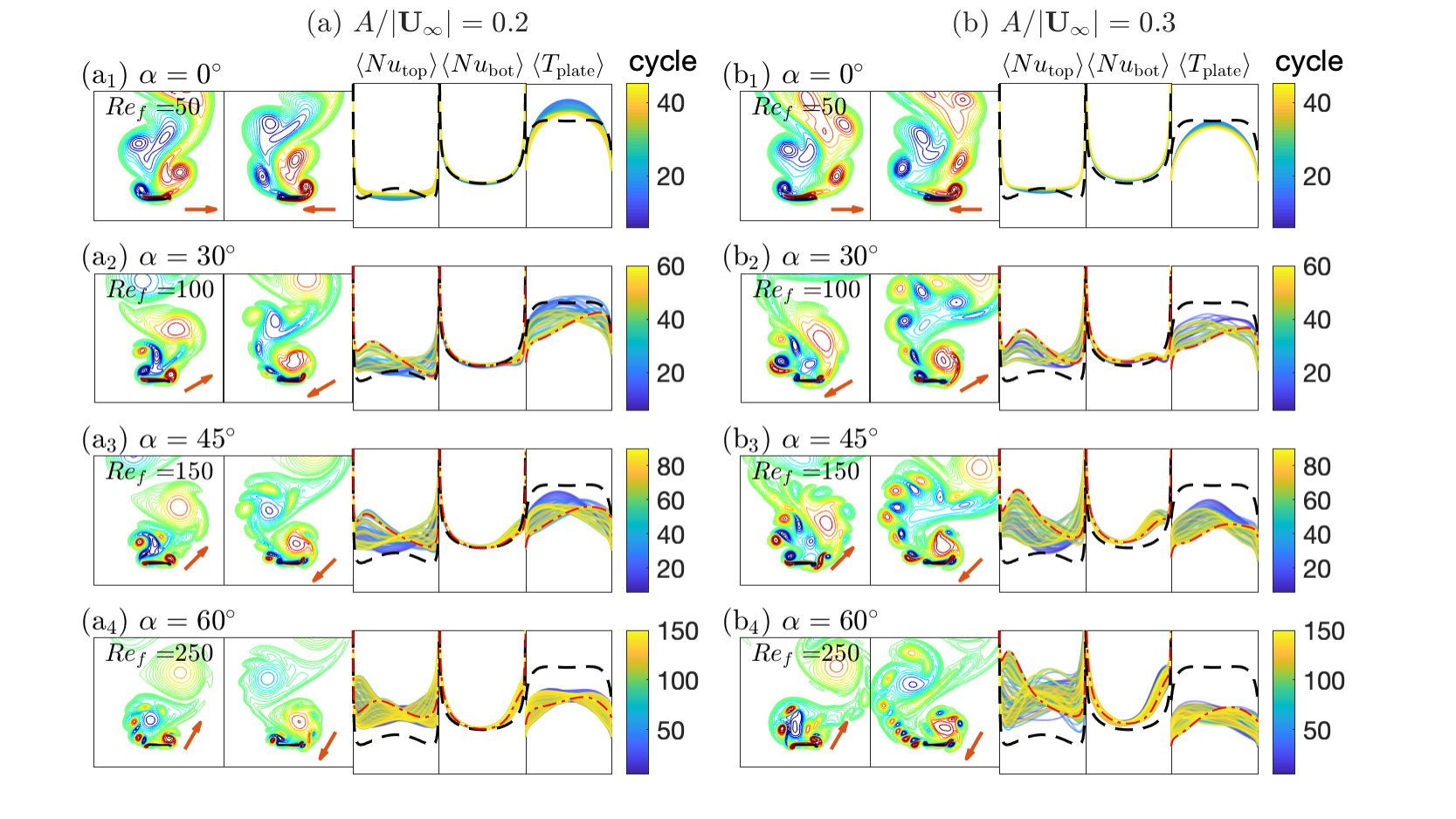}
    \caption{Vorticity fields and cycle-averaged local heat transfer quantities for an oscillating plate
    in a transverse flow ($\gamma=90^\circ$) with $A/|\boldsymbol{U}_{\infty}| = 0.2$ (a) and 0.3 (b) and $\alpha\neq90^\circ$, i.e. oscillation directions
    other than transverse. The red arrow in each vorticity panel indicates the instantaneous plate velocity direction. The red dash-dotted curves show instantaneous distributions of cycle-averaged local heat transfer quantities that are discussed in the main text. Supplementary movies named ``movie16'' and ``movie17'' are available \href{https://drive.google.com/drive/folders/1AUg5BWYRKvcFZnJcTIGqaiImf02z0a8a?usp=drive_link}{here} for panels (a$_1$) and (b$_3$) respectively.}
\label{fig:gamma90vorticityLocalHeatTransfer}
\end{figure}
Examples of flows and heat transfer quantities for the other plate oscillation directions are shown in figure~\ref{fig:gamma90vorticityLocalHeatTransfer}, for $A/|\boldsymbol{U}_{\infty}|=0.2$ (panel (a)) and 0.3 (panel (b)). Subpanels (a$_1$) and (b$_1$) show the in-plane-oscillation case ($\alpha=0^\circ$). As occurred with $\gamma = 45^\circ$, the most noticeable heat transfer enhancement relative to the steady case occurs at this value of $Re_f$, the lowest (see figure~\ref{fig:GlobalAllData}). The graphs of $\langle Nu_{\text{top}} \rangle$ and $\langle T_{\text{plate}} \rangle$ show that the local heat transfer is enhanced mainly at the edges and less enhanced or even worsened in the middle. The overall enhancement in both subpanels (a$_1$) and (b$_1$) is very limited. In these subpanels the cycle-averaged local heat transfer quantities have a left-right symmetry that is broken for the $\alpha$ in the following subpanels.
Here we have vortex grouping at both edges, with the positive vortex group (above the right edge) more prominent. It usually last longer, is more compact and intense, and is closer to the top wall than the negative vortex group. In subpanels (a$_2$), (b$_2$), (a$_3$), and (b$_3$), the negative vortex group is partially or fully separated into individual vortices, while the positive vortex group is a single compact region at the right edge. When the positive vortex group grows to a certain size, it is repelled by the negative vortex group and detaches from the plate. At higher $Re_f$, the negative vortex group also forms into a compact group attached to the plate, as shown in subpanels (a$_4$) and (b$_4$), but the positive vortex group still lasts longer. The second snapshots in the last three rows of subpanels show an instant when the positive vortex group is detaching from the plate and dipoles have been shed from the left edge. At these times, $\langle Nu_{\text{top}} \rangle$ has a local maximum near the left edge and a local minimum near the right edge, with the inverse pattern in $\langle T_{\text{plate}} \rangle$, as shown by the red dash-dotted lines.

\section{Input power}
\label{sec:Power}
% \begin{enumerate}
%     \item Compute the power to oscillate the plate.
%     \item Compute the power to drive the oncoming flow.
% \end{enumerate}
%
In the previous sections we studied the heat transfer of oscillating plates without considering what drives the oscillation and what input power may be required. 
Both natural (e.g. plants) and human-made structures can passively vibrate in ambient air and water flows, in which case no additional power source is required to produce the flow \citep{DH2001,HH2011,WC2022, alben2022, ma2022, ma2021, LLT2013}.
% \textcolor{red}{Dynamics of flags over wide ranges of mass and bending stiffness
% S Alben
% Physical Review Fluids 7 (1), 013903 ;Membrane flutter in three-dimensional inviscid flow
% C Mavroyiakoumou, S Alben
% Journal of Fluid Mechanics 953, A32;Dynamics of tethered membranes in inviscid flow
% C Mavroyiakoumou, S Alben
% Journal of Fluids and Structures 107, 103384;
% Wake states and frequency selection of a streamwise oscillating cylinder
% JS Leontini, DL Jacono, MC Thompson - Journal of Fluid Mechanics, 2013 - }
Many previous studies have examined the benefits of passive fluid-structure interaction for heat transfer specifically \citep{SOS2020,Gosselin2019,LPKR2017,SBS2015,KGSB2020,KC2004,GMA2016}.
% \textcolor{red}{
% Solano, Tomas, Juan C. Ordonez, and Kourosh Shoele. "Flapping dynamics of a flag in the presence of thermal convection." Journal of Fluid Mechanics 895 (2020): A8.
% Mechanics of a plant in fluid flow
% FP Gosselin, Journal of Experimental Botany, 2019
% Lee, J.B., Park, S.G., Kim, B., Ryu, J., and Sung, H.J., Heat transfer enhancement by flexible
% flags clamped vertically in a Poiseuille channel flow, international journal of heat and mass
% transfer, vol. 107, pp. 391-402, 2017
% Soti, A.K., Bhardwaj, R., and Sheridan, J., Flow-induced deformation of a flexible thin structure as manifestation of heat transfer enhancement, International Journal of Heat and Mass
% Transfer, vol. 84, pp. 1070-1081, 2015.
% 111. Kumar, V., Garg, H., Sharma, G., and Bhardwaj, R., Harnessing flow-induced vibration of a
% D-section cylinder for convective heat transfer augmentation in laminar channel flow, Physics
% of Fluids, vol. 32, no. 8, p. 083603, 2020.
% Kim, J. and Choi, H., An immersed-boundary finite-volume method for simulation of heat
% transfer in complex geometries;
% Enhanced forced convection heat transfer using small scale vorticity concentrations effected by flow driven, aeroelastically vibrating reeds
% A Glezer, R Mittal, S Alben
% Technical report, Georgia Institute of Technology Atlanta United States
% }.

In other situations, power is needed to oscillate the plate but the ambient flow is free. Flapping plates are a model for the wings or fins of flying and swimming animals in ambient flows \citep{Dabiri2009}. Under ``free locomotion," the organism reaches a steady state with nonzero forward velocity in a quiescent flow, in which case the ambient flow is simply the negative of the forward velocity of the organism \citep{AS2005, AWBA2012, HCTF2018, GVRT2019, TLHG2016, alben2021b}.
% (cite Coherent locomotion as an attracting state for a free flapping body
% PNAS, 2005; Dynamics of freely swimming flexible foils
% S. Alben, C. Witt, T.V. Baker, E.J. Anderson, and G.V. Lauder
% Physics of Fluids 24, 051901, 2012;Swimming performance, resonance and shape evolution in heaving flexible panels
% AP Hoover, R Cortez, ED Tytell - Journal of Fluid , 2018;On the fluid dynamical effects of synchronization in side-by-side swimmers
% R Godoy-Diana, J Vacher, V Raspa, B Thiria - Biomimetics, 2019;Role of body stiffness in undulatory swimming: insights from robotic and computational models
% ED Tytell, MC Leftwich, CY Hsu, BE Griffith, AH Cohen - Physical Review , 2016;Collective locomotion of two-dimensional lattices of flapping plates. Part 2. Lattice flows and propulsive efficiency
% S Alben
% Journal of Fluid Mechanics 915, A21). 
Other examples are energy-harvesting applications in which flapping bodies are used to harvest energy from the ambient flow \citep{ZP2009, Zhu2011}. 

One may also consider the power needed to oscillate the plate and drive the oncoming flow. Recent works have calculated the flows that maximize heat transfer for a given total power needed to drive the flow, for flows through channels \citep{alben2017improved, APG2024}, between parallel heated walls \citep{HCD2014, TD2017, MKS2018Maximal, MKS2018Optimal, STD2020, Kumar2022, alben2023}, %(cite hassanzadeh2014wall,tobasco2017optimal,motoki2018maximal,motoki2018optimal,souza2020wall,kumar2022three,alben2023transition), 
and in more general geometries \citep{alben2017optimal}. %(alben2017optimal). 

Here we will consider the input power in two situations. In the first, the oncoming flow is freely available from the environment but power is needed to oscillate the plate. In the second situation, power is needed both the drive the oncoming flow past the moving plate and to oscillate the plate.
In the previous sections we found that the best global heat transfer occurs with transverse oscillation ($\alpha=90^\circ$). However, more power may be required to sustain transverse oscillation because it displaces more fluid than other oscillation directions. Varying the oncoming flow direction from transverse to in-plane has a similar effect on the power needed to drive the flow.

First, we express the input power needed to oscillate the plate. The horizontal force exerted on the plate by the fluid is due to viscous shear stress on the two sides of the plate:
\begin{equation}\label{eq:Fx}
    F_x(t) = \displaystyle\frac{1}{Re_f}\int_0^1 [ \partial_y u(x, 0, t)]_-^+ \, \mathrm{d} x,
\end{equation}
where the bracket notation denotes the jump along the plate, the value on the top minus the value on the bottom. The vertical force on the plate is due to the pressure difference between the two sides:
\begin{equation}\label{eq:Fy}
    F_y(t) = \displaystyle\int_0^1 [ -p(x, 0, t)]_-^+ \, \mathrm{d} x.
\end{equation}
The power needed to move the plate is the product of the force the plate exerts on the fluid ($-F_x,-F_y$) with the sinusoidal velocity given in equation~\eqref{eq:BodyVel}, i.e.
\begin{equation}\label{eq:Pbody}
    P_{\text{body}} (t) = -(F_x(t)\times U_b(t) + F_y(t) \times V_b(t)), \quad \tilde{P}_{\text{body}} = Re_f^3P_{\text{body}}.
\end{equation}
Equation~\eqref{eq:Pbody} is the total input power required if the oncoming flow is freely available. In \eqref{eq:Pbody} we rescale the power by multiplying it by $Re_f^3$. The rescaled power corresponds to using $\nu^*/\ell^*$ as the characteristic velocity scale instead of $f^*\ell^*$.
Then the power nondimensionalization does not depend on frequency, which allows us to compare results with different frequencies (i.e. different $Re_f$). 

To consider the power required for the oncoming flow, we can write the energy balance for a fixed rectangular region $R$ containing the oscillating plate \citep{KCD2016Book}:
\begin{align}
    \frac{d}{dt}\iint_R \frac{1}{2} u_i^2 dA = -\oint_{\partial R + \text{plate}} \frac{1}{2}u_i^2 u_j n_j ds + \oint_{\partial R + \text{plate}} u_i \tau_{ij} n_j ds - \iint_R \frac{2}{Re_f}e_{ij}^2 dA. \label{eq:EnergyEqn}
\end{align}
We have used tensor index notation to 
express the terms more easily. The equation says that the rate of change of fluid kinetic energy in the rectangle equals the sum of three terms: the net inward flux of kinetic energy ($n_j$ is the outward unit normal vector), the rate of work done on the fluid in $R$ by pressure and viscous stresses from the oscillating plate and at the outer boundary of $R$ ($\tau_{ij}$ is the stress tensor), and the rate of viscous dissipation in the rectangle ($e_{ij}$ is the rate-of-strain tensor). When we take a long-time-average, the left hand side is zero. Hence the average power dissipated is supplied by the stresses at the plate and outer boundary and by kinetic energy flux at the outer boundary. In heat exchangers, the dominant contribution from the outer boundary is from the pressure stresses there, and is called the ``pumping power" \citep{RK2010Book}. 

We define the average input power required for the entire flow, the oncoming flow plus that induced by the oscillating plate, to be the time average of all the terms at the plate and at the outer boundary in (\ref{eq:EnergyEqn}), in the limit that the outer boundary $\partial R$ tends to infinite distance from the body. By (\ref{eq:EnergyEqn}), the sum of all the terms at the plate and outer boundary equals the average viscous power dissipation rate.
To calculate this term, we consider the flow in the frame that is at rest in the far field by adding $-\bm{U}_{\infty}$ to the fluid velocity. In this frame the body moves upstream while it oscillates and has velocity 
$\bm{U}_{b}-\bm{U}_{\infty}$. For this flow, $e_{ij}$ is the same as before. As the boundary of $R$ tends to infinite distance from the body, all the work is done at the body surface, because then there is no other source of energy input. In effect we have replaced the power to drive the flow past the body with the power to move the body upstream into a quiescent flow.

%In order to calculate this term, we can take our computational domain as $R$ and integrate the last term in (\ref{eq:EnergyEqn}). Because the far-field grid is coarser than near the body, we instead compute it by using 
%quantities computed on the body only. To do this 
 
Thus, the average total power required for the plate oscillation plus the oncoming flow is
\begin{equation}\label{eq:Ptotal}
    \langle P_{\text{total}}(t) \rangle =  -F_x(t)\times(U_{b}(t)-U_{\infty}) -F_y(t) \times (V_{b}(t)-V_{\infty}), \quad \langle \tilde{P}_{\text{total}} \rangle= Re_f^3\langle P_{\text{total}} \rangle,
\end{equation}
which is (\ref{eq:Pbody}) but with the body velocity in the frame where it moves upstream into a quiescent far field flow. In \eqref{eq:Ptotal}, as in \eqref{eq:Pbody}, we rescale the power by $Re_f^3$ to remove the frequency dependence of the nondimensionalization. 

\begin{figure}[ht!]
    \centering
    \begin{subfigure}[b]{\textwidth}
        \centering
        \caption{}
        \includegraphics[width=0.8\textwidth]{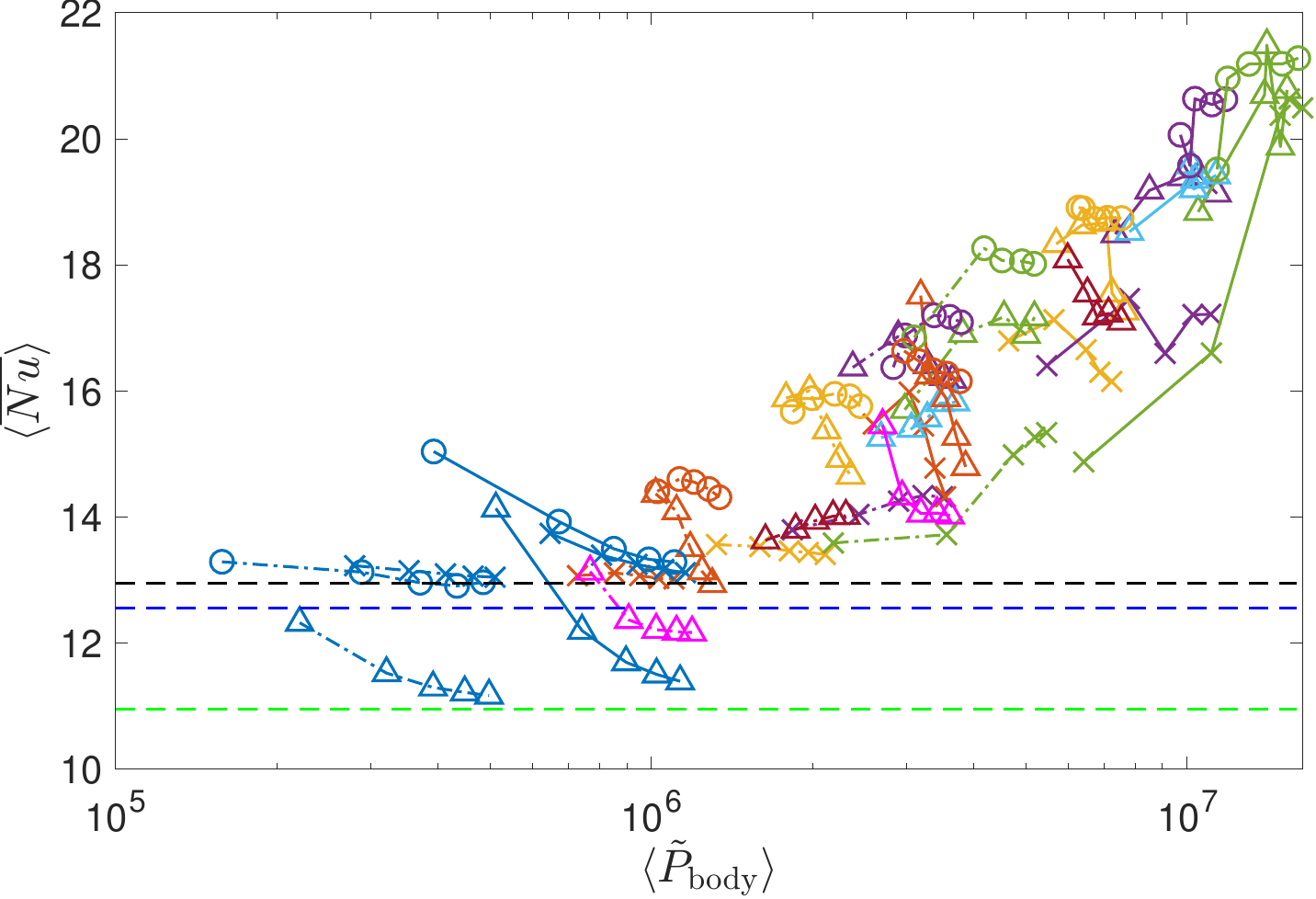}
    \end{subfigure}
    \begin{subfigure}[b]{\textwidth}
        \centering
        \caption{}
        \includegraphics[width=0.8\textwidth]{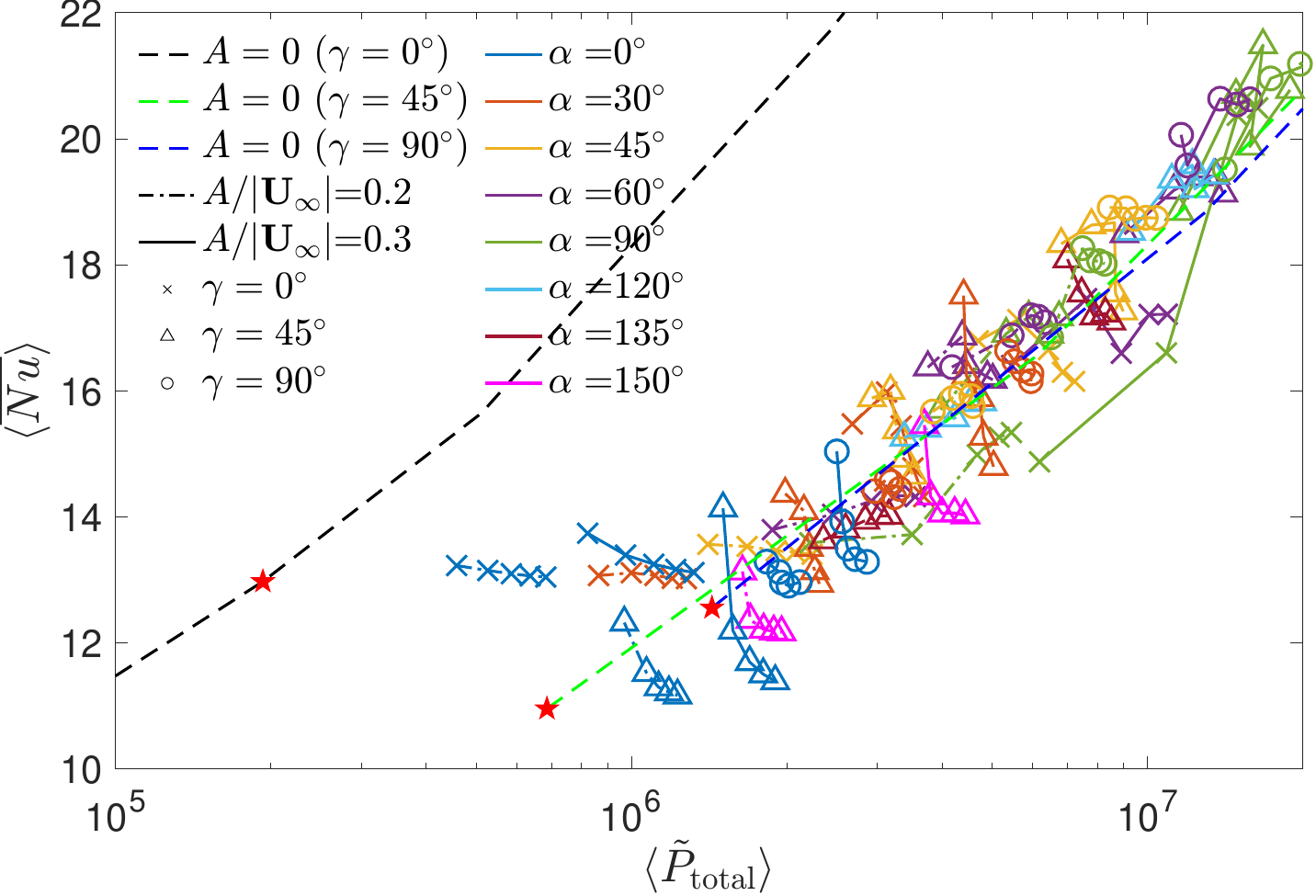}
    \end{subfigure}
    \caption{$\langle \overline{Nu} \rangle$ versus (a) the time averaged power needed to oscillate the plate, and (b) the time averaged total power including the power needed to drive the oncoming uniform flow. The three horizontal dashed lines in panel (a) show $\langle \overline{Nu} \rangle$ at $Re_U = 100$ for the non-oscillating plates, for which $\langle \tilde{P}_{\text{body}} \rangle$ = 0. In panel (b) the data for the non-oscillating plates correspond to $Re_U$ ranging from 100 to 500 for the green and blue dashed lines, and also include $Re_U$ below 100 for the black dashed line. The values with $Re_U = 100$ are marked with red stars.}
    \label{fig:PowerBodyNu}
\end{figure}
In figure~\ref{fig:PowerBodyNu} we plot $\langle \overline{Nu} \rangle$ versus $\langle \tilde{P}_{\text{body}} \rangle$ (panel (a)) and $\langle \tilde{P}_{\text{total}} \rangle$ (panel (b)) for all of the isothermal plate cases. We compute $\langle \tilde{P}_{\text{body}} \rangle$ and $\langle \tilde{P}_{\text{total}} \rangle$ using the same time spans as for $\langle \overline{Nu} \rangle$, discussed in section~\ref{sec:GlobalLocalHeatTransfer} and appendix~\ref{app:GridIndependence}. In each panel, the Pareto optimal cases lie at the upper left of each region of points. For such cases, no other cases provide equal or greater $\langle \overline{Nu} \rangle$ with equal or smaller input power.

For panel (a), the input power is given by equation~\eqref{eq:Pbody} and the oncoming flow is free. Three horizontal dashed lines mark $\langle \overline{Nu} \rangle$ at $Re_U$ = 100 for the non-oscillating plates, for which $\langle \tilde{P}_{\text{body}} \rangle = 0$.
The cases connected with dashed-dotted lines and solid lines have $A/|\boldsymbol{U}_{\infty}|=0.2$ and 0.3 respectively. For each group of connected points, $Re_f$ increases from left to right.  

A few cases with $\gamma=45^\circ$ and $\alpha=0^\circ$ or $30^\circ$ are below the horizontal dashed lines with $\gamma=0^\circ$ and/or $45^\circ$, so the non-oscillating plates with these $\gamma$ values achieve better global heat transfer. Above the dashed lines, the Pareto optimal cases mostly have $\gamma=45^\circ$ and $90^\circ$. 
The Pareto optimal cases with the lowest $\langle \tilde{P}_{\text{body}} \rangle$ (left side) have in-plane oscillations in transverse flow ($\alpha=0^\circ$ and $\gamma=90^\circ$). These cases give a small increase in $\langle \overline{Nu} \rangle$ relative to the steady plate as well as a few oscillating plates with larger $\alpha$ at $\gamma = 0^\circ$ and $45^\circ$. Moving rightward to intermediate $\langle \tilde{P}_{\text{body}} \rangle$,
Pareto optimal cases typically have $A/|\boldsymbol{U}_{\infty}|=0.2$ and $\gamma = 45^\circ$ and $90^\circ$.
In general, $\langle \tilde{P}_{\text{body}} \rangle$ increases strongly with $\alpha$, i.e. with the degree of transverse oscillation. At the far right, where $\langle \tilde{P}_{\text{body}} \rangle$ is largest, the optimal cases have $A/|\boldsymbol{U}_{\infty}|=0.3$ and most have $\gamma = 90^\circ$, transverse oncoming flow. 

Panel (b) shows the results including the power for the oncoming flow, i.e. with $\langle \tilde{P}_{\text{total}} \rangle$ instead of $\langle \tilde{P}_{\text{body}} \rangle$. Now nonzero power is required for the steady plates due to drag from the oncoming flow. The three dashed lines show $\langle \overline{Nu} \rangle$ and $\langle \tilde{P}_{\text{total}} \rangle$ for the steady plates as $Re_U$ ranges from 10 (black dashed line, $\gamma = 0^\circ$) or 100 (green and blue dashed lines, $\gamma = 45^\circ$ and 90$^\circ$) up to 500, with the values at $Re_U$ = 100 marked by red stars. By varying $Re_U$ from 100, we extend the steady cases from the red stars to the dashed curves, so that we can cover the $\langle \overline{Nu} \rangle$ and $\langle \tilde{P}_{\text{total}} \rangle$ ranges of all the oscillating plate data. With this comparison we imagine that we have a fixed amount of input power $\langle \tilde{P}_{\text{total}} \rangle$ that we can use to either move a flow at $Re_U$ = 100 past an oscillating plate or move a flow at various $Re_U$ past a steady plate, to obtain the largest $\langle \overline{Nu} \rangle$. Panel (b) shows that the best option---i.e. the Pareto front---is in-plane flow past a steady plate, the black dashed line. It requires much less power at every $\langle \overline{Nu} \rangle$ value, even compared to in-plane flow with in-plane oscillations (blue crosses, $\alpha = 0^\circ$). The dashed lines for $\gamma=45^\circ$ and $\gamma=90^\circ$
are far to the right, and here pressure drag dominates skin friction for a steady plate. If we constrain the oncoming flow orientation at $\gamma=45^\circ$ or $90^\circ$, then many oscillating cases, particularly at low $Re_f$, lie above and to the left of the corresponding steady curves (the green and blue dashed lines). For these cases, larger $\langle \overline{Nu} \rangle$ is obtained with an oscillating plate at $Re_U$ = 100 than for a steady plate with the same $\langle \tilde{P}_{\text{total}} \rangle$ (which has $Re_U > 100$).

By comparing the groups of points connected by a line of a given color and line type in panels (a) and (b), we can see the effect of adding the power needed to drive the flow. This power is the amount by which the lines are shifted to the right in (b) relative to (a) (with $\langle \overline{Nu} \rangle$, the vertical position, unchanged). Many of the lines are also more compressed horizontally in (b). The horizontal shift and compression is most noticeable on the left side, i.e. for the lower-power cases. Here the power for the oncoming flow makes a proportionately larger contribution to the overall power.
On the right side, the power for the plate oscillation is dominant, and such cases typically have 
$A/|\boldsymbol{U}_{\infty}|=0.3$ and $\alpha$ close to $90^\circ$, i.e. a strong transverse plate oscillation.

We present Pareto fronts for the fixed-flux case in appendix \ref{app:ParetoFixedFlux}. The optimal cases are generally similar to those with fixed plate temperature.

\section{Conclusion}\label{sec:conclusion}

We studied the effect of vortex dynamics on heat transfer for an oscillating plate in an oncoming flow with Reynolds number $Re_U = 100$. We varied the oncoming flow direction ($\gamma$) from in-plane to transverse, the plate oscillation direction ($\alpha$) from in-plane to transverse, and studied the effects of the plate oscillation velocity, frequency, and amplitude.
We classified the resulting vortex dynamics and heat transfer patterns into a small number of main categories.

Steady plates provided benchmark flows, either steady Prandtl-boundary-layer-type flows (at $\gamma = 0^\circ$) or von K\'{a}rm\'{a}n vortex wakes (at $\gamma = 45^\circ$ and $90^\circ$).   
Without a natural wake frequency, many in-plane oncoming flows ($\gamma = 0^\circ$) resulted in periodic dynamics with the period of the plate oscillation. A few cases with transverse oscillation had period 1/2, up-down symmetry, and smaller heat transfer, and a few other cases at larger oscillation velocity ($A/|\boldsymbol{U}_{\infty}|=0.3$) were nonperiodic and had larger heat transfer. 

With natural wake frequencies in addition to the plate frequency, oblique and transverse flows ($\gamma = 45^\circ$ and $90^\circ$) resulted in dynamics that were almost periodic at multiples of the plate oscillation period, or nonperiodic, the latter occurring mainly with more transverse oscillations at larger velocity ($A/|\boldsymbol{U}_{\infty}|=0.3$).

The global heat transfer (figure~\ref{fig:GlobalAllData}) was most strongly affected by the oscillation velocity ($A/|\boldsymbol{U}_{\infty}|$) and direction ($\alpha$), being larger with a faster and more transverse oscillation. In this case the shed vortices were stronger and the dynamics were less periodic, enhancing mixing in the wake region.
The effect of $\gamma$ was relatively small for the cases with better heat transfer ($\gamma \neq 0, \alpha \neq 0$). For a given oscillation velocity ($A/|\boldsymbol{U}_{\infty}|$), increasing the oscillation frequency ($Re_f$) while decreasing the oscillation amplitude had relatively little effect on the global heat transfer, except in a few cases with transitions between types of periodic and nonperiodic dynamics. Local heat transfer was altered by $\gamma$ and $Re_f$, however, through changes in the sizes and locations of shed vortices.

With in-plane flow ($\gamma=0^\circ$) and no plate oscillation, heat transfer is maximum at the leading edge and temperature is maximum near the trailing edge. Adding in-plane oscillation ($\alpha=0^\circ$) gives modest improvements near the trailing edge, but the boundary layers remain intact. With oblique ($\alpha=45^\circ$) and transverse ($\alpha=90^\circ$) oscillations, vortices are shed from the edges, decreasing heat transfer near the leading edge and but enhancing it over the rest of the plate, giving a net improvement. 

With oblique and transverse flow
($\gamma = 45^\circ$ and $90^\circ$), the heat transfer is controlled by the formation of large regions of vorticity, from the merging of vortices shed over multiple cycles. These ``vortex groups" tend to form near one edge and enhance heat transfer mainly at the opposite edge. After many cycles the group detaches from one edge, a new group of oppositely-signed vorticity forms near the other edge, and the cycle repeats. 
The net effect of the vortex groups is to greatly increase heat transfer on the wake side, to a level comparable to that on the side facing the oncoming flow. On the wake side, heat transfer is increased most on the portion farther from the vortex group, which draws colder fluid there. 
We already mentioned that the shed vorticity was stronger with more transverse oscillations ($\alpha$ closer to $90^\circ$). At $\gamma = 45^\circ$, pairs of cases (supplementary angles) have the same degree of transverse motion but one ($\alpha < 90^\circ$) moves more along the oncoming flow and the other across it ($\alpha > 90^\circ$). The former case leads to stronger, more compact vortex groups and somewhat better heat transfer. Vortex grouping was also seen in some cases of in-plane flow ($\gamma=0^\circ$) with transverse oscillation, where it also enhanced heat transfer. 

We considered plates with fixed temperature or fixed heat flux. In general the same parameters are optimal in both cases, with some variability when the maximum (rather than average) plate temperature is considered in the latter case. 

We also considered the input power needed to oscillate the plate and to drive the entire flow including the oncoming flow. The input power can be neglected if the oncoming flow is provided by ambient flows in the environment and the plate oscillation occurs through 
passive fluid-structure interaction. In this case the best case---the one that maximizes $\langle \overline{Nu} \rangle$---has $A/|\boldsymbol{U}_{\infty}|=0.3$, $\alpha = 90^\circ$, and
$\gamma = 45^\circ$ (with $\gamma = 90^\circ$ very close). If power is needed to oscillate the plate, optimality is given by the Pareto front in $\langle \overline{Nu} \rangle$--$\langle \tilde{P}_{\text{body}} \rangle$ space. The front includes the steady plate with $\gamma = 0^\circ$ (with $\langle \tilde{P}_{\text{body}} \rangle$ = 0) as well as many oscillating plates with larger $\langle \overline{Nu} \rangle$ and nonzero $\langle \tilde{P}_{\text{body}} \rangle$. These cases 
have $A/|\boldsymbol{U}_{\infty}|=0.2$ and 0.3, $0^\circ \leq \alpha \leq 90^\circ$, and
$\gamma = 45^\circ$ and $90^\circ$. If power is also needed to drive the oncoming flow, the optimal case (the Pareto front in $\langle \overline{Nu} \rangle$--$\langle \tilde{P}_{\text{total}} \rangle$ space) is
the steady plate with $\gamma = 0^\circ$. If we restrict $\gamma$ to $45^\circ$ and $90^\circ$, then many of the oscillating-plate cases that are Pareto optimal in $\langle \overline{Nu} \rangle$--$\langle \tilde{P}_{\text{body}} \rangle$ space are also Pareto optimal in $\langle \overline{Nu} \rangle$--$\langle \tilde{P}_{\text{total}} \rangle$ space.

\section*{Acknowledgments}
This research was supported by the NSF-DMS Applied Mathematics program under
award number DMS-2204900.

\section*{Supplementary material}
%\begin{mdframed}
Supplementary movies along with a caption list are available at \url{https://drive.google.com/drive/folders/1AUg5BWYRKvcFZnJcTIGqaiImf02z0a8a?usp=drive_link}.
%\end{mdframed}

%\clearpage
\appendix
\section{Grid Independence Study}\label{app:GridIndependence}

In this appendix we give numerical tests to explain the chosen domain and grid sizes in table~\ref{tab:domainBC}. We vary the domain size according to the plate orientation $\gamma$ to avoid boundary effects. For different $\gamma$, we compare the time series of the global heat transfer quantities, $\overline{Nu}$, $T_{\max}$, and $T_{\text{avg}}$ using different grid sizes with different domain sizes. We also investigate the variation of the time averages of the global heat transfer quantities when we average over different numbers of oscillation cycles. 

\begin{figure}[t]
    \centering
    \includegraphics[width=0.9\textwidth]{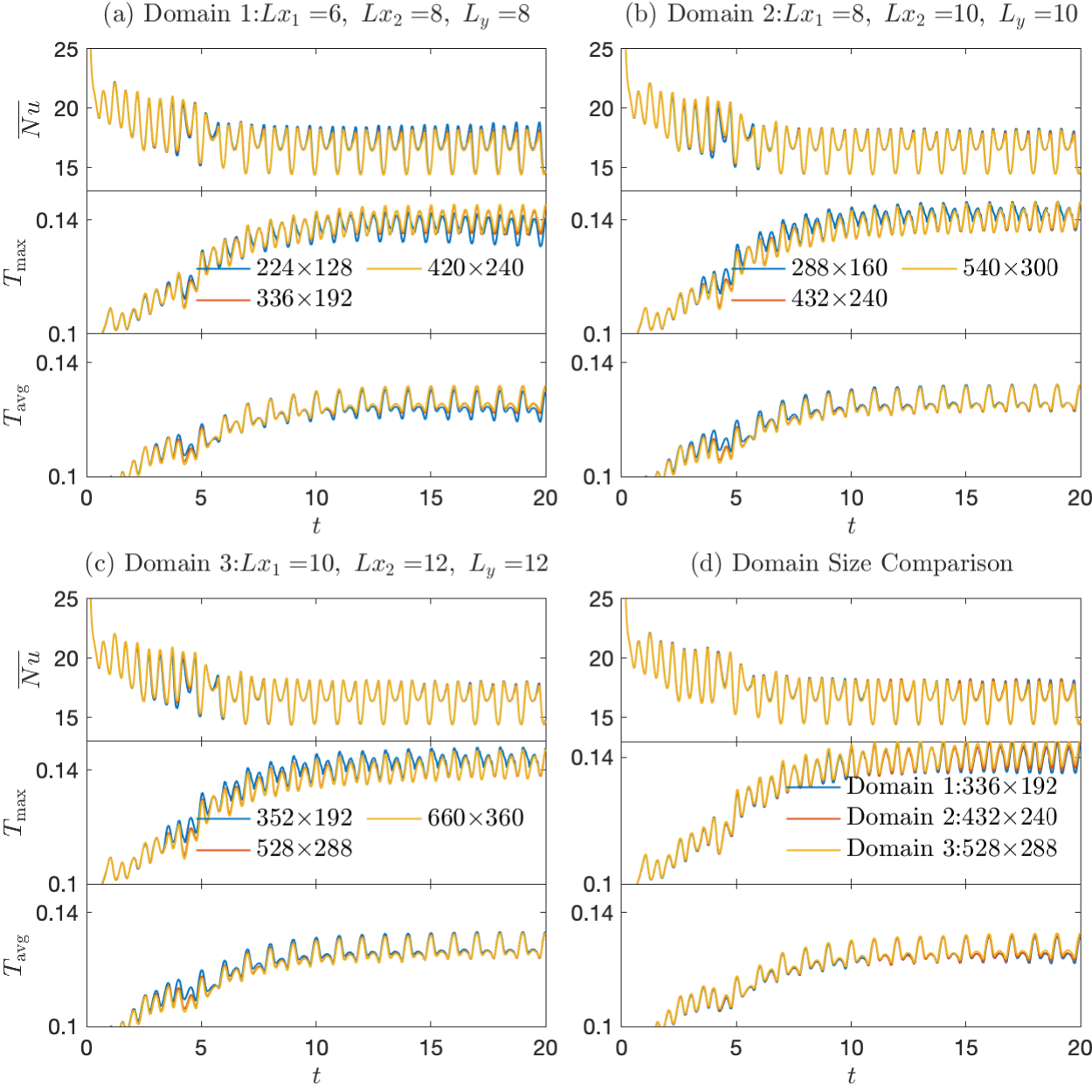}
    \caption{Comparisons of the time series of the global heat transfer quantities, $\overline{Nu}$, $T_{\max}$, and $T_{\text{avg}}$ with different domain sizes and grid sizes when the oncoming flow is horizontal ($\gamma = 0^\circ$). The other simulation parameters are: $Re_f=100$, $Re_U=100$, $\alpha = 90^\circ$ and $A/|\bm{U}_{\infty}|=0.2$.}
    \label{fig:GridGamma0-1}
\end{figure}
\begin{figure}[ht!]
    \centering
    \begin{subfigure}[b]{0.9\textwidth}
        \includegraphics[width=\textwidth]{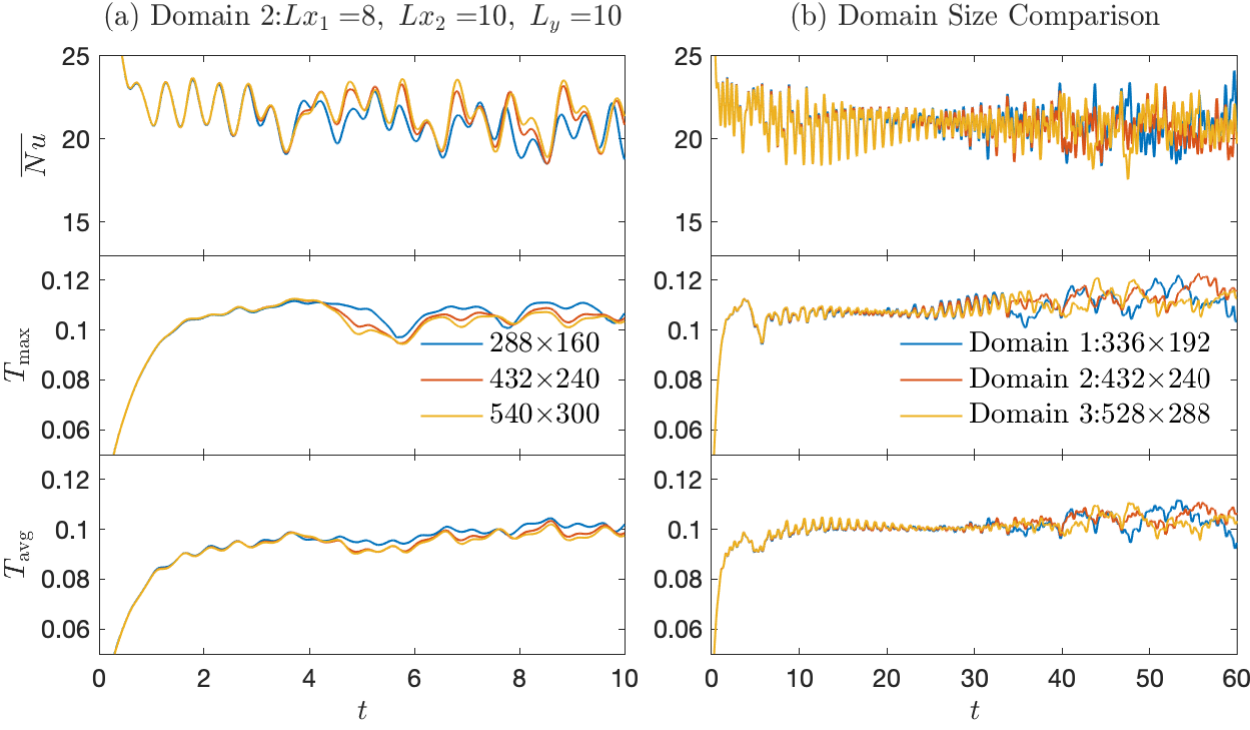}
    \end{subfigure}
    \begin{subfigure}[b]{\textwidth}
    \setcounter{subfigure}{2}
    \caption{}
    \begin{center}
        \includegraphics[width=0.9\textwidth]{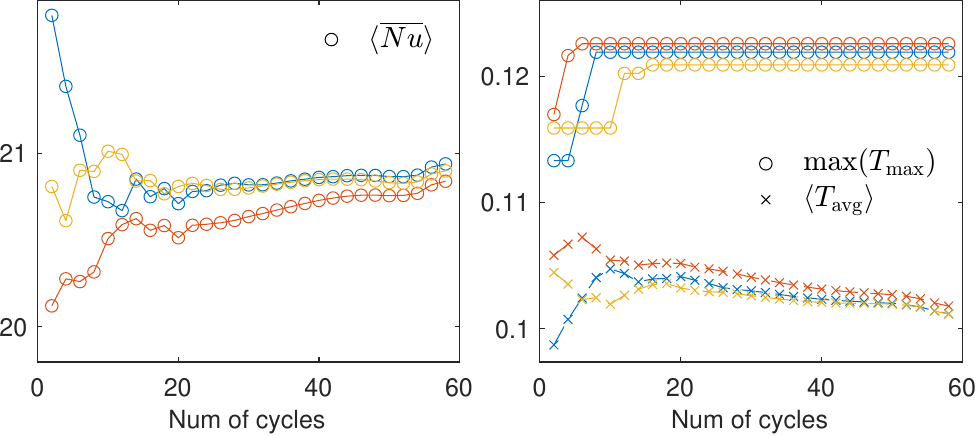}
    \end{center}      
    \end{subfigure}
    \caption{Comparisons of the global heat transfer quantities $\overline{Nu}$, $T_{\max}$, and $T_{\text{avg}}$, and their time averages, using different grid sizes (a) and domain sizes (b) when the oncoming flow is horizontal ($\gamma = 0^\circ$) and $Re_f=200$, $Re_U=100$, $\alpha = 90^\circ$ and $A/|\bm{U}_{\infty}|=0.3$. (c) Time averaged global heat transfer quantities using different numbers of cycles in the three computational domains. The correspondence between the curve color and the domain size is the same as in panel (b).}
    \label{fig:GridGamma0-2}
\end{figure}
%
%
% \begin{figure}
%     \centering
%     \includegraphics[width=0.7\textwidth]{Gamma0DiffDomainDiffCyclesTimeAvgLocal.pdf}
%     \caption{Comparison of the time averaged local Nusselt number (panel (a) and (b)) and the time averaged plate temperature distribution (panel (c) based on different number of cycles in different sizes of computational domains. The last 10, 20, 30, 40 and 50 cycles are used from top to bottom for each column. The simulation parameters are $\gamma = 0^\circ$, $Re_f=200$, $U_{\infty} = 0.5$, $A=0.15$, and $\alpha = 90^\circ$.}
%     \label{fig:GridGamma0-3}
% \end{figure}
%
%
% \begin{figure}
%     \centering
%     \includegraphics[width=0.7\textwidth]{Gamma0DiffCyclesTimeAvgLocal.pdf}
%     \caption{Comparison of the time averaged local Nusselt number and plate temperature distribution based on different number of cycles in domain 2. The simulation parameters are $\gamma = 0^\circ$, $Re_f=200$, $U_{\infty} = 0.5$, $A=0.15$, and $\alpha = 90^\circ$.}
%     \label{fig:GridGamma0-4}
% \end{figure}

When the oncoming flow is horizontal ($\gamma = 0^\circ$), we use a computational domain with horizontal length about twice the vertical length. In figure~\ref{fig:GridGamma0-1}, we compare the time series of the three global heat transfer quantities in three computational domains. The number of grid points is roughly proportional to the domain size. For example, the ratios of $x$ grid points satisfy: $288/224\approx (8+10)/(6+8)$ and $352/288\approx (10+12)/(8+10)$. We find that Domain 2 with the intermediate mesh of $432\times 240$ can produce results with sufficient accuracy. Since it is a periodic flow, the time averages are taken over the last 2 cycles of the plate oscillation. This mesh is also studied in a case with nonperiodic, irregular dynamics in figure~\ref{fig:GridGamma0-2}. Here close agreement with different meshes and domain sizes is only observed in the first several cycles. In panel (c), we show the time averaged quantities $\langle \overline{Nu} \rangle$, $\max{T_{\max}}$ and $\langle T_{\max} \rangle$ using averages over different numbers of cycles. Note that the number of cycles is counted backwards starting from the ending time. For example, if the number of cycles is 2, it means that the last 2 cycles are used to compute the time averages. We see large variations in the time averaged quantities if the number of cycles used is less than 5. Using more than 5 cycles, the time averaged quantities typically vary less than $\pm 5\%$ (note that the range of $y$ axis is quite small). Furthermore, the results do not change much from $20$ to $50$ cycles. It is somewhat coincidental that the time averaged results with the coarsest and finest meshes are closer, because the agreements between the time series in panels (a) and (b) are less close: in panel (a), the finest mesh agrees better with the middle mesh than the coarsest mesh. In this case, we take the last $40$ cycles to compute the time averaged quantities. For the other nonperiodic and irregular cases, we do similar comparisons, and the time span chosen to compute the time average is usually at least half of the total simulated time.

\begin{figure}[ht!]
    \centering
    \includegraphics[width=\textwidth]{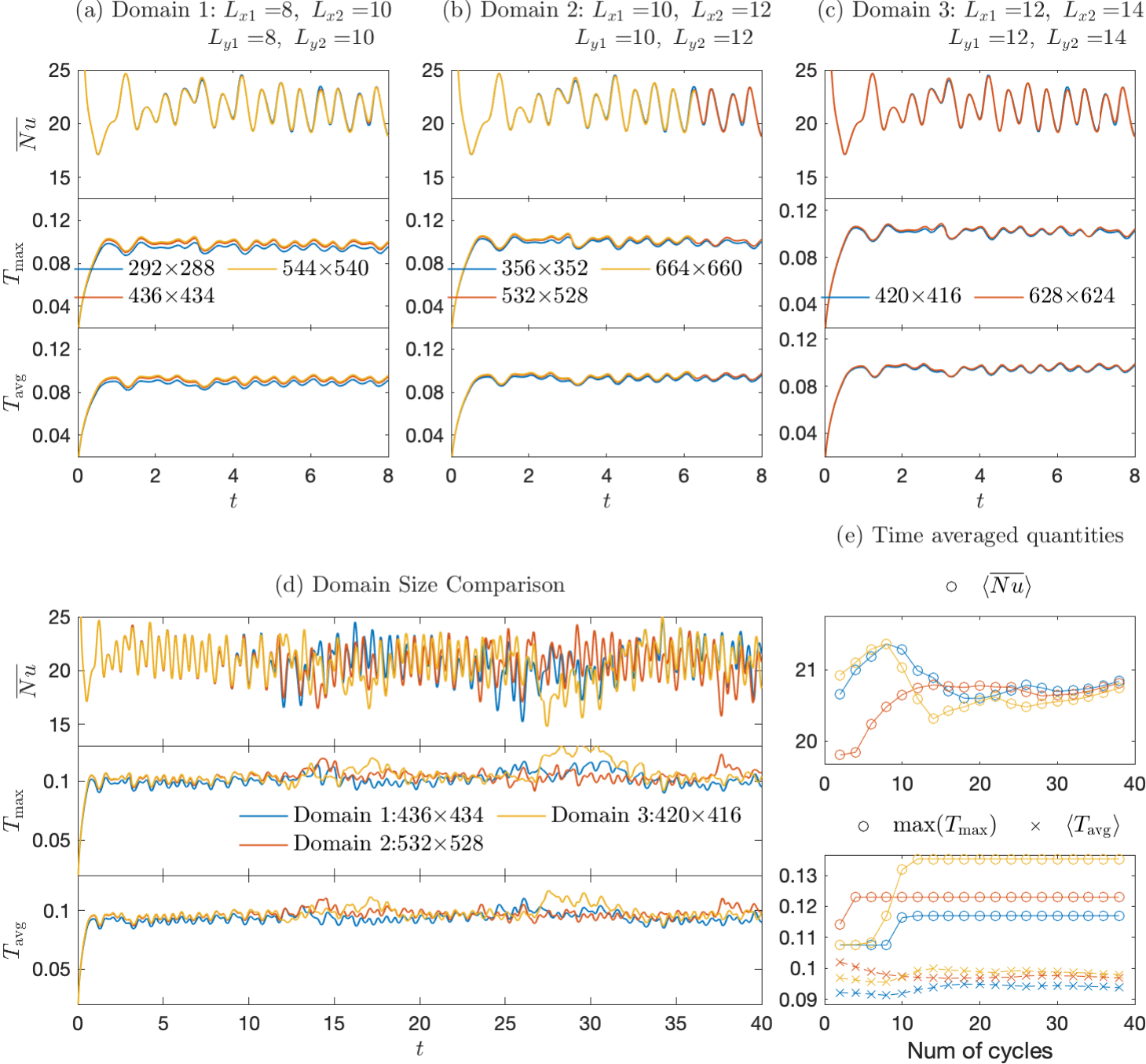}
    \caption{Comparisons of the global heat transfer quantities $\overline{Nu}$, $T_{\max}$, and $T_{\text{avg}}$, and their time averages, using different domain and grid sizes when the oncoming flow is oblique ($\gamma = 45^\circ$) and $Re_f=100$, $Re_U=100$, $\alpha = 90^\circ$ and $A/|\bm{U}_{\infty}|=0.3$. The correspondence between the curve color and the domain size in panel (e) is the same as in panel (d).}
    \label{fig:GridGamma45}
\end{figure}
\begin{figure}[ht!]
    \centering
    \includegraphics[width=\textwidth]{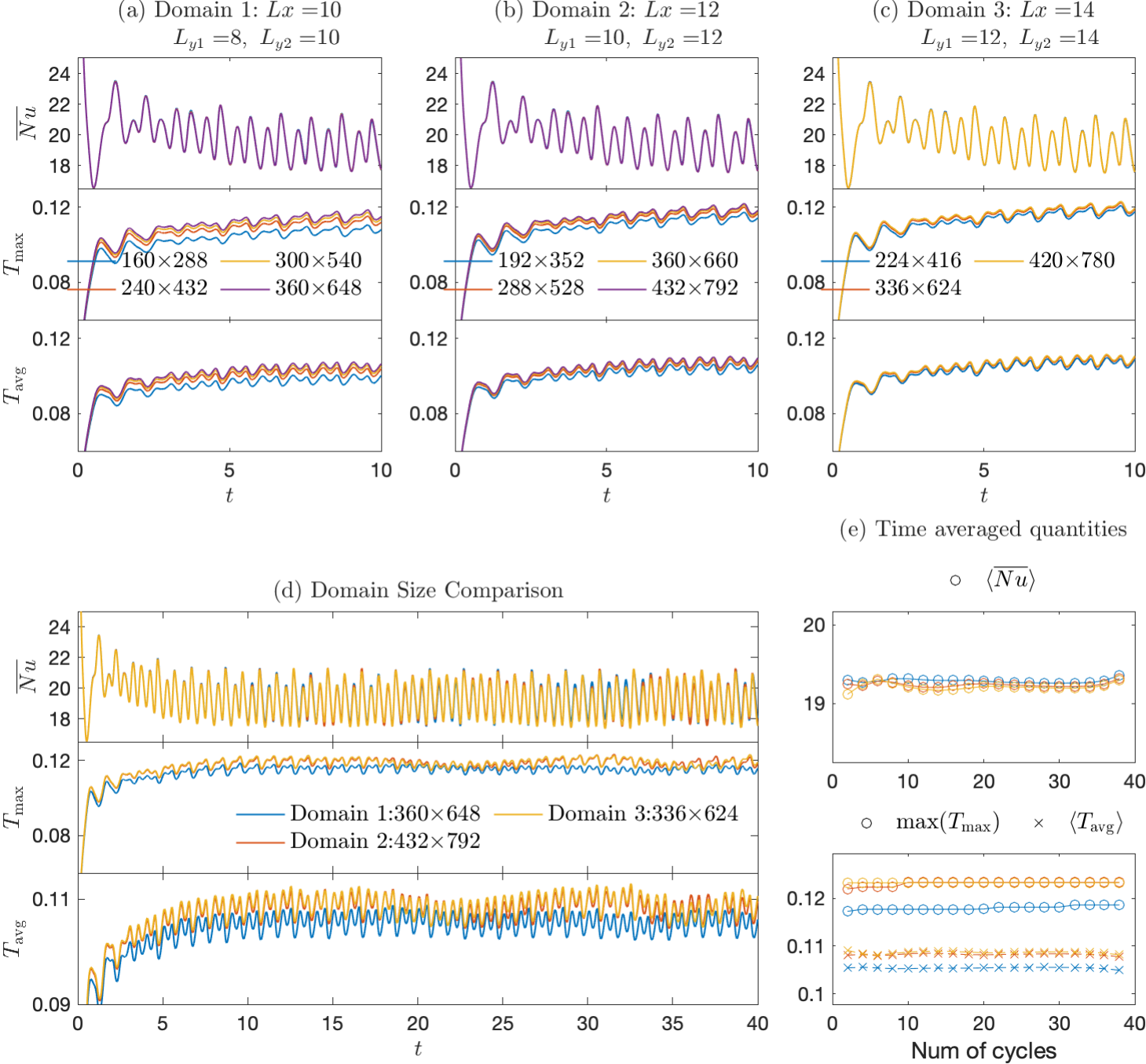}
    \caption{Comparisons of the global heat transfer quantities $\overline{Nu}$, $T_{\max}$, and $T_{\text{avg}}$, and their time averages, using different domain and grid sizes when the oncoming flow is transverse ($\gamma = 90^\circ$) and $Re_f=100$, $Re_U=100$, $A=0.3$, $\alpha = 60^\circ$ and $A/|\bm{U}_{\infty}|=0.3$. The correspondence between the curve color and the domain size in panel (e) is the same as in panel (d).}
    \label{fig:GridGamma90-2}
\end{figure}
For the oblique oncoming flow ($\gamma=45^\circ$), we use a square computational domain since the oncoming flow has equal horizontal and vertical components. Panel (a), (b), and (c) of figure~\ref{fig:GridGamma45} show the time series of the three global heat transfer quantities computed with different domain sizes. In each computation domain, we have good agreement between the results at different grid sizes. Compared to the fixed-temperature results, we find that the fixed-heat-flux results are more sensitive to the grid and domain sizes: in panels (a) and (b), $T_{\max}$ and $T_{\text{avg}}$ on the coarsest mesh are noticeably smaller, unlike $\overline{Nu}$. We compare the results with different computational domain sizes in panel (d). The middle mesh is chosen for Domain 1 and 2 whereas the coarsest mesh is chosen for Domain 3 because its results agree well with those of the finest mesh. In the first several cycles of plate oscillations, $T_{\max}$ and $T_{\text{avg}}$ for Domains 2 and 3 agree well and are smaller for Domain 1. Since boundary effects are more apparent for Domain 1 we use Domain 3 with the $420\times 416$ grid because it is moderately expensive computationally. This case is nonperiodic and irregular, as shown by the time series after 15 cycles. However, the time averaged $\langle \overline{Nu} \rangle$ and $\langle T_{\text{avg}} \rangle$ in the three domains using more than 15 cycles of oscillations agree well, as shown in panel (e). The large variations in $\max(T_{\max})$ are expected because, as the temperature of the hottest spot of the plate at one moment, it is sensitive to the irregular dynamics. 

Finally, we show the convergence study for transverse oncoming flow ($\gamma=90^\circ$) in figure~\ref{fig:GridGamma90-2}. Similar to the case of $\gamma=0^\circ$, the computational domain for $\gamma=90^\circ$ has a vertical dimension about twice the horizontal dimension. The case shown is not periodic but not as irregular as the case in figure~\ref{fig:GridGamma45}, as the time series computed in different domains and with different grid sizes tend to agree for the entire simulated time. We again observe that the results with fixed heat flux are more sensitive to the domain size and grid size. For this case Domain 1 is less accurate for the global heat transfer quantities, as shown in panel (d) and (e). Based on the comparisons in panel (a)--(e), we choose Domain 3 with a mesh of $336\times 624$ for transverse oncoming flow. 
%
%
% \begin{figure}
%     \centering
%     \includegraphics[width=0.7\textwidth]{Gamma45DiffDomainDiffCyclesTimeAvgLocal.pdf}
%     \caption{Comparison of the time averaged local Nusselt number (panel (a) and (b)) and the time averaged plate temperature distribution (panel (c) based on different number of cycles in different sizes of computational domains. The last 10, 20, and 30 cycles are used from top to bottom for each column. The simulation parameters are $\gamma = 45^\circ$, $Re_f=100$, $|\boldsymbol{U}|_{\infty} = 1$, $A=0.3$, and $\alpha = 90^\circ$.}
%     \label{fig:GridGamma45-2}
% \end{figure}
%
%
% \begin{figure}
%     \centering
%     \includegraphics[width=0.7\textwidth]{Gamma45DiffCyclesTimeAvgLocal.pdf}
%     \caption{Comparison of the time averaged local Nusselt number and plate temperature distribution based on different number of cycles in domain 3. The simulation parameters are $\gamma = 45^\circ$, $Re_f=100$, $|\boldsymbol{U}|_{\infty} = 1$, $A=0.3$, and $\alpha = 90^\circ$.}
%     \label{fig:GridGamma45-3}
% \end{figure}

% \begin{figure}
%     \centering
%     \includegraphics[width=\textwidth]{Gamma90GridIndependenceStudy1.pdf}
%     \caption{Comparison of the time series and the time average of the global heat transfer quantities, $\overline{Nu}$, $T_{\max}$, and $T_{\text{avg}}$ with different domain sizes and grid sizes when the oncoming flow is oblique at $\gamma = 90^\circ$. The other simulation parameters are $Re_f=100$, $V_{\infty} = 1$, $\alpha = 60^\circ$ and $A/|\bm{U}_{\infty}|=0.2$.}
%     \label{fig:GridGamma90-1}
% \end{figure}
%
%
\clearpage
\section{Pareto fronts for the fixed-heat-flux case \label{app:ParetoFixedFlux}}

\begin{figure}[ht!]
    \centering
    \begin{subfigure}[b]{\textwidth}
        \centering
        \caption{}
        \includegraphics[width=0.8\textwidth]{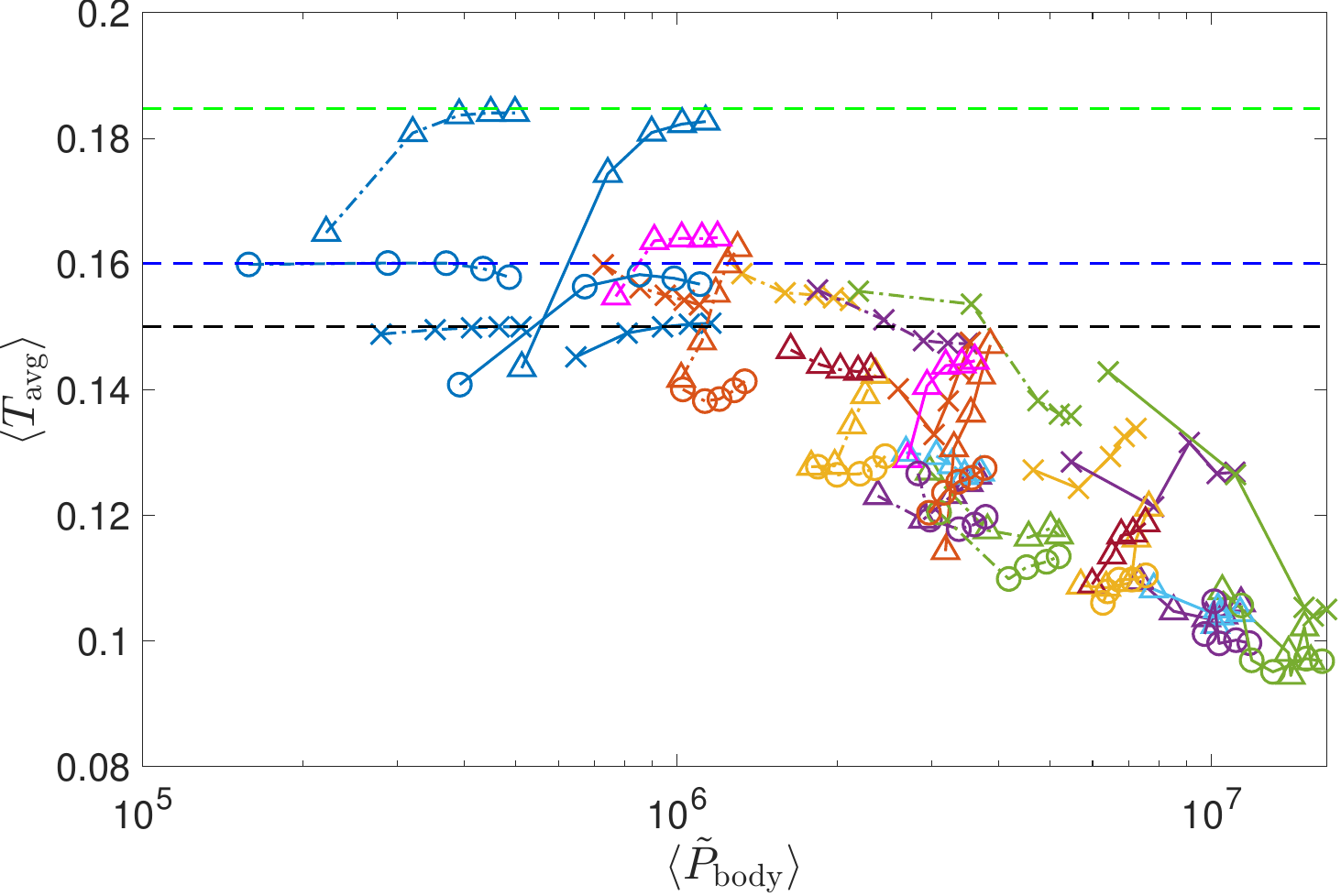}
    \end{subfigure}
    \begin{subfigure}[b]{\textwidth}
        \centering
        \caption{}
        \includegraphics[width=0.8\textwidth]{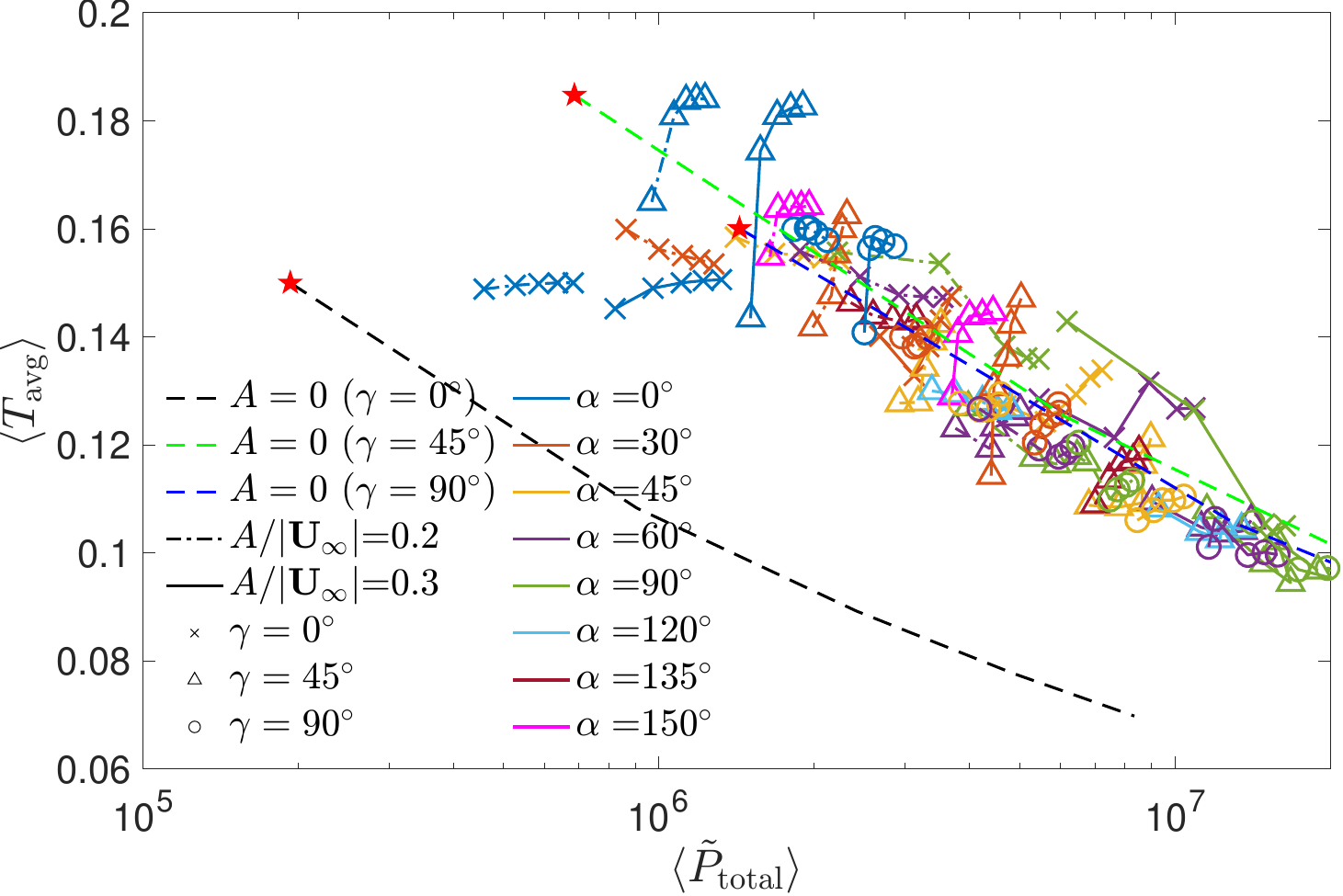}
    \end{subfigure}
    \caption{$\langle T_{\text{avg}} \rangle$ versus the time averaged power needed to oscillate the plate (a), and the time averaged total power including the power needed to drive the oncoming flow (b). The three horizontal dashed lines in (a) mark $\langle T_{\text{avg}} \rangle$ for the steady plate. The three dashed curves in (b) show $\langle T_{\text{avg}} \rangle$ versus $\langle \tilde{P}_{\text{total}} \rangle$ for the steady plate. The values with $Re_U = 100$ are marked with red stars.}
    \label{fig:PowerBodyTavg}
\end{figure}

\begin{figure}[t!]
    \centering
    \begin{subfigure}[b]{\textwidth}
        \centering
        \caption{}
        \includegraphics[width=0.8\textwidth]{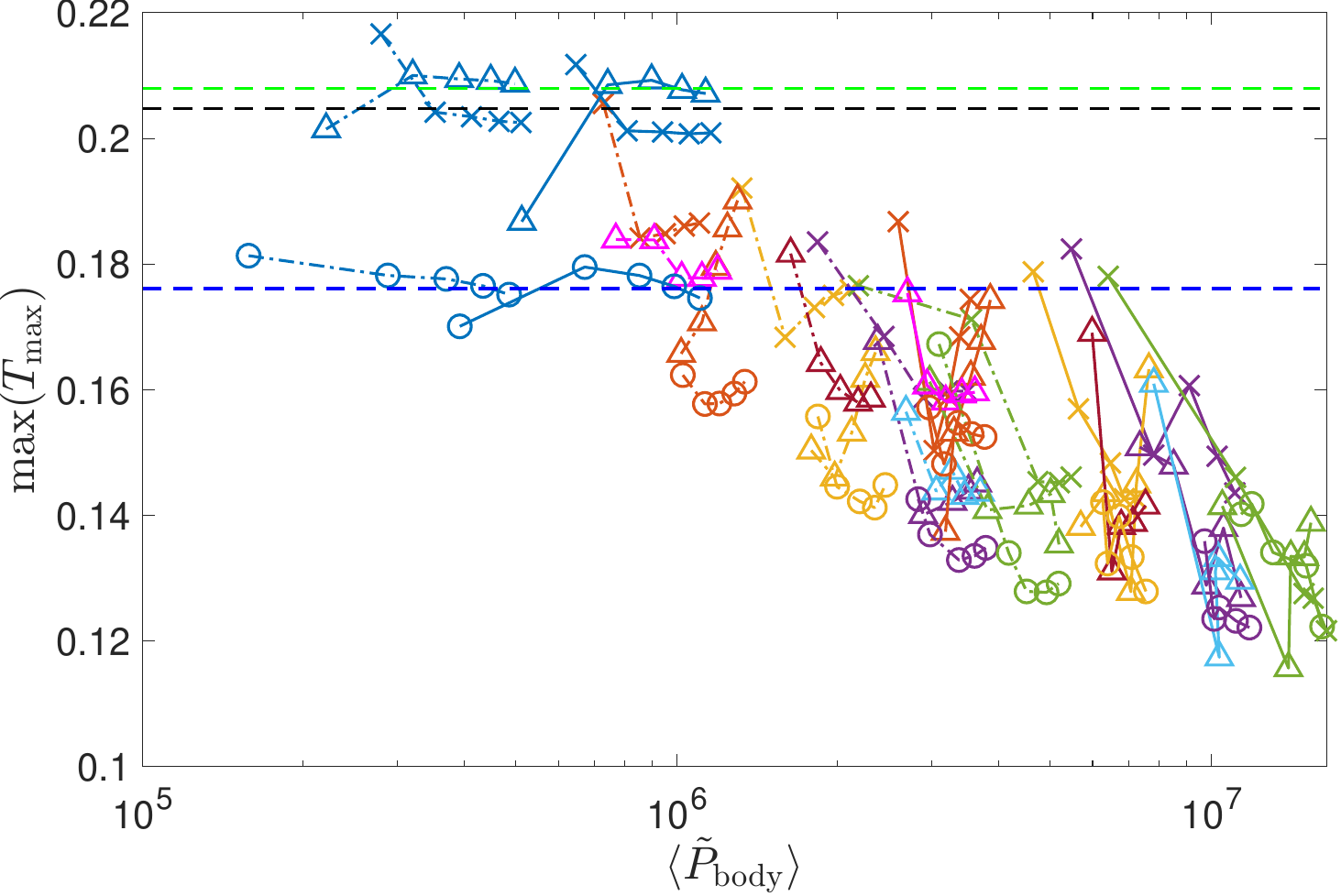}
    \end{subfigure}
    \begin{subfigure}[b]{\textwidth}
        \centering
        \caption{}
        \includegraphics[width=0.8\textwidth]{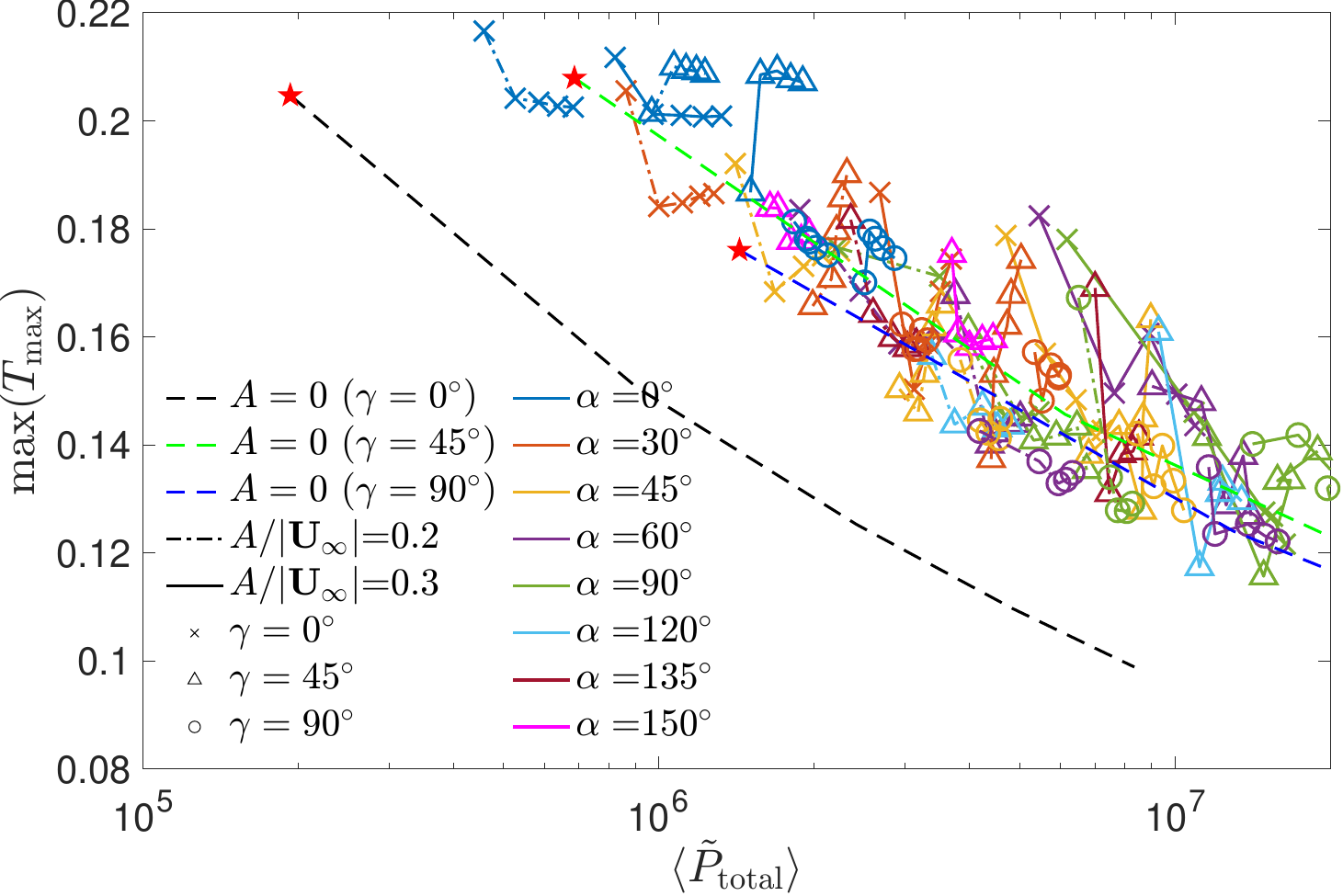}
    \end{subfigure}
    \caption{$\max(T_{\max})$ versus the time averaged power needed to oscillate the plate (a), and the time averaged total power including the power needed to drive the oncoming flow (b). The three horizontal dashed lines in (a) mark $\max(T_{\max})$ for the steady plate. The three dashed curves in (b) show $\max(T_{\max})$ versus $\langle \tilde{P}_{\text{total}} \rangle$ for the steady plate. The values with $Re_U = 100$ are marked with red stars.}
    \label{fig:PowerBodyTmax}
\end{figure}

In this appendix we present $\langle T_{\text{avg}} \rangle$ versus the input powers (figure~\ref{fig:PowerBodyTavg}) and $\max(T_{\max})$ versus the input powers (figure~\ref{fig:PowerBodyTmax}) for the fixed-heat-flux case. For both figures, a case is Pareto optimal with respect to another case if it lies below and to the left of it. The Pareto optimal cases are those that minimize $\langle T_{\text{avg}} \rangle$ or $\max(T_{\max})$ at a given input power, or minimize the input power at a given $\langle T_{\text{avg}} \rangle$ or $\max(T_{\max})$. 

If the oncoming flow is free, the conclusions from figure~\ref{fig:PowerBodyTavg}(a) are similar to those with fixed temperature, as $\langle T_{\text{avg}} \rangle$ is inversely related to $\langle \overline{Nu} \rangle$: the optimal cases have $\gamma=45^\circ$ or $\gamma=90^\circ$. As $\langle T_{\text{avg}} \rangle$ decreases along the Pareto front, the optimal cases transition from $A/|\boldsymbol{U}_{\infty}|=0.2$ to $A/|\boldsymbol{U}_{\infty}|=0.3$. These observations also apply for $\max(T_{\max})$ versus $\langle P_{body} \rangle$, as shown in figure~\ref{fig:PowerBodyTmax}(a). Here, many of the optimal cases correspond to $\gamma=90^\circ$. The reason might be that the transverse flow sets no preferences for vortex grouping at the two edges, which leads to a relatively even temperature distribution, lowering the temperature at the hottest spot.

If the total power in equation~\eqref{eq:Ptotal} is considered, the Pareto frontier is again the steady plate with $\gamma=0^\circ$, as shown in panel (b) of figure~\ref{fig:PowerBodyTavg} and \ref{fig:PowerBodyTmax}. Again, the oscillating plates at low $Re_f$ with $\gamma=45^\circ$ or $\gamma=90^\circ$ outperform the steady plates at $\gamma=45^\circ$ or $\gamma=90^\circ$, but they still consume much more energy than the steady plate at $\gamma=0^\circ$ for given $\langle T_{\text{avg}} \rangle$ or $\max(T_{\max})$.

%\bibliography{references.bib}
% \bibliographystyle{plain}

\end{document}